\documentclass[twocolumn]{aastex631}

\newcommand\aastex{AAS\TeX}
\newcommand\latex{La\TeX}

\newcommand{\ld}{\mathrm{\ell_d}}
\newcommand{\etal} { et~al.\ }  

\shorttitle{Turbulence in Stars}
\shortauthors{Arnett et al.}
\graphicspath{{./}{figures/}}
\begin{document}

\title{3D simulations and MLT: II. Onsager's Ideal Turbulence}

\correspondingauthor{W. David Arnett}

%\author[0000-0002-0786-7307]{Greg J. Schwarz}
%\affiliation{American Astronomical Society \\
%1667 K Street NW, Suite 800 \\
%Washington, DC 20006, USA}
%
%\author{August Muench}
%\affiliation{American Astronomical Society \\
%1667 K Street NW, Suite 800 \\
%Washington, DC 20006, USA}
%
%\collaboration{6}{(AAS Journals Data Editors)}
%
%\author{Butler Burton}
%\affiliation{Leiden University}
%\affiliation{AAS Journals Associate Editor-in-Chief}
%
%\author{Amy Hendrickson}
%\altaffiliation{AASTeX v6+ programmer}
%\affiliation{TeXnology Inc.}
%
%\author{Julie Steffen}
%\affiliation{AAS Director of Publishing}
%\affiliation{American Astronomical Society \\
%1667 K Street NW, Suite 800 \\
%Washington, DC 20006, USA}
%
%\author{Magaret Donnelly}
%\affiliation{IOP Publishing, Washington, DC 20005}
%

\author{W. David Arnett}
\affiliation{Steward Observatory, 
University of Arizona,
933 N. Cherry Avenue,
 Tucson AZ 85721 
}
\author{Raphael Hirschi}
\affiliation{Astrophysics Group, Keele University, Lennard-­Jones Laboratories, Keele, ST5 5BG, UK}
\affiliation{Kavli IPMU (WPI), The University of Tokyo, Kashiwa, Chiba 277­ 8583, Japan}
\author{Simon W. Campbell}
\affiliation{School of Physics and Astronomy, Monash University, Clayton, Australia 3800}
\author{Miroslav Moc\'ak}
\affiliation{School of Physics and Astronomy, Monash University, Clayton, Australia 3800}
\author{Cyril Georgy}
\affiliation{Department of Astronomy, University of Geneva, Ch. Maillettes 51, 1290 Versoix, Switzerland}
\author{Casey Meakin}
\affiliation{Meakin Technologies, Pasadena, CA 91104}
%\collaboration{3}{(Keeke collaboration)}
\author{Andrea Cristini}
\author{Laura J. A. Scott}
\author{Etienne A. Kaiser}
\author{Maxime Viallet}

\email{wdarnett@gmail.com,r.hirschi@keele.ac.uk,simwcampbell@gmail.com, miroslav.mocak@gmail.com,cyril.georgy@unige.ch,
casey.meakin@gmail.com,andrea.cristini89.gmail.com,
l.j.a.scott@keele.ac.uk,e.kaiser@keele.ac.uk,viallet.maxime@gmail.com
}

%% Note that the \and command from previous versions of AASTeX is now
%% depreciated in this version as it is no longer necessary. AASTeX 
%% automatically takes care of all commas and "and"s between authors names.

%% AASTeX 6.31 has the new \collaboration and \nocollaboration commands to
%% provide the collaboration status of a group of authors. These commands 
%% can be used either before or after the list of corresponding authors. The
%% argument for \collaboration is the collaboration identifier. Authors are
%% encouraged to surround collaboration identifiers with ()s. The 
%% \nocollaboration command takes no argument and exists to indicate that
%% the nearby authors are not part of surrounding collaborations.

%% Mark off the abstract in the ``abstract'' environment. 
\begin{abstract}

 We simulate convective turbulence in stars, extending  \cite{alvio1}.  
 Our implicit large eddy simulations (ILES) use the 3D Euler equations with shock capturing \citep{cw84}; we simulate an astrophysically-appropriate inviscid limit (Re$\lesssim$7000)  with causal time stepping
 but no explicit viscosity.   Anomalous dissipation of turbuent kinetic energy occurs as an emergent feature of advection   (``Onsager damping''), a physical process caused by the moderate shocks, 
which terminate the turbulent kinetic energy spectrum.   This differs from incompressible turbulence in which
nonlinear fluid effects involving vorticity are supposed to control dissipation \citep{gitaylor,perry}.
 In strongly stratified stellar convection the asymptotic limit  for the global damping length of turbulent kinetic energy is  $\ell_d \sim \langle u^3 \rangle /\langle \epsilon \rangle$. This ``dissipative anomaly'' \citep{onsager}
fixes the value of the ``mixing length parameter", 
$\alpha = \ell_{\rm MLT}/H_P =\overline{\langle\Gamma_1\rangle}$, which is $\sim\, 5/3$ for complete ionization.  The estimate is numerically robust, agrees to within $\sim$10\% with estimates from stellar evolution with constant $\alpha$. For  weak stratification $\ell_d$ shrinks to the depth of a  thin convective region.
For stellar atmospheres computed  with ``hyperviscosity stabilization'' and variable $\Gamma_1$ in regions of partial ionization, the agreement is  $\sim$20\% \citep{magic2016}.    Our ILES are filamentary, produce surfaces of separation at boundary layers, resolve the energy-containing eddies, and develop a turbulent cascade down to the   grid scale (which agrees with the $4096^3$ direct numerical simulations (DNS) of \citealt{kaneda2003}).  The cascade converges quickly \citep{onsager}, and satisfies a power-law velocity spectrum similar to   (\citealt{kolmg41}). 
Our ILES exhibit  intermittency,  anisotropy, and interactions between coherent structures, features missing from K41 theory.  We derive a dissipation rate from Reynolds stresses which agrees with (i)  our ILES,  (ii) experiments \citep{warhaft}, and  (iii) high Reynolds number DNS of the Navier-Stokes equations (\citealt{iyer4096,sree}).  
 
\end{abstract}

%% Keywords should appear after the \end{abstract} command. 
%% The AAS Journals now uses Unified Astronomy Thesaurus concepts:
%% https://astrothesaurus.org
%% You will be asked to selected these concepts during the submission process
%% but this old "keyword" functionality is maintained in case authors want
%% to include these concepts in their preprints.
%\keywords{Classical Novae (251) --- Ultraviolet astronomy(1736) --- History of astronomy(1868) --- Interdisciplinary astronomy(804)}

%% From the front matter, we move on to the body of the paper.
%% Sections are demarcated by \section and \subsection, respectively.
%% Observe the use of the LaTeX \label
%% command after the \subsection to give a symbolic KEY to the
%% subsection for cross-referencing in a \ref command.
%% You can use LaTeX's \ref and \label commands to keep track of
%% cross-references to sections, equations, tables, and figures.
%% That way, if you change the order of any elements, LaTeX will
%% automatically renumber them.
%%
%% We recommend that authors also use the natbib \citep
%% and \citet commands to identify citations.  The citations are
%% tied to the reference list via symbolic KEYs. The KEY corresponds
%% to the KEY in the \bibitem in the reference list below. 

\section{Introduction} \label{s-intro}

We continue our study of stellar convection (Paper I, \citealt{alvio1}). 
We solve the {\em three-dimensional (3D), compressible Euler equations} (conservation of mass, momentum, and energy; see \S2,   Appendix \S\ref{a-E-NS}; \citealt{llfm}),  with a finite volume scheme and using the piece-wise parabolic method (PPM) for shock capture \citep{cw84}. No viscosity is assumed; these astrophysical simulations are inviscid.
Minimal assumptions are made concerning flow patterns or boundary layers.
These implicit large eddy simulation (ILES) methods were originally developed for supernova and pre-supernova simulations \citep{fma89,ba98}, with realistic microphysics for stellar interiors. 

 \cite{bv58} used the ``mixing length" theory of Prandtl \citep{anderson,biermann25}, with an adjustable parameter 
to  estimate dissipation, to close the theory of stellar convection; also see \S\ref{BVtheory}. 
When ILES \citep{amy09vel} were  applied to this problem 
a 3D cascade of turbulent energy developed automatically, exhibiting damping in agreement with  \cite{kolmg41},  but with {\em no viscous dissipation defined, and no adjustable parameter needed}.   This  curious result was due to an ``energy-dissipation anomaly'' predicted by \cite{onsager}, which constrains all convection at high Reynolds number (Re).  This is Onsager's ``ideal'' turbulence.

The standard approach to turbulence uses the Navier-Stokes (NS) equations, which were designed for laboratory-scale fluid flow (\citealt{batchelor60,llfm} \S15). Like the Euler equations they enforce conservation of mass, momentum, and energy, but with two added assumptions:  incompressibility and  molecular dissipation. 
Both can fail for stars, where large sizes and density stratification occur.  %, and vigorous events occur. 

Turbulent flows are classified by a dimensionless ratio, the Reynolds number, Re$= \ell \Delta u /\nu_{\rm NS} $, which is the ratio of the inertial term to the dissipation term in the NS equation.
Turbulence increases with Re.
The characteristic length $\ell$  (the ``integral scale'', or the size of the largest eddies) and the characteristic turbulent velocity $\Delta u$ are macroscopic properties, while the Navier-Stokes (NS) viscosity $\nu_{\rm NS}$ is a measure of molecular dissipation  \citep{pope}, a microscopic quantity.  At large scale, nonlinear coherent processes can dominate local, collisional ones. 
As the inviscid limit is approached ($\nu_{\rm NS} \rightarrow 0$),  the NS equations become singular (as $1/\nu_{\rm NS} $, requiring steep velocity gradients), but not the Euler equations. 
For solar conditions, we have {Re\ $\gtrsim 10^{14}$} (\citealt{hanasoge-gizon}),  which implies a tiny effective viscosity.  A typical laboratory Re {\em would be far 
too small (too viscous) to be realistic for stars,} which are almost ``inviscid''.

We find that {\em dissipation occurs as an emergent feature of nonlinear fluid dynamics (a ``dissipative anomaly")},  as we approach  stellar conditions.     
As this limit is approached (\S\ref{blowup}; \citealt{eyink-comp}),  turbulent dissipation 
is dominated by  mild  shocks related to the turbulence \citep{hoffman,perry}.

 Our ILES 
develop a cascade in turbulent kinetic energy  which is an extension of the incompressible case \citep{gitaylor,batchelor60}.
Experiments (\citealt{warhaft}) and  direct numerical simulation (DNS, \citealt{sree,iyer4096})  have shown that anomalous features are essential to real turbulence: intermittency, coherent structures,  
and a fluctuating turbulent cascade.
We find that these develop automatically in our ILES through the nonlinear advection term.
Similar conclusions apply in principle to the NS equations (\citealt{dls13,buckmaster}; see also  
\citealt{eyink-comp}, their  Appendix A.2.c, who use  a completely different approach involving renormalization-group invariance). 

In \S2 we examine the theoretical approaches used in (i) our ILES of  the Euler equations, and (ii) large eddy simulations (LES) and (iii) DNS of the NS equations.  
We estimate the effective Reynolds numbers (Re) we attain, and compare specific turbulent kinetic energy (TKE) spectra from simulations. 
 
In \S3 we extract from our ILES the flux of turbulent kinetic energy (TKE) in our turbulent cascade, the ``dissipative anomaly''. 
 We derive an {\em analytic} rate of dissipation from a Reynolds stress, 
which agrees quantitatively with the simulations, and with the conventional  dissipation in a turbulent cascade \citep{kolmg41}. This ``dissipative anomaly'' is a dynamic constraint on convective flow, requiring a fixed value for the average damping length. For a given driving, turbulence decays (grows) until balanced on average. 
We examine intermittency due to this approximate balance in TKE around a quasi-steady state. 

 In \S4 we summarize resolution studies \citep{andrea,321D} for the case of  weakly stratified convection for  shell burning of O and of C, both cooled by neutrino pair emission.  The simulations develop a dissipation length  which agrees  with the Reynolds stress value derived in \S3,  in \cite{amy09vel}, and is insensitive to resolution error.

In \S5 we examine  strong stratification, 
and find significant acceleration by pressure dilatation (see also \citealt{eyink-comp}). 
Sufficiently large stratification 
leads to a significant negative flux of TKE, unlike MLT.  We derive an asymptotic value for the TKE dissipation length in strong stratificationt  at high Re, independent of the dissipation mechanism.

In \S6 we discuss astronomical calibration of MLT, 
from consistency of stellar evolution with the color-magnitude %Hertzsprung-Russell 
diagram, and conclude that {\em the apparent ``universality'' of the mixing-length  \citep{alvio87} is a consequence of anomalous dissipation in deeply stratified convection, and our predicted value of $\alpha_{\mathrm MLT}=\ell_{\mathrm MLT}/H_P = \overline{\langle\Gamma_1\rangle} \approx 5/3$ fits the data surprisingly well. }
 Incomplete ionization and dissociation cause deviations in surface structure and 
$\overline{\langle\Gamma_1\rangle}$; these effects are estimated to be $\pm20$\% from simulations stabilized by hyperviscosity \citep{magic2016}.  
 
In \S7 we summarize the implications from this and the previous paper \citep{alvio1}.

\section{Methods}\label{s-methods}
 Among fluid dynamic algorithms for turbulent flow, a key difference lies in the treatment of dissipation.

\subsection{Euler equations and ILES}\label{s-euler}
Our library\footnote{See \cite{ma07a,ma07b,amy09vel,viallet2013,miro2014,321D,andrea,andrea2,alvio1}, and references therein.} of simulations, previously interpreted within the context of  \cite{kolmg41}, have a better  interpretation, which we have found 
in response to reviews regarding ``anomalous behavior'' (\citealt{warhaft,sree}). 
We consider the surprising  independence of TKE 
dissipation  on viscosity \citep{onsager,eyink}. If the  viscous coeficient $\nu_{\rm NS}$ is set to zero, 
reducing the NS equations to the Euler equations, 
dissipation still occurs, but for physical reasons associated with the nonlinear advection term, not numerical error (\S\ref{STKE}). The reason is related to what mathematicians call the ``Onsager conjecture'' \citep{dls13,buckmaster,isett}.

\subsubsection{New results: Onsager's conjecture}\label{sLars}

 In his note on statistical hydrodynamics, \cite{onsager} considered weak solutions\footnote{ ``Weak solutions'' allow surfaces of discontinuity, such as shocks.}
 of velocity $u$ in the fluid equations, satisfying the H\"older condition
$| u(x, t) - u(x', t)| \le C|x - x'| ^{\theta}, $ where the constant $C$ is independent of $x,x'$. He conjectured
that:
(a) any weak solution $u$ satisfying this with $\theta > {1 \over 3}$ conserves the turbulent kinetic energy, and
(b) for any $\theta$ $\le {1 \over 3}$ satisfying this, there exist weak solutions $u$ which do not conserve this energy. 
 
 {\em This implies a definite rate for dissipation of turbulent kinetic energy in a region with a ``rough'' (i.e., turbulent) velocity field.  This ``anomalous dissipation'' is independent of viscosity, being a property of turbulence not microphysics (an ``Onsager dissipation''). }
 
Further work supports this conjecture (e.g., \citealt{constantin,duchon,dls13,eyink-2018,eyink-comp,eyink-rel,buckmaster,isett}), as do our ILES (Fig.~3, in \citealt{321D}).
The precise mechanism of the turbulent energy cascade is becoming clearer. The \cite{gitaylor} suggestion of ``vortex-stretching'' (also favored by \citealt{onsager}), is partially correct, but incomplete. \cite{perry} finds that ``strain-rate self-amplication'' (the steepening of compressive strain-rates via nonlinear self-advection, i.e., shocks) also contributes and may dominate the inter-scale transfer of energy in turbulence.
  
Onsager's prediction of the non-conservation of turbulent kinetic energy (TKE)
is striking:  in our ILES, turbulent dissipation occurs 
by mild shocks\footnote{The shock jump conditions provide a dissipation independent of viscosity.}
, which convert TKE into internal energy 
 (\S\ref{cascade}). 
{\em For a given driving, the flow depends upon dissipation which in turn depends implicitly on the flow itself. }
 This is a statistical constraint with negative feedback and time delay, reflecting both the nonlocality of the process and its connection to intermittency (see e.g., \S\ref{BVtheory}, Eq. \ref{mlt}). We find that no explicit term need be added to the Euler equations; this damping is provided directly by mild shocks where the velocity field is ``rough'' (turbulent).

  \cite{iles07} stressed the importance of shocks in ILES. 
At the zone level, we solve the Riemann problem by a shock-capturing algorithm, 
the piecewise parabolic method (PPM) of \citealt{cw84,woodward,leveque})\footnote{The pioneering work by Paul Woodward and his collaborators is not widely known in the solar literature; see however \cite{porterRG,porterwoodward,porterILES,woodward,sytine,woodward09}. We thank Woodward  for providing access to unpublished results which preceded and paralleled ours.}.  

The ``effective numerical viscosity'' of PPM has been shown  (\citealt{sytine,woodward,porterILES,kritsuk}) to be much smaller, for the same number of zones, than for a pseudo-viscous method or a direct numerical simulation  (DNS); i.e., $\sim$4 times higher Reynolds number.\footnote{Most dissipation in PPM for Euler equations is provided by ``flattening'', to eliminate oscilation behind shocks (monotonicity). There can still be small errors behind some strong shocks (``$\sim$0.03 in density, which disappear downstream''). A small amount of additional dissipation may then be  needed,  $\leq$0.1 of that required for ``stabilization".  In this sense our ILES are ``inviscid''.  See \cite{cw84}.  }
DNS and our ILES  converge to the same limit with zone refinement \citep{sytine}. 
Had it been available when we began,  the piecewise parabolic boltzmann scheme (PPB)  might have allowed an additional factor of 3 reduction in numerical dissipation \citep{ppb}. However, the insensitivity of our overall dissipation to our zoning suggests that this would make no qualitative difference in the results reported here. 
 
\subsubsection{Emergent dissipation and ``Blowup"}\label{blowup}
 Viscosity smooths gradients of velocity; 
 shocks have an opposite tendency. Shocks imply  discontinuities, 
a consequence of the nonlinearity of the compressible fluid dynamics equations \citep{llfm}. Shocks convert some of their kinetic energy into internal energy, a dissipation process.  This increases the sound speed behind the front,  so that pressure waves can overtake it, accumulate, and steepen the gradients there.  While the internal structure of the front is affected by the dissipation mechanism,  the jump conditions and total dissipation are not.

As the inviscid limit  is approached in a compressible turbulent medium, viscosity is less able to smooth gradients, so that shocks develop. 
Formally, shocks have infinite gradients of velocity
at the front; these could be smoothed by viscosity, real or numerical, or replaced by the shock jump conditions. 
The Euler equations can develop singularities in a finite time, a phenomena that is sometimes termed ``blowup''  (\citealt{frisch,hoffman,benzi}). 
Our numerical version of the Euler equations also develops singularities (shocks).  Shock capture
 could be approximated by an appropriate  laplacian viscosity term if desired. However our method gives a physically correct dissipation at the grid level (e.g., \S\ref{tkespectrum}, \S\ref{kolmog-damping}, \S\ref{Ssubgrid}), with fewer zones required.

As a gradient in velocity grows, PPM
replaces the developing discontinuities by a nonlinear solution of the Riemann problem at the grid level,
regularizing the simulations, and providing local support. Here ``local support'' implies monotonicity, i.e., no ringing at discontinuities (Gibbs phenomena); such errors plague both Fourier and high order polynomial methods of fluid dynamics. 

To resolve dissipation in the incompressible NS equations, the computational zone size is constrained by 
viscosity \citep{pope}. 
For compressible flow, the Euler equations support shocks, so dissipation can occur by microscopic processes in the shock front, where gradients are large
but the width of the shock can be small. \cite{perry,perry2} has shown that strain-rate self-amplification (shocks) are a major mechanism of energy transfer from large to small scales in turbulence.
Replacing a shock by jump conditions allows us to {\em capture dissipation without resolving shock structure,} avoiding the linear size restriction of DNS for high Re \citep{pope}.

At low viscosity (high Re), almost identical initial velocities can evolve on different paths, a behavior characteristic of turbulence. 
In our view this is not due to ``incorrect'' initial conditions, but to {\em insufficient dissipation 
to keep close trajectories from developing steep velocity gradients (shocks). }
A similar problem arises with a Fourier representation of the velocity field by a finite number of frequencies.

In our ILES, shock structure is replaced by a discontinuity between two zones, so in this sense our algorithm 
minimizes the number of zones required.  Microscopic details of shock structure are replaced by macroscopic discontinuities \citep{llfm}, allowing  {\em fluid dissipation with no explicit viscosity.}
This does not mean that microscopic dissipation does not happen (it does in the narrow shock front), only that {\em we need not resolve it. It is globally constrained by the Euler equations. }

 Even for mild convection (down  to Mach numbers as small as ~0.01, which we simulate),  shocks give dissipation  in our ILES. 
For  lower Mach numbers, which we do not simulate, other mechanisms are available  \citep{gitaylor,onsager,batchelor60,hoffman,perry}.   We do not attempt to determine this transition precisely. 
 Dissipation is not  simply a microscopic phenomenon, but an emergent feature of the nonlinear advection term   which is common to the Euler  and NS equations (\S\ref{STKE}). 
 
 Turbulence gives a multi-scale behaviour which allows
 shocks to  regularize our ILES.  This kind of ``coarse grained'' approach is appropriate to stellar physics. It allows us to use  ``generalized" Euler equations, with a dissipative anomaly for TKE \citep{onsager} to give the rate of energy flow in the turbulent cascade.  
 Dissipation results from the Euler equations themselves, and can be described by Reynolds stresses generated by flows associated with the advection term (\S\ref{sReystress}, \S\ref{Ssubgrid}).  
 
 Alternatively, \cite{eyink-comp,eyink-rel} have discussed these same points (and more) in a non-perturbative renormalization group analysis of the compressible NS equations, a very different viewpoint. Their more general  approach  gives interesting parallels \citep{eyink-2018} to our 
 discussion,  and the encouragement of no obvious contradictions.

\subsection{The Reynolds number}\label{s-setup}

Turbulent flows are classified by a dimensionles ratio, the Reynolds number, 
${\rm Re} = \ell \Delta u /\nu_{\rm NS} $.
Large $\ell$ gives large Re and extreme turbulence, provided that atomic-scale collisions are frequent enough for the fluid approximation to be valid, which is generally true in stars. If we use the Euler equations, for which $\nu \equiv \nu_{\rm NS}\equiv0$ so ${\rm Re}\rightarrow \infty $,  how do we define a Reynolds number?

\cite{llfm} in their  Eq.~32.6 express Re as
 \begin{equation}
{\rm Re} \sim   ( \ell / \lambda_0 )^{4 \over 3 },
\end{equation}
for $\ell \gg \lambda_0$,  so Re is a measure of the ratio of 
 the  integral length scale $\ell$ (largest eddies)  to  an internal scale $\lambda_0$ (smallest eddies as defined by  a dissipation scale); 
see also \cite{warhaft}.
 
 Shocks are a collective nonlinear behavior.
 With shocks the effective dissipation scale is the shock thickness.   
 The grid scale is the zone size $\Delta r$,  which is also the effective dissipation scale for the shock-capture algorithm in our ILES. If $ \Delta r \gg \lambda_0 $, shock capture allows dissipation to be realistically simulated for systems much larger than the shock thickness,  for the same computational Re constraint (number of zones).

{\em An effective Reynolds number ${\rm Re}^*$ can be determined by the available zoning,} rather than a microscopic viscosity.  {\rm Finer zoning allows smaller effective viscosities (i.e., reduces gradients) and so permits  larger Reynolds numbers. }
A general calibration of Re$^*$ relative to Re is nontrivial (\citealt{llfm}, \S32; \citealt{haugen2004}).
A useful approximation at large Re for our problem is
 \begin{equation}
  {\rm Re}^* \sim (n/2)^{4 \over 3}, \label{eq-Re}
 \end{equation}
 where $n/2$ is the number of computational cells across the turbulent domain in our $n^3$ simulation. This is easily evaluated for both Euler simulations (which have a prefactor of order unity) and NS simulations (which have a prefactor about 4 times smaller,  assuming that the NS algorithm uses 4 zones to deal with shocks). % in DNS). 
 
 As long as the physical shock thickness is much less than the zone size,  the shock jump conditions are a physically correct dissipation mechanism,  so this is a promising approach to fluid flow for large scales (large Re).
 
In our ILES the range of $n$ is $128$ to $1536$ so that the estimate of ${\rm Re}^*$ ranges from $256$ to  $7 \times 10^3$. The morphology of the simulations can be 
bracketed by photographs of experiments  which are identified with 
Reynolds numbers.  We approach the ``mixing transition'' at Re$\,\sim10^4$  of \cite{dimotakis}, who presents flows in this range  (his Fig. 1 gives Re of $1.75\times 10^3$ to $2.3 \times 10^4$, and Fig. 2 gives Re of $2.5 \times 10^3$ to $1.0 \times 10^4$). 
See also \cite{vandyke}. Thus {\em Re$\,\approx$\,Re$^*$ (for Re*$\gg 1$)}.\footnote{We find that the critical values of Re$^*$ for {\em onset of turbulence}  tend to be lower than Re for some common engineering cases, possibly because of fluid-fluid boundaries in stars,  or our use of the Euler equations.}
 
Early experiments established that  solutions to the NS equations possessed a Reynolds-number similarity:
if the values of Re are the same then the scaled flows are the same.
The Euler equations are invariant with respect to scale transformations in $\ell$,  $u$, and mass density $\rho$  \citep{pope}, and represent the inviscid limit of the NS equations.
  Having small size relative to stars, laboratory-scale turbulence is strongly affected by finite $\nu_{\rm NS}$ in the Navier-Stokes equations \citep{batchelor60}.

\subsubsection{ILES results}\label{S-ILES}

  Paper I \citep{alvio1} is accompanied  by movies
 of (i) the evolution in time (``Very high resolution movie of the C-shell''), and of (ii) a fly-through of the computations at a given instant in time (``Carbon shell ($1024^3$) simulation: fly-through movie")\footnote{The time evolution and the fly-through  movies may be found at   \url{http://www.astro.keele.ac.uk/shyne/321D/convection-and-convective-boundary-mixing/visualisations}.}.
The simulations are approaching a statistical steady state in time, and statistical homogeneity in space, 
as the similar visual appearance of the movie and the fly-through suggests.

\begin{figure*}[h]
\figurenum{1}
\label{fig34}
\includegraphics[totalheight=190pt,bb=-10 0 200 400]{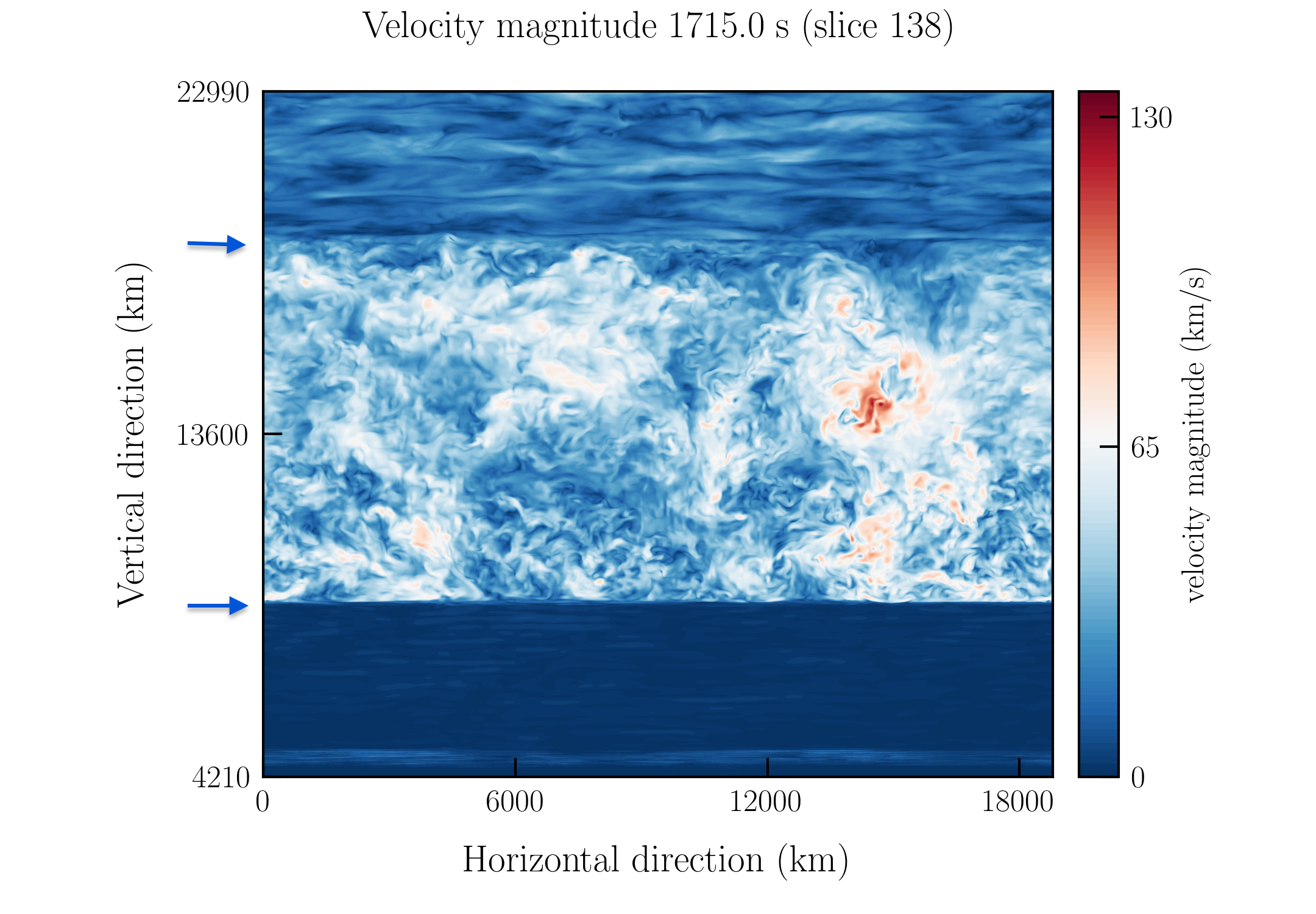}

\caption{
Vertical slice of velocity magnitude (Mach $\sim0.02$) from a compressible Euler simulation (using  PROMPI), of a C+C burning shell (after $1024^3$  case in \citealt{andrea2} and Fig.~1, \citealt{alvio1}).
According to Eq.~\ref{eq-Re}, this ${\rm Re}^*\sim 6000$
(compare to Fig.~2 in \citealt{dimotakis}). 
Weak pressure waves propagate, and steep gradients are captured by a Riemann solver (PPM, \S{\ref{S-ILES}}). This is a typical frame from the movie at time $t=1715$ s (slice 138).
Blue arrows at left axis point to sharply defined top and bottom boundary layers, which form naturally and support surface waves. Here Onsager's $\theta=1/3$ implies a singularity in the $u$ gradient.   The top and bottom edges of the grid are beyond the edges of the figure.  Wave motion is noticeable above the top boundary layer; 
a distant damping layer prevents reflection of waves back into the convective region. The velocity field shows filamentary fine structure (``roughness") due to the turbulent cascade, as well as coherent structures (rolls and plumes) which form and dissolve. Intermittency is suggested in this snapshot, and is obvious in the movies.   On average smaller scale structures are weaker and evolve faster, but occasionally large scale structures are torn apart by powerful mutual interactions. 
}
\end{figure*}

\placefigure{1}

Our ILES allow us to separately monitor the conservation of turbulent kinetic energy (TKE), which is not conserved (it dissipates to become internal energy), and the conservation of total energy, which is almost exactly conserved in our simulations.  
This allows us to directly measure dissipation due to nonlinear fluid effects \citep{perry}.
Multiple runs with different grids 
allow us to probe the dissipation rate and the TKE spectra with changing resolution.

Fig.~\ref{fig34} shows the complex filamentary structure that develops in ILES of the turbulent velocity field \citep{andrea,andrea2}; the magnitude of velocity is shown for a cross-section of the C+C burning shell. This is  frame 138 from the movie; 
these contours of velocity magnitude show ``roughness''    
 \citep{tennekes,onsager}. 
Turbulence appears in patches in space-time, a feature associated with an anomalous behavior:  intermittency.

The initial state was a relaxed model mapped from a $512^3$ onto a $1024^3$ grid. The first steps at increased resolution show changes at the smallest scales first, as expected, which then work up to the largest scales. We find that the ``texture'' of the velocity field becomes {\em finer rather than smeared,} and  above the grid scale sharp gradients are maintained ($\theta \leq {1 \over 3}$, \S\ref{sLars}).

The top and bottom regions support waves and are in a quasi-steady state.
At the fluid boundaries these waves  connect to nonlinear solutions of the Riemann problem, giving monotonic support for shocks and rarefactions (H\"older continuity). These arise naturally in high Re turbulence
($\theta \leq {1 \over 3}$, middle layer). 
We have  horizontal wave-damping layers at the distant edge of the grid \citep{ma07b} to avoid reflections. Similarly, radial waves are not reflected. The bottom layer is also a domain where waves dominate, but its higher density gives lower amplitude waves, which are barely visible with this choice of color palatte. The wave domains and turbulent regions are separated by thin boundary layers which develop naturally (blue arrows in Fig.~\ref{fig34}).

Our sequence of ILES have changes only in resolution \citep{andrea,alvio1}.
Like \cite{onsager}, we find a rapid convergence to a cascade. 
The energy is contained by 
a few modes of low-order.\footnote{For the O+O shell (\S\ref{O+Oshell}), the lowest mode has $\sim 45$~percent of the turbulent energy, with $\sim 75$~percent in lowest 4 modes, during the 400 s examined by \cite{ma07b}. 
The C+C shell (\S\ref{C+Cshell}) is simpler, with no significant ingestion of a new fuel and the lowest 3  modes dominate. \label{fnote1}}
 These results agree with
\cite{holmes}, who stress the importance of low order modes 
to the energy-containing scales of turbulence  (see their Chapter~3).

 \subsection{Navier-Stokes methods}
 
\subsubsection{DNS}\label{Sdns}
\begin{figure*}[h]
\figurenum{2}
\label{fig33}
\includegraphics[angle=0,scale=0.6,bb=-100 0 0 670]{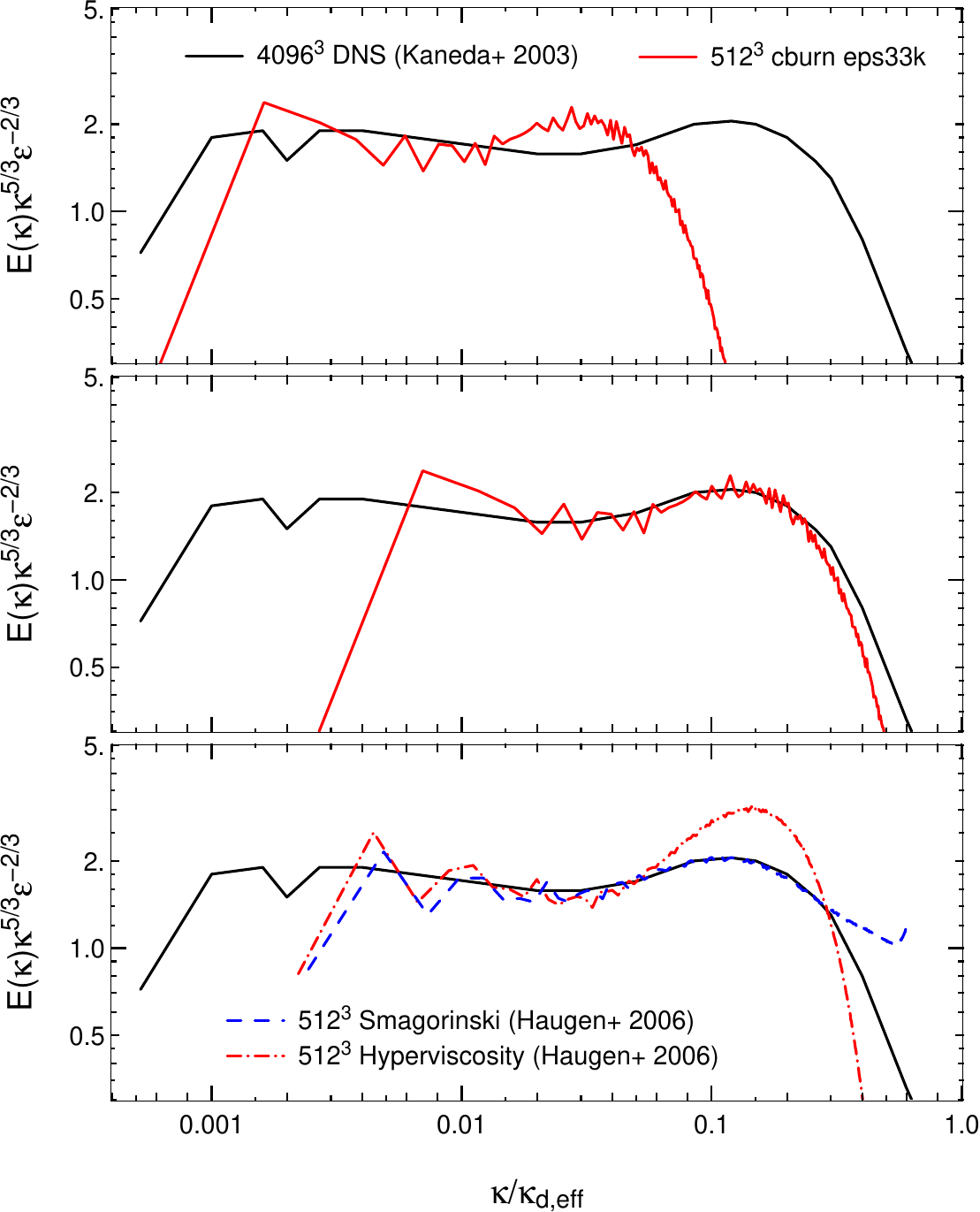}

\caption{
Cascades:  spectra of specific kinetic energy (after Fig.~1, \citealt{haugen2006}). {\bf (Top)} DNS showing a reference solution ($4096^3$, Re=1201, NS equations) from the Earth simulator ( solid line, \citealt{kaneda2003}), 
and  Euler ILES (PROMPI, \citealt{andrea2}, Fig.~C1, with a $512^3$ grid,  ${\rm Re}^*\sim1600$ from Eq.~\ref{eq-Re}).
 The spectra are ``compensated" so that the inertial range is horizontal. The integral scale (the energy-containing eddies) lies at the left end  of the curves, and the dissipation range at the right end. The DNS and the {\bf (red)} eps33k ILES (high luminosity boost) spectra are in a quasi-steady state. 
 The ``molecular'' dissipation scale is the Kolmogorov scale $\eta$, (where $\eta^4=\nu_{\rm NS}^3/\epsilon$). For stars this is small, formally lying far beyond the right end of the ILES curve, so that  
 {\em physical dissipation is dominated by shocks, independent of viscous parameters}, adjusting to whatever heating rate is imposed (\S\ref{kolmog-damping}). The effective dissipation scale is resolved; it is much larger than the width of the simulated shock, i.e., the width of a zone.  For the converged DNS the dissipation scale is also resolved, but using the NS viscosity, which could be reasonable for laboratory scales. 
{\bf (Middle)} Same as ({\bf Top}), except ILES is shifted to the right so the dissipation ranges may be compared.
The solid curve in the dissipation region of the benchmark DNS \citep{kaneda2003}  matches our ILES \citep{andrea2} surprisingly well, despite very different physical mechanisms and mildly different Re.
DNS and ILES seem to converge to the same limit for high Re; see also \cite{porterILES}. The ILES dissipation is insensitive to zoning, from grids of $128^3$ to $1024^3$.
The ILES inertial range lengthens with finer zoning.
{\bf (Bottom)} Same as ({\bf Middle}), but replacing the (red) ILES spectrum with two spectra from \cite{haugen2006} 
of $512^3$  LES spectra of the NS equations, which are also shifted to the right.  The hyperviscosity (red dash-dotted line) and the Smagorinsky viscosity (blue dashed line) are normalized to the reference DNS by 1.1 and 0.95,  corresponding to stronger and weaker dissipation than in the converged DNS reference (black solid line) or the ILES above. 
Hyperviscosity gives a distorted peak (the ``bottleneck'') and an abruptly truncated spectrum, while the Smagorinsky pseudo-viscosity gives errors when approaching the grid scales. DNS and ILES are similar in decrease at high wave-number.
}
\end{figure*}

In contrast to ILES,\footnote{Our ILES grid need extend only from the integral scale $\ell$ down to a size 
where the total dissipation by  shocks in the turbulent cascade is captured (our grid scale $\Delta r$),
not to a Kolmogorov molecular dissipation scale $\eta$ as usually assumed in DNS (see Fig.~\ref{fig33}). 
  This requires no changes in our code, only in our interpretation. This extends the incompressible picture of the cascade \citep{gitaylor,batchelor60} to compressible flow.}
direct numerical simulation (DNS) uses the NS equations 
and attempts to simulate all relevant scales of that flow; it is the standard against which other methods are compared in the fluid dynamics and applied mathematics communities.

DNS usually involves incompressible flow and molecular dissipation. For astrophysical scales these assumptions become untenable.  
The ratio of the NS dissipation to the other (Euler) terms scales as 1/Re;  see \S\ref{a-E-NS}.
For large Re, other (nonlinear) dissipation mechanisms will dominate (\S\ref{blowup}); this is %generally 
the case for stars. The standard theoretical framework (e.g., based on \citealt{batchelor60,pope}), requires modification in  this limit. 
For standard DNS of turbulence, most  of the computational effort 
 ($>$99\% according to \cite{pope}, p.~357, \S9) is used to resolve the small scales, i.e., 
 the Kolmogorov dissipation scale \citep{tennekes,pope}, $\eta = (\nu_{\rm NS}^3/ \epsilon)^{1 \over 4 }$, where 
 $\nu_{\rm NS}$ is the kinematic viscosity and $\epsilon$ is the effective rate of turbulent dissipation per unit mass. 
 This leaves correspondingly few computational resources for  the large scales, which contain most of the energy and drive the flow, or for longer temporal evolution, which is vital for statistics.
 Despite these constraints, high resolution DNS are a useful theoretical check for astrophysical problems.  

  Experiments \citep{warhaft} and high resolution DNS  \citep{sree,falkovich} demand features beyond those in the standard picture of \cite{kolmg41}.
  The DNS of \cite{iyer4096}, have a Taylor-scale Re of 650 using a $4096^3$ grid, and show a ``ramp-cliff'' structure in turbulent velocity $u$, 
  with ramps being relatively placid regions between cliffs.  Following %a suggestion by 
  \cite{iyer4096}, who find that ``the scalar intermittency is dominated by the most singular shock-like cliffs in the scalar field,'' {\em we interpret these cliffs as shocks} (\S\ref{blowup}).   A strict DNS approach would require that these shocks be resolved after viscous spreading on the grid, taking multiple zones per shock, as did \cite{iyer4096}.

\placefigure{2}

\subsubsection{LES} \label{LES}
 A less demanding class of computational options, the large eddy simulations (LES) of the NS equations\footnote{We follow the terminology of \cite{pope}.}, do not attempt to resolve the details of dissipation. 
 For a given number of zones this allows  larger length scales to be simulated, at the expense of  approximation errors in  the dissipation rate and turbulent cascade (\S\ref{tkespectrum}).
  
 We discuss only two of the many options: (i) a Smagorinsky pseudo-viscosity for shock capture \citep{haugen2006,pope}, and (ii) hyperviscosity stabilization \citep{haugen2006,kritsuk,sn89s}.
  
To avoid numerical issues with shocks, an artificially large viscosity (small Re) is adopted to smooth any sharp front in velocity.  This converts kinetic energy into internal energy %in steep velocity gradients 
(Fig. 12.3 \& 12.4, \citealt{richtmeyer}). Usually several zones are required for the pseudo-viscous representation of a shock.
To obtain an average of the turbulent behavior, space-time integrations over representative volumes of space and several turn-over times are needed.

 The hyperviscosity method of \cite{sn89} is different.
 They note: 
``Since this is compressible flow, shocks can form. {\em These instabilities are removed and the code is stabilized by applying artificial diffusion to all the fluid equations.}"  This seems to mean that artificial diffusion is increased (Re decreased) until the flow is ``stabilized'',  i.e., {\em no longer turbulent}, but laminar. This  is then used to approximate the space-time average of a turbulent flow.

Turbulence is a physical instability, not a numerical artifact, so arbitrary damping may have unintended consequences (\S\ref{s-art2}).

Due to the conflation of artificial viscosity with nonlinear fluid dynamics, LES do not provide a clear estimate of turbulent driving and damping, as we present for ILES (\S\ref{cascade}, \S\ref{Ssubgrid}, \S\ref{Sstratified}).
Eq.~\ref{eq-Re} may then {\em overestimate} the effective Re.

 \subsection{Specific TKE spectra\footnote{We agree with \cite{kritsuk} that we should not attempt a serial ordering of quality for well tested numerical codes used to attack a variety of multi-faceted problems. Our method may have some advantages for turbulence study.}}\label{tkespectrum}
 
On average in our ILES, the turbulent kinetic energy  flows in a cascade with a Kolmogorov-like spectrum, beginning at the integral scale $\ell$ and ending at the effective dissipation scale, which is at the resolution of our simulations  (the grid scale). 
Our presentation was inspired by  Fig.~1 in \cite{haugen2006}.
We place our ILES in context  by examining spectra of  specific turbulent kinetic energy (TKE)
in Fig.~\ref{fig33}. The spectra are scaled 
so that the inertial range becomes a horizontal line, which connects the energy containing eddies (the integral scale) at the left end of the curve, to the dissipation range at the right end.
The inertial range is supposed to be a 
conduit for TKE, with the  physics of acceleration contained in the integral scale and the physics of damping in the dissipation range.  We can evaluate each from our ILES.

Shocks are 3D objects, having a microscopic dimension across the front, and two macroscopic dimensions defining a surface. In turbulence, interacting shocks provide communication between large and small scales, and thus  contribute to  Fig.~\ref{fig33} in multiple places. In the oversimplified picture of the classical cascade, a turbulent ``patch'' has a single wavelength; it would be created on the left, propagate through the inertial range, to be dissipated on the right in the dissipation range. Instead we find a ``leaky'' conduit \citep{perry}, where interracting  shocks can affect both large and small scales at once, and ``accelerate''  the dissipation as required  \citep{onsager}. 

Such a construction necessarily favors 
intermittency; the separation of scales for driving and damping make synchronization impossible. The mechanisms also can differ for different problems, e.g., damping can be due to
molecular collisions, shocks, plasma damping, or charge separation in collisionless plasma, suggesting relevance to a range of topics from cosmic-ray acceleration to generation of bursts of radio noise.

All panels show a benchmark DNS (solid black line) of $4096^3$ zones  \citep{kaneda2003} with no magnetic field\footnote{Traditionally magnetic fields are ignored deep in stellar interiors because of their relatively low energy densities. With significant rotation and turbulence this approximation should be re-examined; magnetic fields can resist shear and amplify \citep{parker}. The predicted turbulent spectrum (Fig.~\ref{fig33}) may affect this.}. This benchmark is a converged DNS of the NS equations, which represents the solution to which all correct NS algorithms converge for this Re.  
Eq.~\ref{eq-Re} gives an estimated Reynolds number of Re$^*$$\sim$1600 for the $512^3$ grid  of our ILES, which is slightly less viscous than the Taylor microscale Re$\sim$1201 for \cite{kaneda2003}, which our energy spectrum resembles.

The top panel compares the Kaneda DNS to our ILES  for a $512^3$ grid 
with boosted heating (eps33k, red line). 
The largest length scales are at the left, and differ because the boundary conditions %acceleration algorithms 
differ slightly between DNS and ILES.
The very largest eddies show strong fluctuations which are slow to average out.
The eps33k (red) case relaxes to establish an inertial range, %but 
which does not extend so far to the right as the Kaneda case because the coarser grid has larger zones ($512^3$ vs. $4096^3$). 
The dissipation scale occurs where turbulent kinetic energy is converted into internal energy, and its position depends upon the mechanism which makes that conversion.
Dissipation is less efficient for molecular collisions (DNS) spread over a uniform medium  (the Kolmogorov scale) 
than for shock dissipation (ILES)  occuring in a thin front, so the DNS requires more inertial range between the integral scale and the dissipation scale for this range of Re, as shown. %(the grid scale).
%Both simulations capture a dissipation scale (falloff at far right). 

The middle panel differs from the top panel only in a horizontal shift of our ILES, made to aid comparison of dissipation regions. It  is interesting how well the shape of DNS dissipation (due to molecular viscosity) agrees with that of ILES dissipation (due to shocks), considering the %very 
different mechanisms invoked. The implied length scales differ by orders of magnitude, from molecular viscosity scales for NS equations to computational zone size (shock width) for the Euler equations.
The turbulent cascade develops whatever dissipation rate is needed.
Use of the shock jump conditions means that the computation %al zone size 
need not resolve the physical shock structure, and supports our horizontal shift through the neutral inertial range (the ``cascade'').

The bottom panel  compares  DNS to two lower resolution solutions of the NS equations, hyperviscosity LES (red dot-dashed line) and Smagorinsky LES (blue dashed line), both with $512^3$ zones and no magnetic field \cite{brandenburg}. The two LES are normalized to the DNS results. Horizontal shifts were performed as in the middle panel. We expect the hyperviscosity case to be indicitive of the
%a best case  (unstabilized) 
STAGGER code \citep{sn89,sn89s}.

The hyperviscosity LES show a %prominent 
``bottleneck'' feature, which is larger than for DNS, ILES, or Smagorinski LES. The hyperviscosity LES has more dissipation; see also \citealt{porterILES1}, Fig.~7.2 and 7.3. This bump (near scaled wavenumber 0.1), is due to a design choice: the higher viscosity  required for the abrupt termination of the cascade. 
To the right of this bump the hyperviscosity spectrum gives a corresponding error of opposite sign, a deficit of short wavelength motion prior to that cutoff. 

The hyperviscosity LES is sixth order in space and third order in time.  As we have emphasized in \S\ref{blowup}, turbulent velocities (and shocks) are not well represented by high order polynomials. The piecewise cubic representation of our ILES is better suited to a H\"older continuous flow (\S\ref{sLars}). 
The combination of a higher order dissipation and a sharp cutoff at the grid scale probably enhance the bottleneck error \citep{haugen2004} by approximating shocks by a high order polynomial (Fig.~\ref{fig33}).

The Smagorinsky LES shows an opposite tendency, a slower fall at high wavenumber than the benchmark DNS, which is attributed to ``strongly nonlocal interactions in wavenumber space", and is likely to be related to the choice of  pseudo-viscosity.  

The bottleneck becomes smaller with finer resolution in both DNS and ILES \citep{porterILES1}. 
It is striking that a $512^3$ ILES can resemble a $4096^3$ DNS in many respects (Fig.~\ref{fig33}), despite 1/8th as many zones per coordinate; see also \S\ref{Sdns}.

 \section{Dynamics of the Turbulent Cascade}\label{cascade}

The rate of turbulent dissipation  $\epsilon$ is not uniform in space and time in our ILES, but must balance driving on average. 
 
 \subsection{Global damping}\label{kolmog-damping}

Based upon the assumption of universality of isotropy at small scales,
 \cite{kolmg41} derived an average value for the rate of dissipation over the cascade,
\begin{equation}
\langle \overline{ \epsilon } \rangle \sim (\Delta u)^3/\ell_d, \label{kolmogorov}
\end{equation} 
where $\Delta u$ is the velocity scale of the largest eddies and $\ell_d$ the damping length for TKE (roughly the size of the turbulent region). The angle brackets  denote a spatial average over the turbulent region, and the overline a temporal average over several turnover times. \cite{onsager} has commented that Eq.~\ref{kolmogorov} was independently derived at least twice. For a review of Eq.~\ref{kolmogorov}, see \cite{vassilicous}.

Do not confuse the average property of the global flow, $\ld$, long established experimentally
as a global feature of turbulence (\citealt{batchelor60}, \S6.1), 
with the Kolmogorov scale, $\eta$,  which is local, based on molecular viscosity.
Always $ \ell_d \gg \eta$.  The average rate of dissipation $\langle \overline{ \epsilon } \rangle$ is determined  by $\ld$, which depends on  (i) the flow structure, and (ii) the velocity scale of the largest eddies $\Delta u$. In contrast $\eta = (\nu_{\rm NS}^3/ \epsilon)^{1 \over 4 }$  is an estimate of the scale at which a NS molecular viscous dissipation would smooth the flow (Fig.~\ref{fig33}). 
For DNS, these are separated by the inertial range (\citealt{batchelor60}), with an $\eta$ which  depends on the NS viscosity, and goes to $0$ as 1/Re does for inviscid flow. For ILES the size of the smallest possible eddy is  the grid scale, which is always finite.

 According to Kolmogorov\footnote{However see \cite{perry} on ``leaky'' cascades.}, the cascade proceeds down to smaller scale eddies, losing information about the large scale  structure of the flow, according to 
\begin{equation}
\langle \overline{ \epsilon } \rangle \sim ( v_{\lambda})^3/\lambda, \label{kolmogorov2}
\end{equation} 
where $v_\lambda = u - \overline{u} $
 for each eddy size $ \lambda \le \ell$ (\citealt{llfm}, \S31).
The turbulent kinetic energy is assumed to flow to smaller scales through each level of the cascade \citep{batchelor60},
down to the dissipation range.

Instead (in Paper I and \citealt{amy09vel}), we find lumpy, intermittent dissipation, shocks, 
and direct interactions between large and small scales of the flow, although {\em the average value of the rate of dissipation}
  $\langle \overline{ \epsilon } \rangle$ is similar to Eq.~\ref{kolmogorov}. 
We find $(\Delta u)^3 \sim u_{\rm rms}^3 $, where $u$ is the turbulent velocity. 
\citealt{amy09vel} found  (their Fig.~2),
\begin{equation}
 \langle \overline{ \epsilon_{nuc} } \rangle  \approx 
  \overline{ u_{\rm rms}^3  }/\ld ,\label{kolm-ave}
\end{equation} 
where $\ld$ defines {\em a global damping length for turbulent kinetic energy}, and $\epsilon_{nuc}$ is the nuclear heating which drives the convective shell, so $\overline{\langle \epsilon_{nuc} \rangle} = \overline{\langle \epsilon \rangle}$. 
This is similar to Eq.\,6.1.1 in (\citealt{batchelor60}), with
the empirical scaling factor, $A  \approx  1 $.
{ \em
Eq.~\ref{kolm-ave} gives the value of the TKE dissipative anomaly,  the flux of TKE in the cascade, and  a specific scaling for total energy conservation in Fig.~\ref{fig33}. }

\subsection{Acceleration of the cascade}\label{s-accel-cascade}

 \cite{onsager} suggested that cascade redistribution of energy proceeds by steps typically in a geometric series, by a factor of order $ 1 \over 2$ per step. 
 {\em The first few steps in the cascade limit the over-all speed of the process,} 
 which means that ``subsequent steps must be accelerated''. The actual first stages are complicated, but our ILES treats them without problem. The subsequent stages occur quickly, supporting Onsager's picture.
 If no other process intervenes, shocks will develop on a sonic time scale to provide whatever dissipation is necessary for the cascade (\S\ref{blowup}). 
{\em There is no need to resolve the scales of molecular viscosity with our ILES. }
 
There is a subtle question here of energy flow versus energy content. 
\cite{eyink-2018} quotes \citealt{llfm}, \S31: ``We therefore conclude that, for the large eddies which are the basis of any turbulent flow, the viscosity is unimportant and may be equated to zero, so that the motion of these eddies obeys Euler's equation. In particular, it follows from this that there is no appreciable dissipation of energy in the large eddies.'' 
\cite{eyink-2018} suggests:
``The latter statement is only
true, however, for the direct viscous dissipation of kinetic energy at inertial-range scales, whereas the energy in 
eddies at those scales, in fact, must be dissipated. The rate of decrease of energy for free-decay %\citep{amy09vel} 
or rate of power-input for forced turbulence are objective facts that cannot depend upon the resolution of eddies in the inertial-range.''
The energy cascade is in a steady state if flow in from larger scales is balanced by flow out to smaller scales, with no accumulation.

\cite{eyink-2018} explicitly writes the coarse-grained Euler equations plus the anomaly (``an energy flux from deformation work of the large-scale strain against the small scale stresses''). The mechanism of TKE loss is just the energy cascade \citep{onsager}, and also  applies to our ILES of the Euler equations (\S\ref{sReystress}).
At these large scales the TKE is not thermalized (yet) but flows ``away from our notice'' to smaller scales (Fig.~\ref{fig33} and \S\ref{tkespectrum}).

Our ILES  form such a turbulent cascade, generated at the integral scale. This TKE cascades down through the inertial range, to the dissipation range (our grid scale).  Our methods allow us to monitor the average dissipation due to the Riemann solver, which will be shown %in Fig.~\ref{fig3} and Fig.~\ref{fig4} 
in \S\ref{Ssubgrid}. This is the nonconservation of TKE for ($\theta \leq 1/3$) discussed by Onsager (\S\ref{sLars}).
{\em This dissipation of TKE is not controlled by an imposed parameter, but a self-consistent result of the ILES of the Euler equations, which implicitly  constrain the flow.}  

The ILES develop a %numerical 
cascade which we find to dissipate at the rate that Kolmogorov derived assuming local isotropy for turbulence (Eq.~\ref{kolm-ave}). 
However, like results from experiment  \citep{warhaft}  and DNS \citep{sree,iyer4096}, 
our ILES have an ``anomalous''  higher order behavior. 
Our previous interpretation of our numerical results is subtly altered and strengthened.
Further research is needed to precisely determine the degree of overlap between DNS and ILES methods.

\subsection{Time dependent B\"ohm-Vitense theory (MLT)}\label{BVtheory}

To better understand turbulent time dependence, and to connect to stellar evolutionary calculations, we consider the familiar and simpler MLT. 
MLT is inconsistent: it assumes flow, but is a local theory. It artfully averages over a ``mixing length'' to deal with this problem. This makes comparison with 3D ILES conceptually challenging. %As a start we
We replace the algebraic formulation \citep{biermann25,bv58} with a differential equation \citep{wda68}, using B\"ohm-Vitense's estimate of driving ${\cal B}^{bv}$, and damping  ${\cal D}^{bv} = v |v|/\ell_d$ from \S\ref{kolmog-damping} (the relaxation timescale is $\ld / |v|$), so 
\begin{equation}
dv/dt = {\cal B}^{bv} - {\cal D}^{bv}, \label{eq-wda68}
\end{equation}
where $v$ is the flow speed and $\ld$ the MLT mixing length, 
 {\em which is directly related to dissipation length. }
 If we multiply this equation by $v$ we obtain a primitive % turbulent kinetic energy 
 TKE equation. 
In a hypothetical  quasi-steady state (MLT), we would have
\begin{equation}
 \langle \overline{ {\cal B}^{bv} } \rangle \sim \langle \overline{ {\cal D}^{bv} } \rangle, \label{mlt}
 \end{equation}
 which is the simplest verson of the ``dynamic constraint''  (\S\ref{sLars}).
 While ${\cal B}^{bv}$ is local, ${\cal D}^{bv}$ is a property of the turbulent cascade, so they represent different length scales, and can not remain in phase. 
 Unless ${\cal B}^{bv}$ and ${\cal D}^{bv}$ are in phase, Eq.~\ref{eq-wda68} has dynamic solutions
 which must be averaged. %are ignored.
This demands  intermittency \citep{tennekes}, not as a superficial feature, but as necessary for the existence of a quasi-steady state in both this simple and in more realistic convection models. 
 
In MLT,
 \begin{equation}\label{eq-vbB}
\langle  \overline{ {\cal B}^{bv} } \rangle \equiv \langle \overline { g \beta_T \Delta \nabla } \rangle,
\end{equation}
where  $g$ is the gravitational acceleration, $\Delta \nabla= \nabla - \nabla_e $ is the super-adiabatic excess,
and $\beta_T $ is  the thermal expansion coefficient \citep{kippen}. 
The entropy excess $\Delta \nabla$ may contain  contributions from composition differences, which  are ignored in MLT, but are required at boundaries of evolving stars. To evaluate these, the problem of mixing must be solved consistently with that of convection  
%, complicating the problem 
\citep{321D,woodward2015,miro}.

\subsection{TKE equation}\label{STKE}

 Some of the behavior in our library of 3D ILES can be captured with a more general turbulent kinetic energy (TKE) equation  \citep{ma07b,miro,alvio1}, which we improve here to include insights from \cite{onsager}.
% Our approximations spring from our simulations, not visa versa.
   
Using (i) the momentum conservation equation for turbulent flow in a hydrostatic background, (ii) the mass conservation equation, and (iii) taking a dot product with the velocity fluctuation $\bf u'$ relative to a co-moving frame \citep{alvio1}, gives a turbulent  kinetic energy (TKE) equation (compare Eq.~\ref{eq-wda68}),
\begin{equation}
D_t (\rho{\bf u' \cdot u' /2} ) = -{\bf u' \cdot } \nabla P' + \rho' {\bf u' \cdot  g}  \label{tke1}
\end{equation}
where the primes denote turbulent fluctuations. The left-hand side is a co-moving time derivative.
 
 Taking averages and
 requiring  $ \langle \overline{ \rho' {\bf u' \cdot  g} } \rangle \rightarrow 0$  in order to remain in the comoving frame after fluctuations,  
 \begin{equation}
\langle \overline{D_t (\rho{\bf u' \cdot u' /2} ) }\rangle = -\langle \overline{ {\bf u' \cdot } \nabla P' }\rangle   \label{tke2}
\end{equation}

 \subsubsection{Turbulent damping by Reynolds stress}\label{sReystress}
 
\begin{figure*}[h]
\figurenum{3}
\label{figppert}
\includegraphics[angle=0,scale=0.65,bb=-70 320 200 500]{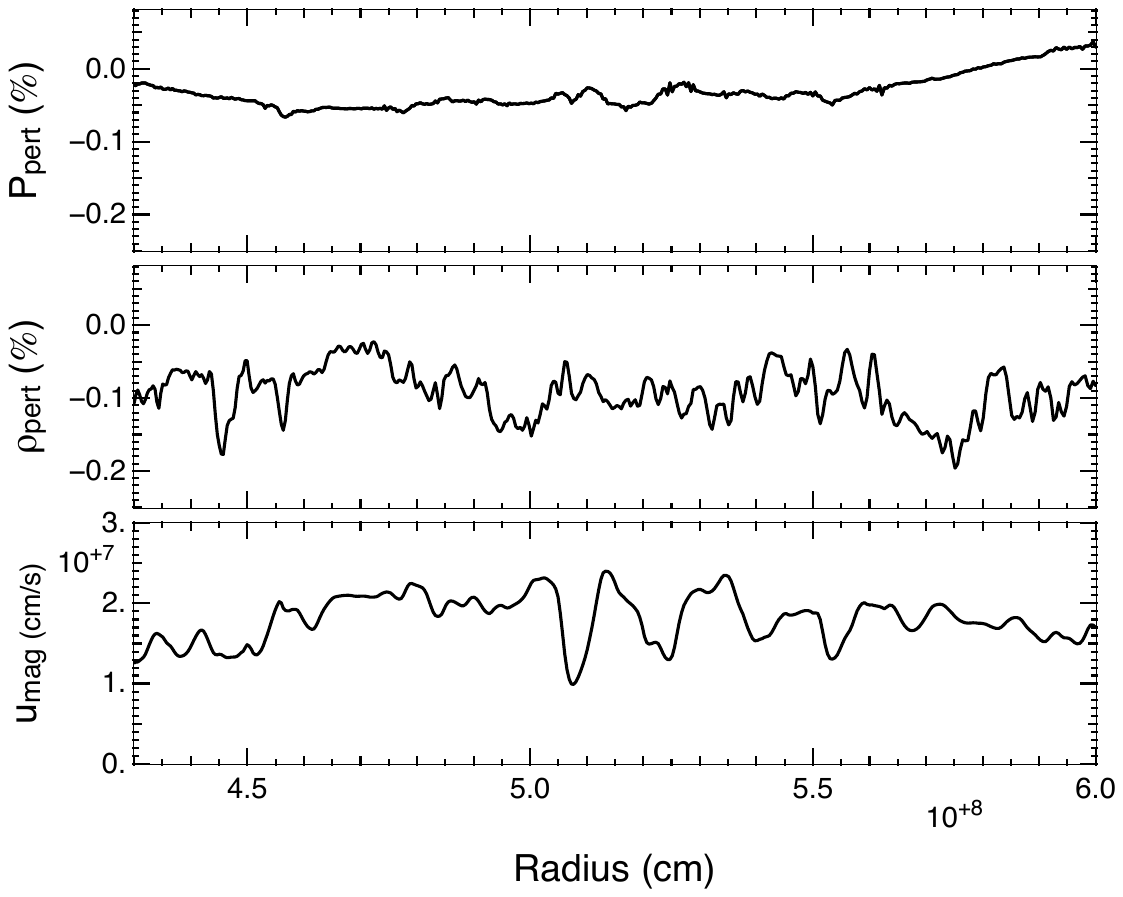}

\caption{ 
({\bf Top}) pressure fluctuations,  ({\bf Middle}) density fluctuations, and  ({\bf Bottom})  turbulent velocity magnitude, all versus radius, from the interior of a PROMPI ILES of the O+O convective shell \citep{321D}. These patterns  suggest the presence of weak shocks. Presssure perturbations are weakened by expansion but the density perturbations are more robust, implying entropy perturbations also. The velocity structure is intermediate between these two, being  ``rough''  at scales which are much smaller than the depth of the convective region (most of the convective region is shown; boundaries are omitted for simplicity). In our ILES,  the PPM algorithm \citep{cw84}  applies ``flattening'' (i.e., ``local support'') to suppress Gibbs ringing behind these weak shocks; this causes a ``dissipative anomaly'' to occur throughout the convective region, even with no viscosity \citep{onsager}.  }  
\end{figure*}

\placefigure{3}

In our  ILES, {\em
 high-frequency oscillations in the turbulent  velocity are responsible for a non-trivial Reynolds stress,  giving a finite dissipation in the Re~$\rightarrow \infty$ limit }
 \citep{dls13}. 
 This corresponds to the domain of $\theta \leq {1 \over 3}$ in \S\ref{sLars}, for which there is a thermalization of TKE by damping.  Such convective velocity fluctuations are shown in Fig.~2 of \cite{321D}.

If we average $D_t (\rho{\bf u' \cdot u' /2} ) $ from Eq.~\ref{tke2} and assume a steady state, there is a residue  in the $\theta \le {1 \over 3}$ region which is just the desired damping term $\rho \epsilon_K$ (see Eq.~\ref{kolm-ave}). 
We have
\begin{equation}
   \rho \epsilon_K =  \langle \overline{ {\bf u' \cdot } \nabla P' } \rangle  \label{eq-tke2}
\end{equation}
 but what is $ \epsilon_K$?
  
 This nonlinear mathematics is key to understanding turbulent cascades. The term is common to the Euler and NS equations, so they can converge to the same value in the inviscid limit.

The rate of increase in TKE is the product of an acceleration and a velocity, so
\begin{equation}
{\bf u \cdot} \big ( \partial_t {\bf u} +{\bf u \cdot \nabla u } \big ) \rightarrow
\partial_t  u^2/2 +\partial_r u^3/3 , \label{eq-epsRe}
\end{equation}
because the driving acceleration is parallel to gravity.
For a quasi-steady state (a few turnover times), $\langle \overline{ \partial_t  u^2/2 } \rangle \approx 0 $,
with a net cancelation.
The other term is finite.

In the $\theta \leq {1 \over 3}$ region we have ``rough'' (i.e., turbulent) flow. % , so any initial preference for radial motion is forgotten.   
Fig.~\ref{figppert} shows the fluctuations of pressure, density and the velocity magnitude from the middle of the O+O burning shell from the Perth simulation ($1536 \times 1024^2$) using PROMPI \citep{321D}.

The fractional pressure perturbations are smaller than those for density; the density perturbations are less smooth, indicating significant entropy variations.  
The variations in turbulent velocity are indicated by velocity magnitude ($u_{mag} \equiv\sqrt{\bf u \cdot u}$), which samples all three dimensions; this curve is smoother than the other two, but still ``rough'', suggesting weak shocks (Mach$\sim$0.05). For comparison, Fig.~\ref{fig34} shows a slice of velocity magnitude through the C+C shell.

This rough velocity field gives a residue  from the $\partial_r u^3/3$ term
 which is robust against zoning error, being dominated by the largest scales;
it varies little from $256^3$ to $1024^3$ grids \citep{andrea,andrea2,321D}. It is damped by ``flattening'', which is consistent with a field of weak shocks spread over the convective region. As the amplitude of ``roughness'' increases, so does dissipation, giving negative feedback and a nonlinear bound to growth.
 %, and whose net effect is a quasi-steady state with significant intermittency. Our ``flattening'' prevents Gibbs ringing behind shocks (provides
Without this ``flattening'', a wave would remain sinusoidal, conserving kinetic energy.

This is essentially the same physics, seen from a different viewpoint, as 
shown in Fig.~1 of \cite{iyer4096} for a DNS realization using ``molecular viscosity'' for shocks, showing ``ramp-and-cliff'' structure \citep{sree79}. DNS uses molecular viscosity to thermalize the post-shock regions, requiring several zones to span each shock (a cliff followed by a ramp is the simplest combination). ILES handles this more compactly.

%{\color{green} SIMON, please check the scales to make sure I get it right. Ppert $\pm 0.0003$, 
%rhopert $\pm 0.001$?  vmag $(1.5 \pm 0.5) \times 10^7 $ cm/sec so Mach $\sim \sqrt{0.0003} \sim 0.017$ ? }

Upon Reynolds averaging, the spatial derivative from Eq.~\ref{eq-epsRe} leaves a residue, which we evaluate numerically from our ILES.  On average,
\begin{equation}
 \epsilon_K = - \langle \overline{ \partial_r  (u^3/3) } \rangle  \approx -\langle \overline{u_r^3} \rangle /\Delta_{\rm cz}, \label{eq-Rstress}
\end{equation}
which is Eq.~\ref{kolm-ave}, and the turbulent damping length is $\ell_d \approx \Delta_{\rm cz}$, the size of the turbulent region  for weak stratification. 
%This is equivalent to a shock, which on average dissipates at a rate $\langle \overline{ u_r^3 }\rangle$ per unit distance traversed,  traversing the convection zone.

\cite{amy09vel} verified this numerically for weakly stratified convection (see their Fig.~2 caption).  
We numerically reconfirm it (Tables \ref{table2} and \ref{table4} in \S\ref{Ssubgrid}).
Here we have obtained a similar result from our estimate of the Reynolds stresses in the turbulent flow (Eq.~\ref{eq-Rstress}). This provides a normalization consistent with unity for our ILES in Fig.~\ref{fig33}.

The TKE equation (Eq.~\ref{eq-tke2}) 
does not assume any particular flow or explicit dissipation \citep{ma07b}. 
With weak stratification\footnote{Here ``weak'' means the depth of the convection region $\Delta_{\rm cz} $ is no more than a few pressure scale heights. For ``strong'' stratification, see \S\ref{Sstratified}.} 
our ILES develop narrow downflows and broad upflows, but little net TKE flux due to significant up-down cancellation.
The assumed symmetry of up-down flow in MLT requires the net flux of
TKE to be zero, but finite TKE becomes necessary as increasing stratification further breaks symmetry.

This damping as a function of radius will be shown in detail in Fig.~\ref{fig3} and \ref{fig4}; its integral is the TKE anomaly (\S\ref{sLars}).
For strong stratification we have an asymptotic value for the damping length, so $\Delta_{\rm cz}$ is replaced by 
$\ell_d \sim H_{\rho}$
 in Eq.~\ref{eq-Rstress}; see \S\ref{strongstrat}.

\subsubsection{Phase lags, driving and damping}\label{phaselag}

\begin{figure*}[h]
\figurenum{4}
\label{fig35}
\includegraphics[totalheight=200pt,bb=120 -50 00 400]{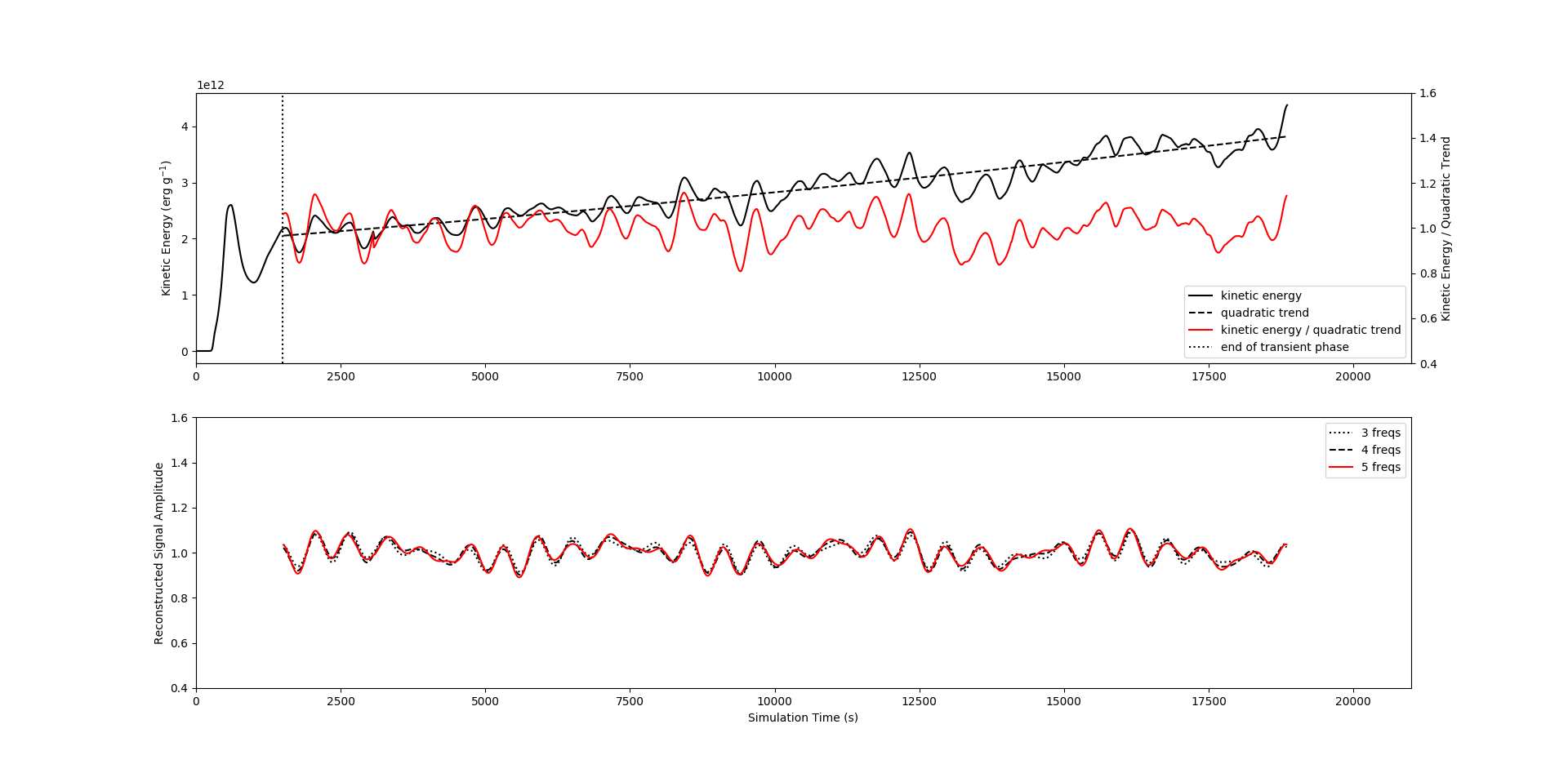}

\caption{
Evolution with PROMPI of the turbulent kinetic energy in a small section of the C+C burning shell (top panel), and low order decomposition into modes of 3, 4, and 5 frequencies (bottom panel), after \cite{andrea2}.  In the top panel the evolution is split into an initial transient phase, to the left of the vertical dotted line, and a longer evolutionary phase, to the right. Here a secular increase is seen, which is due to consumption of fuel and agrees with the 1D model.  If this is allowed for, the result is the red line, which is not static, but has a stable average over 30 pulses. The bottom panel shows the dominant modes corrected for the quadratic trend, as three approximations (3, 4, and 5 frequencies).
Finite variability is maintained because the computational domain is sufficiently small \citep{ma07b,amy09vel} to resolve patches of turbulence.  The curves lie closely together, indicating that the pulses in TKE are dominated by these 3 low order modes, consistent with \cite{holmes} and \S\ref{s-accel-cascade}. The initial transient phase provides a direct estimate of the cascade ratio in the simulations (see \S\ref{phaselag}; \citealt{onsager}). The initially flat pre-rise phase (far left top panel) shows no motion until thermal burning builds a superadiabatic excess to accelerate matter. This is a measure of the ``turn-over'' time.
}
\end{figure*}

\placefigure{4}

Neutrino-cooled stages of burning are accesible to ILES (e.g., \citealt{ma07b,wda96}). Deep convection in photon-cooled stages may be more challenging computationally because slower evolution is accompanied by flow at very low Mach number, giving subtle conservation law problems \citep{sproof}.
 
If we ignore the necessary phase lags between driving and damping, 
 then on average in a volume segment of the turbulence we have Eq.~\ref{mlt},
which is something like MLT (see \citealt{kippen}, Eq.~7.6). 

If we do not ignore phase lags, we get 
Fig.~\ref{fig35}, which  shows the TKE, for a $256^3$ ILES from onset through 30 pulses. In this approximation to a section of the whole shell, a quasi-steady state of bounded pulses is formed.
By their interaction the pulses suggest the presence of several low order modes. Following  \citep{holmes}, we find by a Karhunen-Lo\`eve decomposition of the time series, that three low order modes are indeed dominant. For this particular case there is also a slow evolution due to increasingly active thermonuclear burning, which agrees well with predictions of 1D evolutionary models. There is no indication of any growing  instability on a shorter or similar timescale.

\subsection{Other dynamic terms}\label{s-added}

Our ILES include all terms relevant for turbulence, but this is not true for MLT: it 
omits some  acceleration terms needed for the strong stratification case (which cannot be 
 isotropic, for example).  Despite these and other objections, MLT works in some sense \citep{alvio87}. Why? Part of the reason was given in Paper I \citep{alvio1}; here we examine the issue further.
 
In a steady state and in a co-moving frame, the work done by fluctuations is balanced by turbulent dissipation (Eq.~\ref{eq-tke2});  see also (\citealt{ma07b,nsa}).
For weak stratification this can be reduced to the MLT balance between drag and buoyancy alone (\S\ref{BVtheory}),  
but there are additional terms in general, which we derive.

Using the identity
\begin{equation}
 P' {\bf \nabla \cdot u'} + {\bf u' \cdot } \nabla P'  = {\bf \nabla \cdot  (\bf u'} P' ),
\end{equation}
we obtain after averaging 
\begin{equation}
-\langle \overline{ {\bf u' \cdot } \nabla P' } \rangle  = W_B+W_P+{\cal A}. \label{eq-bandp}
\end{equation}
We give the definitions below.

\subsubsection{The buoyancy term $W_B$}

In MLT the net mass flow is zero, so  inflowing density excesses and deficits are balanced.
If the flow is strongly subsonic, the perturbations in temperature, density and composition give small  pressure perturbations (e.g., Fig~\ref{figppert}). This is the ``buoyancy'' limiting case, and the work done by buoyancy $W_B$ is
\begin{equation}
W_B =   \langle \overline{ (\rho'{\bf u' \cdot g}) \beta_T \Delta \nabla } \rangle, \label{buoy}
\end{equation} 
which uses the acceleration due to the entropy excess (Eq.~\ref{eq-vbB}) in MLT \citep{viallet2013}.

We have randomly perturbed a hydrostatic initial model by 
mass-conserving density variations of low amplitude which, for no change in pressure or composition, imply entropy variations. Interesting behavior results: stable regions ($\theta > {1 \over 3}$) quickly damp these small blemishes by sound wave radiation, while unstable regions ($\theta \le {1 \over 3}$) begin growing them exponentially, at first attempting to form small convective rolls from the variously distributed torques. Because of the choice of initial conditions, the power of the fluctuations is at the grid scale at first, but nonlinear evolution continues until a global turbulent region develops, powered by burning at the integral scale (Fig.~\ref{fig34}).

\subsubsection{Pressure dilatation term $W_P$}\label{sW_P}

The next important term in Eq.~\ref{eq-bandp} is
the work done by turbulent pressure fluctuations $P'$ on the turbulent flow, the ``pressure dilatation''.
For weak stratification {\bf and low Mach number}, $W_P \approx 0$ by up-down symmetry  in the mass flux,
but for strong stratification,
\begin{equation}
W_P =  \langle \overline{ P' {\bf \nabla \cdot u' } } \rangle 
      \sim \langle \overline{\rho} \rangle \langle \overline{ u_r^3 }\rangle / H_\rho , \label{pdil}
\end{equation} 
where we use $P'  \sim \langle \overline{ \rho } \rangle u_r^2$, and ${\bf \nabla \cdot u' } \sim u_r / H_\rho$  \citep{viallet2013}.

When the convective Mach number is small,  so is the turbulent pressure fluctuation $P'$.
In this limit the flow has little effect on the background structure, and is nearly anelastic. As the flow becomes more stratified (more plume-like), $W_P$ becomes nonlocal and more important relative to $W_B$.

\subsubsection{Acoustic and turbulent kinetic energy fluxes} \label{c-fluxes}

The KE equation involves two turbulent fluxes: the TKE flux $f_k = \langle \overline{ \rho }\rangle \langle \overline{ u'_r \cdot { 1 \over 2 } (u'_r)^2 } \rangle $ and the acoustic flux $f_p =\langle \overline{ u'_r P' } \rangle$, so
\begin{equation}
{\cal A} = {\bf - \nabla_r \cdot ( f_p + f_k )  }.
\end{equation}

There is some freedom in the formulation of the kinetic energy equation, especially in the expression  which splits  the sum of the driving terms; see  discussions in \citep{nsa} and \citep{am11turb}. We split the driving into 
$W_P$ and $W_B$ as these are thermodynamically the most fundamental quantities: $W_B$ measures the reversible exchange between kinetic energy and potential energy, $W_P$ measures the reversible exchange between kinetic energy and internal energy \citep{viallet2013}.

The 3D simulations for C, Ne and O burning are more vigorous than for H or He burning (mildly compressible flow;  see \S\ref{strongstrat}), so the acoustic flux term is also small but nonzero (see Fig.~2, Paper I).
The divergences of the acoustic flux, 
and of the turbulent kinetic energy flux,
are omitted from MLT. 

\cite{andrea2} discuss a series of ILES of the C shell in which the driving by nuclear burning is successively scaled from weak to strong. As the driving increases the new terms (pressure dilatation and divergences of acoustic and kinetic energy) become important. 
Due to the increase in acceleration, the shell expands, pulse amplitude increases, and mass is ejected from the grid for the most vigorous burning, connecting mass loss to vigorous convection.

 \section{Weak stratification}\label{Ssubgrid}
 
Except for surface convection zones (atmospheres), convection in stars is weakly stratified. This includes all convection driven by quasi-static nuclear burning.
Cooling by neutrino emission accelerates the stages of C, Ne and O burning so that, unlike photon-cooled stages (H and He), more powerful methods of fluid flow simulation become feasibile: we choose our time step with a causality (Courant) condition \citep{leveque}, allowing neutral stability of the fluid. 
Both shell and core burning have weak stratification.

Guided by Onsager's conjecture (\S\ref{sLars}),
we use Eq.~\ref{kolm-ave} to extract ``effective'' damping lengths $\ld$ from our 3D simulations with no new assumptions. This provides an absolute scaling for the ILES in Fig.~\ref{fig33} and the anomalous damping implied by our ILES.   

Despite having no imposed viscosity, our ILES reproduce a statistically steady convection zone for 30 turnover times, as Fig.~\ref{fig35} showed. 
Our initial models and 3D simulations are already consistent, so $\overline{ \langle \epsilon_{nuc} \rangle }= W_B / \langle \overline{ \rho } \rangle $. 

For weak stratification we
follow \cite{viallet2013}, keep only the buoyancy term (Eq.~\ref{buoy}) in Eq.~\ref{eq-bandp}, and use Eq.~\ref{eq-Rstress} in Eq.~\ref{eq-tke2}. We assume acoustic and TKE fluxes are small.
If driving equals damping on average,
\begin{equation}
 \overline{ \langle \epsilon_{nuc} \rangle } = W_B / \langle \overline{ \rho } \rangle = -  \langle \overline{u_r^3} \rangle /\Delta_{\rm cz},
\end{equation} 
where each term may be independently evaluated from the ILES.
We define a consistency factor
\begin{equation}
  {\cal C} \equiv -\langle \overline{u_r^3} \rangle / \overline{ \langle  \epsilon_{nuc} \rangle }   = \ld/\Delta_{\rm cz} ,
\end{equation} 
which would be unity if $\ld = \Delta_{\rm cz}$, i.e., if the ILES are consistent with Onsager ideal turbulence in weak stratification.

Interior convection seems insensitive to uncertainties related to atmospheric physics, and thus is primarily a turbulence problem. Entrainment remains a key issue \citep{laura}. 
High resolution ILES give reasonably consistent results for the short term dissipation rate 
\citep{porterwoodward,ma07b,amy09vel,321D,jones17,andrea}, as do experiment \citep{vassilicous}.

 \subsection{The O+O shell}\label{O+Oshell}

\begin{figure*}[t]
\figurenum{5}
\label{fig3}
\includegraphics[angle=0,scale=0.9,bb= -50 0 100 300]{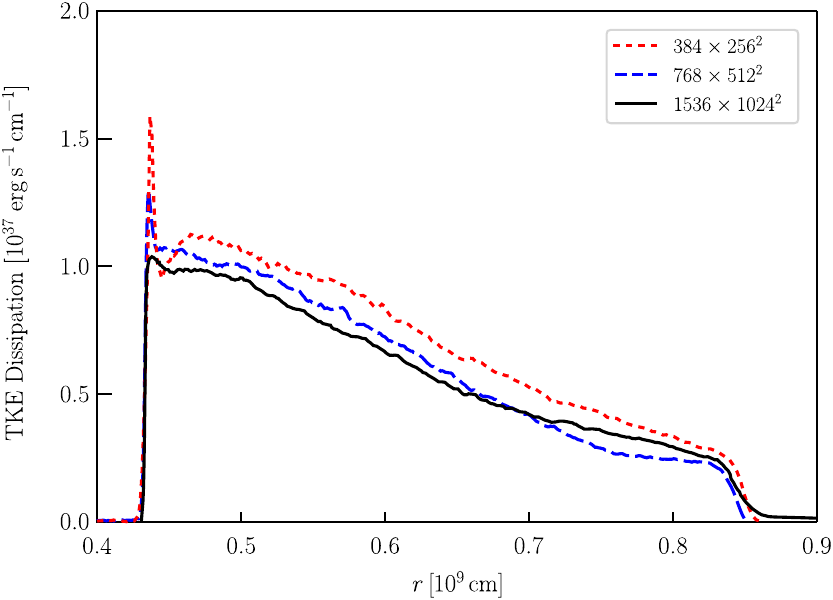}

\caption{
The anomalous dissipation of turbulent kinetic energy $\overline{\epsilon_K}$, plotted as  
$ \overline{\epsilon_K} \times(\overline{ 4 \pi r^2 \rho} )    $
versus radius, for the oxygen burning shell; see \S\ref{sReystress}.  $\overline{\epsilon_K}$ is balanced by generation of kinetic energy by buoyancy, giving a quasi-steady cascade of turbulent kinetic energy as described in \S\ref{tkespectrum} and \S\ref{cascade}. 
The ``med-res'' ($384 \times 256^2$, red dotted line) and ``hi-res'' cases ($768\times 512^2$, blue dashed line)  of \citep{viallet2013}, and a ``very-hi-res'' case ($1536\times1024^2$, black line) (the ``Perth'' simulation, see \citealt{321D}) are shown. Using Eq.~\ref{eq-Re}, the Reynolds numbers are
Re$^*\sim$1100, 2800 and   7000. 
The ``spike'' at $r \sim 0.43 \times 10^9$ cm is due to inadequate spatial resolution of the lower (thinner) boundary layer, and begins to vanish at the highest resolution. The upper boundary layer seems well resolved in space. 
The remaining dissipation is the ``slope'' feature between $ 0.44 \times 10^9$cm $\leq r\leq 0.85\times 10^9$cm. It is statistically steady but not static, and is due to dissipation by mild shocks 
 (see Eq.~\ref{kolm-ave}; \ref{sW_P} ). It is affected by the density gradient,  
and found to have a Kolmogorov-like intensity on average (\citealt{amy09vel}). The dissipation is not due to numerical error, but rather to an emergent and robust feature of the nonlinear  Euler equations: the turbulent cascade. Onsager's prediction of $\theta \leq 1/3$ for boundaries is confirmed; see \S\ref{sLars}. Ingestion of $^{20}$Ne also causes heating and evolution of the background \citep{miro}; this is absent in the C+C shell (Fig.~\ref{fig4}). The number of turnovers taken to construct the curves is limited by computational cost; med-res, hi-res and very hi-res have 4, 2.1 and $\lesssim2$ turnovers,
}
\end{figure*}
\placefigure{5}

Fig.~\ref{fig3} shows the time-averaged 
rate of anomalous dissipation  (\S\ref{sLars}) of turbulent kinetic energy,  weighted by the increment in spherical mass coordinate,
versus radius,
$\overline{\epsilon_{K} }\times   \overline{ \ 4 \pi r^2  \rho}   $
for the oxygen burning shell  at several different numerical resolutions.
Here  the time average $\overline{\epsilon_K  }$ must be supplemented by a space average to obtain $\langle \overline{ \epsilon_K } \rangle $ in Eq.~\ref{kolm-ave}. 
 The dissipation due to the turbulent cascade is relatively flat; most  of the variation in Fig.~\ref{fig3} is in the mass increment factor. Convection automatically choses its boundaries, which are sharply defined at $\theta = 1/3$ (\S\ref{sLars}). 

 The ``slope'' feature in Fig.~\ref{fig3} (between $r \sim 0.43$ and $0.85 \times 10^9$ cm)  has a constant gradient which is not sensitive to zoning in space, from $128^3$ to $1536\times 1024^2$, and has a Kolmogorov-like value  over the averaging window in time, which agrees with Eq.~\ref{kolm-ave}.
 
 This is a result of our ILES, confirming \cite{amy09vel}, and not an imposed assumption.  It is (i) a numerical realization of  damping ({\S\ref{sLars}) by emergent nonlinear effects (\S\ref{STKE}), and (ii) a confirmation of our Reynolds stress expression (Eq.~\ref{eq-Rstress}).

The amplitude of the ``slope'' shows a mild fluctuation with zoning, despite being averaged over a few turnover times. 
Dissipation goes to zero abruptly at the lower boundary interface which separates the convective from the non-convective layer. 
The flow in the upper and lower sandwiching layers is ``smooth'' ($\theta>{1 \over 3}$) being wave-like, instead of turbulent  (``rough'', $\theta <{1\over 3}$); see \S\ref{sLars}).

 Inadequate resolution at the lower boundary is easily identified as a different ``spike'' for each choice of zoning; these spikes are small, begining to  disappear as zones are added.  The spikes occur at the radius of the thin boundary layer (Fig. 1 in Paper I), and are due to  3D waves which
 sweep past the radial grid coordinate. This error decreases with finer zoning, and corresponds to an inability to resolve the surface waves in the boundary layer on the grid used. The bottom spike is more prominent because the stratification  in pressure (and density)  is steepest at the bottom of the convective region.
 
 \begin{deluxetable*}{llllllll}
\tablewidth{250pt}
\tablecaption{Dissipation lengths: O+O shell, $\Delta_{cz} \sim 2 H_P$    \label{table2}}
\tabletypesize{\small}

\tablehead{\colhead{stage}  & \colhead{zoning} & \colhead{$\Delta_{cz}$} &\colhead{$\ld$}  & \colhead{$\ld / \Delta_{cz} $} & \colhead{$\ld/H_P$} & \colhead{$v_{\rm rms}$} & \colhead{$t_{\rm ave}/t_{\rm to}$} \\
 & depth $\times$ area & $10^8$cm  & $10^8$cm &$={\cal C}$ & & $10^7{\rm cm\,s^{-1}}$ &
}
\startdata
O shell$^a$  & $400 \times 100^2$ & $4.0$ & $3.6$ & 0.90 & 1.48  &$ 0.97$  & $4$  \\
O shell$^b$   & $192 \times 128^2$ & $4.2$ & $4.9$ & 1.16 &2.03 & $1.07$ & 4 \\
O shell$^b$   & $384 \times 256^2$ &$4.2$ & $5.5$ &  1.31  &2.27 & $1.09 $ & 4\\
O shell$^b$   & $786 \times 512^2$ &$4.2 $ & $4.7$ &  1.12  & 1.94 & $1.09 $ & 2.1\\
O shell$^c$   & $1536 \times 1024^2$ & $4.2 $&      \nodata       &     \nodata    &   \nodata   &    \nodata    &  $\lesssim 2$  \\
\enddata
\tablenotetext{a}{\cite{ma07b}; oxygen burning with $^{20}$Ne entrainment. }
\tablenotetext{b}{\cite{viallet2013}; oxygen burning with $^{20}$Ne entrainment.} 
\tablenotetext{c}{\cite{321D}; oxygen burning with $^{20}$Ne entrainment (``Perth'').} 

\end{deluxetable*}

\placetable{1}

%{\color{green} SIMON, Please fill in the Perth data if you can. }

Dissipation in the upper boundary 
was converged at the range of resolutions shown.  This milder condition for numerical convergence on the upper boundary is to be expected because of the larger scale heights in all of the structural and flow fields there.

Unlike the stable regions at $r<0.43 \times 10^9$cm and $r>0.85  \times 10^9$cm, 
the convective region has fluid trajectories which, due to mild shocks, give 
an average dissipation length of  $\ld \approx \Delta_{cz} $ (Eq.~\ref{eq-Rstress}).

Table~\ref{table2} summarizes the oxygen-burning shell results in which 
only the zoning is changed. 
Here $\alpha_{\rm MLT}= \ld/H_P \approx 2 \approx \Delta_{\rm cz}/H_P$. 
Structural variables, such as depth of convection zone $\Delta_{cz}$,  and density scale height $H_\rho$, are insensitive to zoning.
The rms velocity $v_{\rm rms}$ fluctuates, and the number of turnover times ($t_{\rm ave}/t_{\rm to}$) used in  time averages are shown. 
All of our ILES show convection as episodic (\citealt{ma07b}; \S\ref{phaselag}),  with a fluctuating behavior, but steady on average.

The oxygen burning simulation begins with 
entrainment of $^{20}$Ne into the $^{16}$O shell, giving
two  burning zones in the same convective shell 
\citep{miro}\footnote{This was also seen in carefully modelled proton ingestions into He-burning convection zones \citep{mocak10,fujimoto00,campbell10}.}. 
 \cite{321D} discuss our highest resolution simulation (``Perth'') of the oxygen-burning shell.

{\em ILES numerical simulations for weakly-stratified convection produce a dissipation which  agrees 
 with Kolmogorov theory for homogeneous, isotropic turbulence}.
The estimates  of the consistency factor $\cal C$ are roughly unity for all resolutions: 
\begin{equation}
{\cal C} = \ell_d/\Delta_{cz} \rightarrow 0.90, 1.16, 1.31, 1.12,
 \end{equation}
despite the Ne ingestion episode,  which seems to affect these numerical results slightly ($\pm 15\%$); see also  \citep{miro}, and below.

\subsection{The C+C shell}\label{C+Cshell}

\begin{figure*}[t]
\figurenum{6}
\label{fig4}
\includegraphics[angle=0,scale=0.9,bb=-10 0 200 370]{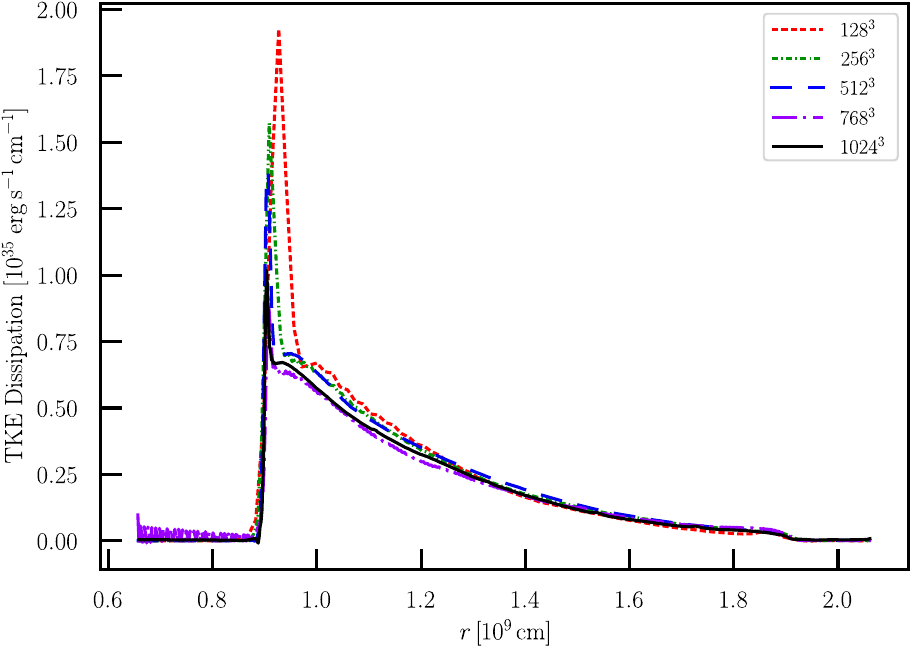}

\caption{
The anomalous dissipation of turbulent kinetic energy $\overline{\epsilon_K}$, plotted as
$\overline{\epsilon_K} \times \overline{ 4 \pi r^2 \rho } $
versus radius, for the carbon burning shell. 
 $\overline{\epsilon_K}$ is balanced by generation of kinetic energy by buoyancy, giving a quasi-steady cascade of turbulent kinetic energy (\S\ref{cascade}). 
Simulations of $128^3$, $256^3$, $512^3$, $768^3$,
and $1024^3$  \citep{andrea,andrea2} are shown. 
Using Eq.~\ref{eq-Re}, the Reynolds numbers range from Re$\sim$256 to 4096.
Strong intermittency is associated with the pulses in turbulent kinetic energy (Fig.~2 of \citealt{alvio1}), and requires time averaging (see text).
 Unlike the O+O simulation, 
 the simpler burning associated with C+C has no strong ingestion of new fuel, requiring readjustment of the structure, but the effect of finer zoning  seems similar. The sharp ``spike'' is due to numerical error, and monotonically decreases in size with increased resolution. However the broad ``slope'' in the dissipation profile is an emergent and robust feature of the nonlinear Euler equations (the turbulent cascade), and converges with resolution toward an asymptotic form  $\overline{\epsilon_K}$  for damping (see Eq.~\ref{kolm-ave}; \ref{sW_P} ).
 Onsager's prediction  of $\theta \leq 1/3$ for boundaries is confirmed; see \S\ref{sLars}. }
\end{figure*}
\placefigure{6}

Figure~\ref{fig4} shows the time-averaged anomalous dissipation over the carbon burning shell as a function of radius, 
in the same format as Fig.~\ref{fig3}; see \cite{andrea,andrea2}. Unlike the previous simulations, there is no ingestion of a new and different fuel ($^{20}$Ne), so the evolution is smoother  (see \citealt{miro} for further discussion of $^{20}$Ne ingestion).
The  very weak energy generation is boosted  by $10^3$ (to be merely weak).  This gave a shorter burning time (still many turnovers), no noticeable structural change, but more computational efficiency in this C+C simulation; see \cite{andrea}. 
The higher ``spike'' errors, seen in C+C relative to O+O, 
are due to our choice of pseudo-cartesian coordinates here, which were computationally faster but less accurate than the true spherical ones used for O+O.

These results are consistent with the oxygen shell simulations, but simpler due to the lack of a $^{20}$Ne ingestion episode. Again, the boundary dissipation spike (\citealt{andrea}, Fig.~10) tends to disappear at the highest resolutions, while the ``slope'' (anomalous dissipation) is insensitive to zoning (\citealt{andrea}, Fig.~9,).

Table~\ref{table4} shows the results of ILES for C+C shell burning.
There are ($4.0, 4.4, 10.6, 1.5$)
turnovers  for zoning of ($128^3, 256^3, 512^3, 1024^3$), so the highest resolution in space corresponds to the poorest statistics in time because of budgetary limitations and the need to disentangle the behavior caused by 
intermittency
The cases with highest spatial resolution  have fewer turnovers, and lose some statistical accuracy. 
We have
\begin{equation}
 {\cal C } = \ld/\Delta_{cz} \rightarrow 1.05, 1.01, 1.005,  1.08,
 \end{equation}
 so the dissipation length is close to the depth of the convection zone. 
 
As with O+O, runs of greater duration at higher resolution are desired, but there already seems to be a clustering around $  \ld/\Delta_{cz} \approx 1.0$. 
The C+C shell is slightly more stratified than the O+O shell, with a depth of 3 rather than 2 $H_P$. Deeply stratified convection zones have smaller values of $ \ld/\Delta_{cz}$ (see \S5), so
this C+C result seems consistent with the previous O+O one.

\subsubsection{Resolution issues}\label{cresolution} 

 Figures~\ref{fig3} \& \ref{fig4} provide insight into the zoning required to spatially resolve the thin lower boundary layer.  The ``Perth" simulation of the O+O shell, at $1536 \times 1024^2$, already seems to be reducing the ``spike'' into a merger with  the ``slope'', although we have few turnovers for such a short duration in time (see Table~\ref{table2}).
 Errors in the dissipation rate due to coarse zoning (the ``spikes'' in Figs.~\ref{fig3} \& \ref{fig4}) appear to be converging toward zero with higher resolution, while the ``slope'' (anomalous dissipation due to the cascade in the $\theta \leq 1/3$ region) is robustly nonzero. 

The composition profiles also converge to a fixed shape \citep[][their Fig.~16]{andrea2}. This is due to dissipation, but
{\em a dissipation which is constrained by the 
turbulent cascade} (Fig.~\ref{fig33}). 
A turbulent boundary layer develops between convective and non-convective regions, which ILES seems to represent robustly.

\begin{deluxetable*}{lllllllll}
\tablewidth{250pt}
\tablecaption{Dissipation lengths: C+C shell$^a$, $\Delta_{cz}  \sim 3.0 H_P$   \label{table4}}
\tabletypesize{\small}
\tablehead{\colhead{stage}  & \colhead{zoning} & \colhead{$H_P$} & \colhead{$\Delta_{cz} $} &\colhead{$\ld$} & $\ld / \Delta_{cz} $  &$\ld/H_P$ & $ v_{\rm rms}$ & $t_{\rm ave}/t_{\rm to}$ \\
 & depth$^3$ & $10^8$cm & $10^9$cm & $10^9$cm & $={\cal C}$ & &  $10^6 {\rm cm\, s^{-1}}$
}
\startdata
C shell  & $128^3$   & $3.22$ &  $1.030$  &  $1.08 $ & 1.05  &3.35  & $3.76 $ &  3.99\\
C shell  & $256^3$   & $3.47$ & $1.033 $&  $1.04 $  & 1.01  &3.00 & $4.36 $ & 4.37 \\
C shell  & $512^3$   & $3.41$ & $1.025 $ &  $1.03 $  & 1.005 & 3.02 & $4.34 $  &  10.6\\
%C shell  & $768^3$  & $1.027 $ & $ 2.54 $ &    \nodata      & \nodata        & $ 4.42 $& 1.5 \\
C shell  & $1024^3$ & $3.43$ &  $1.009 $&  $1.09 $  & 1.08 & 2.94 & $3.93 $ &  2\\
%C shell$^d$  & $1536^3$ &  $1.048 \times 10^9$&  $2.90 \times 10^9$  & \nodata & \nodata & \nodata &  0.6\\
\enddata
\tablenotetext{a}{\cite{andrea}; boosted carbon burning,  pseudo-cartesian grid.}
%\tablenotetext{e}{\cite{andrea2}, carbon burning, pseudo-cartesian grid.}
%\tablenotetext{g}{\cite{porterwoodward} deep atmosphere. Note: hard boundaries, $\gamma=5/3$.}

\end{deluxetable*}

\placetable{2}

 A related problem is that of the resolution of entrainment \citep{ma07b,andrea,laura}, which may involve the turbulent boundary layer.
 \cite{woodward2015} have shown that the entrainment rate in a simulation depends upon the zoning, and with due caution,
 suggest that their $1536^3$ may already capture the Kelvin-Helmholtz waves (which they and we believe drive the entrainment).  
 This seems consistent with our ``Perth'' simulation, which does not have the factor of 3 zoning
 advantage of piecewise-parabolic boltzmann advection of  \cite{woodward2015}, but does have the improved brute-force resolution of a ``box-in-star'' grid \citep{321D},  which allows local enhancement of zoning in the box relative to the whole star.

\subsubsection{Dynamic constraint and dissipation anomaly}\label{s-dyncon}

The Onsager conjecture (\S\ref{sLars}) implies that the high frequency fluctuations (turbulence) will cause the spatial part of the advection operator to act as damping {\em on average} (\S\ref{sReystress}). 
This is equivalent to $\ell_d \approx \Delta_{cz}$ in Table~\ref{table4}.  Even with no explicit dissipation term ($\nu_{NS}\equiv 0$), our ILES generates a spontaneous turbulent damping which is equivalent to Eq.~\ref{eq-Rstress}; see also \cite{dls13,eyink-comp}. This adjusts to balance turbulent  driving, on average, in a steady state.

The dissipation is  not an adjustable parameter, but  due to shocks in the nearly inviscid flow.
This gives an added statistical constraint on the self-consistency of the flow: on average the damping and driving must balance. It also implies a constraint for the velocity spectra in Fig.~\ref{fig33}, which 
fit the \cite{kaneda2003} curve.

It is striking that by purely numerical experiments with the Euler equations, without using any viscous term, we recover Eq.~\ref{kolmogorov}, which is the Kolmogrov result for dissipation in a turbulent cascade.  

  As we refine our zoning the global dissipation rate converges quickly, as \cite{onsager} anticipated; for $256^3$ zoning and higher, the total dissipation rate changes little. %}
There is no kinematic viscosity $\nu_{\rm NS}$. %except implicitly through the condition that t
The effective value of the Reynolds number 
(\S\ref{eq-Re}) must be sufficiently large (\citealt{onsager,eyink}), so that %which is consistent with the idea that  
shocks provide the disssipation in the inviscid limit, which they do.

\section{Strongly-stratified convection}\label{Sstratified}

The most strongly stratified convection occurs near stellar surfaces (atmospheres). 
This brings additional complications, some of which may not be negligible.

The base of a convective shell has higher sound speed and smaller scale height, so the Courant condition on the time step is more restrictive. 
%Neutrino cooling may be too  weak to significantly accelerate the evolution for H and He burning. 
These effects cause fluid dynamic simulations to become more challenging. While our previous ILES are well suited for multi-processing, scaling well on multiple cpus as well as graphics cards at a convective Mach number $\geq 0.005$, this is not yet true for lower Mach number flow.

Do the neglected acceleration terms (\S\ref{s-added}) become important with strong stratification?
Strong stratification breaks the up-down symmetry (assumed in MLT), so that a strong negative flux of TKE  occurs,  giving a strong nonlocal, radial coupling in the flow.

In a co-moving frame, fast, dense, narrow downflows (which move into higher pressure) must be balanced by slow, less-dense, wide upflows (which move into lower pressure), as was the case for weak stratification.
However, as the up-down asymmetry grows, acceleration by ``pressure dilatation''  can not be ignored, giving negative fluxes of TKE.
LES of  strongly stratified atmospheres \citep{sn89,nsa,alns00} showed such negative TKE fluxes.
\cite{ma10apss} showed that this was a feature of strong stratification, not of top versus bottom driving. {\em MLT ignores the fluxes of turbulent kinetic energy, not because they are small relative to enthalpy fluctuations (they are not; see Eq.~\ref{mlt}), but because they are assumed to cancel by up-down symmetry }\citep{bv58,bv89}. 

\subsection{RG simulations}\label{RGsims}

We base our discussion of strongly-stratified convection on the ``red giant'' (RG) simulations of \cite{viallet2013}, 
which were done with the MUSIC code \citep{viallet2011}.
The Euler equations were solved with implicit time stepping. 
MUSIC allows larger time steps in the deep, most nearly static regions, but at the expense of higher computational cost per step. MUSIC allows small flow speeds but does not ignore sound waves (i.e., no anelastic approximation). 
Time steps were chosen based on a causality (Courant) constraint on {\em fluid flow}, instead of {\em sound speed} as in PROMPI.  
Despite the completely different solvers, these simulations agree well with ILES from PROMPI where they overlap.  

Simulations  of convection in stellar atmospheres \citep{alns00} focus on the outer layers where the observed photon spectra are formed. \cite{viallet2011,viallet2013} focused instead on the deep layers\footnote{This might be where a tachocline would develop.} of the surface convection zone. % A complete picture requires both.

\subsubsection{The initial state}
 The stellar structure code used to construct the initial model is described in \cite{baraffe}.  MLT was used (with $\alpha = 1.7$) to treat convection, and the extent of the convective region was based on the Schwarzschild criterion. The structure was integrated from the photosphere down to 20\% of the stellar radius, avoiding the nuclear burning region. This initial stratification\footnote{Such a model would correspond to a $5 M_\odot$  star at the end of central He burning, finishing its blue loop and evolving toward lower $T_{\rm eff}$ and away from the red edge of the Cepheid instability strip \citep{baraffe}.} became a starting model for the 3D hydrodynamic code. 
 
 Convection extended down to $r \sim 2.3 \times 10^{12}$ cm, nearly half of the stellar radius. 
 The nuclear burning region was not included in the computational domain, instead a radiative flux corresponding to the stellar luminosity was imposed at the inner boundary.
 The surface layers were characterized by strong superadiabatic stratification, which would be explicitly resolved in stellar atmospheres simulations; as in \cite{viallet2011}, a proxy was used. % for the surface layers. 

%In contrast to simulations of stellar atmospheres, 
The RG simulations explicitly calculated a dynamic lower boundary for turbulent convection. 
A boundary layer was formed, similar in character to that at the lower arrow in Fig.~\ref{fig34}.
%Although limited by 
Despite coarser zoning, the RG lower boundary resembles that in Fig.~4, \citealt{321D}, which shows vertical and horizontal velocities, composition gradient, and wave emission into the convectively stable region.
Below this is a non-convective radiative region. 

The steep gradient approaching the atmosphere was replaced by an approximate boundary condition (Newtonian cooling), giving a depth of 7.8 pressure scale heights of turbulence on the grid\footnote{
For the solar convection zone, a depth $\Delta_{\rm cz}\gtrsim 20 H_P$ would be better, but $7.8 H_P$ is sufficient  to show some of the qualitative modifications due to stratification.}.
The initial 1D model had a stratification of 14.3 pressure scale heights, with the additional 4.5 pressure scale heights absorbed into the Newtonian cooling region,  reducing the total stratification which was active dynamically. 

%Time steps in the implicit code \citep{viallet2011} were constrained by the fluid velocity (accuracy) rather than sound speed. 
These choices may be considered to complement  those used for 3D simulation of stellar atmospheres, emphasizing physical effects which are characteristic of the deeper layers, rather than the atmosphere, while avoiding artificial dissipation. 

\subsubsection{Flaws in Atmosphere and RG simulations}\label{s-flaws}

There are weaknesses in both the RG simulations (at the upper boundary), and the LES of atmospheres (at the lower boundary). In order to ``stabilize the numerical algorithms'' \citep{sn89,sn89s}, adjustment of the dissipation in stellar atmospheres seems\footnote{This appears from a visual comparison to demonstrably turbulent flows (e.g., \citealt{ma07b}).  It might also be expected from a ``bottleneck'' spectrum; \S\ref{tkespectrum}.  An authors' estimate  of the Re  in the bottom layers of the atmospheric LES, would be helpful.   From Eq.~\ref{eq-Re} we estimate $  {\rm Re^* } < 90 $ for \cite{tramp-stein},  a factor of $ 4$ higher (360)  for \cite{magic2013}, but smaller for ``stabilized''  flows.} to result in more laminar flow at the bottom of the computational volume, even though the plasma there  should be highly turbulent. Such a numerical effect could distort fast downflows. In \S\ref{LES}  and Fig.~\ref{fig33}, hyperviscosity codes were shown to have a defect (the ``bottleneck'') in the velocity spectra relative to ILES and higher resolution DNS. This flaw seems to have little effect on the predicted stellar spectra \citep{nsa,magic2013}, but may affect the deep flows \citep{hanasoge-gizon}.

In the RG simulations, the Newtonian cooling approximation has a similar effect of smearing fast downflows. %, but by the choice of boundary conditions. 
These ILES  should have a deficit, relative to reality, of narrow fast downflows  at the top of the convective region, which in stellar atmospheres would already appear in the 4.5 scale heights absorbed in  the Newtonian cooling boundary, and which may propagate to the lower boundary as even narrower downflows for high Re.
The two types of simulation may be wrong in the same sense. %New simulations at higher resolution would be prudent.

 The integral Re for the solar convection region, Re$\,\sim10^{14}$, is far higher than our attainable computational one, Re$^*\,\lesssim 1.7 \times 10^4$ (see \S\ref{s-methods}, and \citealt{hanasoge-gizon,am16key}). Our estimated Re for the anelastic ASH code \citep{miesch09}, and for the hyperviscosity STAGGER  code  \citep{nsa}, is far lower than for the $1024^3$ \citep{andrea} and $1536^3$  \citep{woodward2015} ILES.
 %{\color{red}, or even the MUSIC simulations.}

\subsubsection{Numerical Results }

With these issues in mind, we now examine strongly stratified convection from the MUSIC  simulations.

TKE fluctuates strongly; compare \citealt{alvio1} (Fig.~2) and \citealt{viallet2013} (Fig.~2). 
The lower boundary quickly relaxes to a turbulent boundary layer\footnote{The thermal relaxation time for integral scale radiative diffusion is much longer for laminar flow than for turbulence, due to the turbulent cascade to smaller scales.}, as seen in the weak stratification simulations \citep[see also][]{ehsan}.

 Table~\ref{table3} summarizes the RG simulations, to be compared with those in \S\ref{Ssubgrid} 
 for weak stratification. 
The lower resolutions shown, relative to Table~\ref{table2} \& \ref{table4}, reflect the greater computational load\footnote{MUSIC would be superior to PROMPI at the lowest Mach numbers, but neutrino cooling speeds the evolution for C, Ne and O burning, avoiding these lower velocities.} of the implicit hydrodynamic solver in the MUSIC code (see discussion in \citealt{ma07b,viallet2013,woodward2015}).  We estimate an effective Reynolds number Re$^*$ as in \S\ref{s-setup}.

\begin{deluxetable*}{lllllllll}[htb!]
\tablewidth{300pt}
\tablecaption{Dissipation lengths: Red Giant, $\Delta_{cz}  \sim 7.8 H_P$ on grid  \label{table3}}
\tabletypesize{\small}

\tablehead{\colhead{stage}  & \colhead{zoning} & \colhead{$\Delta_{cz}$(cm)} &\colhead{$\ld$(cm)} & $\ld / \Delta_{cz} $   & $v_{\rm rms}$(cm/s) & $t_{\rm ave}/t_{\rm to}$ &Re$^*$ \\
 & depth $\times$ area 
}
\startdata
RG$^a$   & $216 \times 128^2$ & $2.0 \times 10^{12}$ & $7.0\times 10^{11}$   & 0.350  & $2.27\times 10^5$ &3.3 & 256 \\
RG$^a$   & $432 \times 256^2$ & $2.0 \times 10^{12}$ & $7.7\times 10^{11}$ & 0.385  & $2.34 \times 10^5$ & 3.3 & 645\\
\enddata
\tablenotetext{a}{\cite{viallet2013}; red giant, only bottom of convection zone (see text). } 
\end{deluxetable*}

\placetable{3}

Differences caused by stratification are:
(i) the TKE flux is nonzero,
(ii) acceleration by pressure dilatation is nonzero, 
iii) waveforms change from rolls to plumes\footnote{Compare these 3D simulations \citep{viallet2013} to 2D MUSIC simulations of \cite{pratt}.}, and
(iv) the dissipation length $\ld$ (\S\ref{cascade})  decreases to less than the depth of convection $\Delta_{\rm cz}$. 
Strong stratification implies a more spherical geometry, so that large wavelength global modes may be important (e.g. \citealt{woodward2015,jones17}).
 Highly stratified convection with a localized source of turbulence will have a significant flux of turbulent kinetic energy, and thus is  nonlocal, unlike MLT.
 The radial flux of turbulent kinetic energy is negative and a significant fraction in magnitude ($\sim$35${\%}$) of the enthalpy flux.
Convection is driven to a significant extent ($\sim$40$\%$) by ``pressure dilatation'', $W_P$,  in addition to buoyancy. 

\cite{viallet2013} show that $W_P$ is negligible if $H_\rho \gg \ld$  and
becomes comparable to the dissipation rate for $H_\rho \ll \ld$. 
The ratio of turbulent ram pressure to local pressure, $\rho u'^2/P$, ranges from $\sim 2 \times 10^{-3}$ at the bottom, to $\sim 2 \times 10^{-2}$ at the top of the convection zone
 \citep[Fig.~3 and \S4.1.1]{viallet2013}.
In coordinates co-moving with the matter,  the quadratic dependence on $u'$ of turbulent pressure fluctuations act on the flow to give a {\em net acceleration  downward. The cubic dependence on $u'$ causes the fastest motion to dominate the TKE, so the tendency of buoyancy to produce negative TKE flux is strengthened. 

Because of the large heat content, the total energy flux is still positive.
The ratio of kinetic energy flux to enthalpy flux\footnote{Consider an idealized case: a parcel of gas is dropped, changing potential energy PE into kinetic energy KE. This is then thermalized, and slowly raised to its initial height. The ratio of kinetic energy to enthalpy is $1/(\Gamma-1)$.} is  $\sim (1/\Gamma - 1) = -0.4$ for $\Gamma = 5/3, $ which is approximately the same as the averaged numerical result for the deep interior of the red giant ($-0.35$).

The luminosity has a more complex dependence on adiabatic excess $\Delta \nabla$ than in MLT: it includes the luminosity due to turbulent kinetic energy $L_{\rm TKE}$, which is negative, and may affect 
 the inferred stellar luminosity.  In standard stellar evolution, this error must be absorbed  (if possible) into the calibration of  the mixing length.

\subsubsection{Asymptote for strong stratification}\label{strongstrat}

Our ILES have all the relevant acceleration terms (although modest resolution); we now explore approximations which capture the physics.
The turbulent kinetic equation 
 for a quasi-steady state is, 
\begin{eqnarray}
-\langle \overline{ {\bf u' \cdot } \nabla P' } \rangle  =&  W_B +W_P  - \rho \epsilon_K   \nonumber \\
& - {\bf \nabla_r \cdot (f_{k} + f_{p} ) .}
 \label{mlt_acc2}
\end{eqnarray}

A transition region from weak to strong stratification occurs 
as $ \Delta_{\rm cz} \gtrsim H_{\rho} $.
Each limit tends to be associated with a flow morphology.
If $\Delta_{\rm cz} \lesssim H_\rho$, $W_P$ is small, and we recover the weak stratification result (\em convective rolls}, \S\ref{Ssubgrid}).
In the strong stratification case,  $W_P$ adds to the driving of {\em convective plumes} \citep{viallet2013}. 

Averaging over several turnovers is required for a sufficiently smooth curve to be 
 an extension for the time independent cubic equation of \cite{bv58}.
$W_P$, $W_B$ and $ \epsilon_K $ track each other, in a cycle of potential energy, kinetic energy, and internal energy maxima \citep[\S5]{viallet2013}. 

The simplest case is a steady state with no convective zone growth,  thin boundary layers, 
 and small wave flux (\citealt{viallet2013}, Eq.~32, Fig.~7), so 
\begin{equation}
W_B + W_P \approx \langle \overline{ \rho}\rangle \epsilon_K  . \label{eq-wbwpess}
\end{equation}
%For example, 
The buoyancy work is
\begin{equation}
W_B =   \langle \overline{ (\rho'{\bf u' \cdot g}) \beta_T \Delta \nabla } \rangle 
\approx  \langle \overline{ \rho } \rangle\langle \overline{ v {\cal B} } \rangle
 %\label{buoy}
\end{equation} 
%involves a integration over angles, $\langle (\cdots) \rangle$, and an average over several turnover times,
%$\overline{(\cdots)}$.
Since each term in Eq.~\ref{eq-wbwpess} has been averaged in space-time, this defines an envelope of values  for a quasi-steady state. %Fig.~\ref{fig35} shows the evolution of the turbulent kinetic energy for the C+C burning shell, as well as approximations of 3, 4 and 5 frequencies \citep{andrea2}.

For strong stratification\footnote{See discussion of ``pressure-dilatation defect'' in \citep{eyink-comp}.}, $\Delta_{\rm cz} \gg H_\rho$,  and using Eq.~\ref{pdil}, we have
\begin{equation}
W_P =  \langle \overline{ P' {\bf \nabla \cdot u' } } \rangle 
      \sim - \langle \overline{\rho} \rangle \langle \overline{ u_r^3 }\rangle / H_\rho , %\label{pdil}
\end{equation} 
so that a generalization of the MLT cubic equation is
\begin{equation}
  \langle \overline{ v {\cal B} } \rangle \approx
    \overline{ \langle u_{\rm r}^3  }\rangle /\Delta_{\rm cz}
    +  \langle \overline{ u_r^3 }\rangle / H_\rho \label{eq-asymp1}
\end{equation} 
where {\em both buoyancy and pressure dilatation affect the acceleration of flow, but with opposite signs. }  Turbulence resists being compressed into a smaller volume.\footnote{This may be important for core collapse supernovae.}
 The $v {\cal B}$ term is important for buoyancy braking at boundaries \citep{321D,alvio1}.

If $\Delta_{\rm cz} \gg H_\rho$, this gives the effective damping length $\ld$ in the strong stratification limit (convective plumes), so we obtain a strikingly simple limiting case,
\begin{equation}
\alpha \equiv \ld / H_P \approx H_\rho/ H_P   = \Gamma_1  , \label{gamma1}
\end{equation}
where $\Gamma_1$ is the adiabatic index of \cite{chandra}: 
\begin{equation}
\Gamma_1 = \beta + {(4-3\beta)(\gamma - 1) \over \beta +12(\gamma-1)(1-\beta)} ,
\end{equation}
and where $\gamma = c_P/c_V$ is the ratio of specific heats and $\beta$ is the ratio of gas pressure to total pressure. 
The appearance of $\Gamma_1$ is due to the historical choice of using the {\em pressure} scale height $H_P$ for measuring mixing length.

Even the averaged value of
Eq.~\ref{gamma1}
 can be  complex in regions of partial ionization and dissociation, so a detailed atmosphere may be required for precision (but see \S6; \citealt{magic2015}).

\subsubsection{Length scales in strong stratification}

It is challenging to resolve a lower boundary for both strong stratification 
and vigorous turbulence. The numerical data trend toward the asymptotic limit we have derived, even though our largest stratification on the grid is only $\Delta_{\rm cz}/H_P \sim 7.8$, much less than the Sun ($\Delta_{\rm cz}/H_P > 20$).

{\em For strong stratification, the ``effective mixing length''  asymptotically approaches the density scale height, due to the action of  the pressure dilatation term (not in MLT).} This confirms a ``saturation'' behavior noticed by \cite{amy09vel} based upon cruder simulation data.
For strong stratification and complete ionization, the ``mixing length''  ($\ld/H_P \sim 5/3$) really does resemble a universal constant
(see \S\ref{alpha-hrt}),
a behavior unexplained  in MLT \citep{alvio87}.

\section{HRD estimates of $\alpha_{\rm MLT}$}\label{alpha-hrt}

As we saw in \S\ref{BVtheory}, specifying a mixing length in MLT is equivalent to specifying a global damping length for turbulence, to the extent possible in an incomplete theory like MLT.  
In \S\ref{Ssubgrid} and \S\ref{Sstratified} we evaluated damping lengths {\em self-consistently with the predicted flow, and accounted for anomalous dissipation \citep{onsager},} for both weak and strong stratification, without using  any parameterized viscosity. 
We now compare our estimates to   published values of ``the stellar mixing length parameter"  from a representative variety of methods. 
    
\subsection{Stellar Evolution alone} \label{s-art1}

Table~\ref{table1} shows  several  published values of the constant $\alpha_{\rm MLT} = \ell_{\rm MLT} / H_P$ from MLT calculations of stellar evolution; these have been successfully used to fit the observed HR diagram (HRD). 
For the main sequence to the giant branch, the stellar evolution values are essentially tests of the deep properties of surface convection zones of stars of mass $M \gtrsim 1 M_\odot$; apparently they have strong stratification and $\alpha_{\rm MLT} \approx \Gamma_1 \sim 5/3$, as expected from \S\ref{strongstrat}. 
Why should a single preferred value of  $\alpha_{\rm MLT} $  work for ``essentially all stellar evolution sequences'' (\citealt{alvio87})?
The weak stratification would not give a such agreement 
because of changes in the depth of convection zones with stellar mass and evolution, but strong stratification does. 
    
The stellar evolution sequences are in reasonable agreement despite the fact that these values of 
$\alpha_{\rm MLT}= \ell_d/H_P \sim \overline{ \langle \Gamma_1 \rangle} $ conflate several distinct issues.
They represent space-time averages over the convective region rather than local values.
For example, $\Gamma_1$ may vary significantly through a region of partial ionization, but 
such effects tend to be diluted by averaging. 
The effects of ignoring (i) levitation, due to convective pressure fluctuations \citep{ludwig-rg},  
(ii)  the flux of turbulent  kinetic energy $f_{\rm k}$, and turbulent heating in the background energy equation \citep{porterwoodward,viallet2013}, and (iii) entrainment due to turbulence at the boundaries \citep{ma07b}, all make the stellar models incomplete. 
These errors are lumped together, 
as are errors in composition \citep{asplund}, whenever the free parameter $\alpha_{\rm MLT}$ is adjusted.

\begin{deluxetable*}{llll}[t]
\tablewidth{880pt}
\tablecaption{Evolutionary estimates of $\alpha_{\rm MLT}=\ell_{MLT}/H_P$ \label{table1}}
\tabletypesize{\small}

\tablehead{ \colhead{reference to code calibration} & \colhead{$\alpha_{\rm MLT}^a$} &\colhead{($\delta_\epsilon$)$^b$} 
}
\startdata
  This paper, strong stratification & $\Gamma_1 \sim 5/3 \sim 1.67$ & $\sim 0$ \\
\\   
Some stellar evolution sequences:\\
  \cite{ekstrom} &1.6467($M\leq1.25 M_\odot$) & $\sim -0.02$ \\
  \cite{ekstrom} &1.65  ($M>1.25 M_\odot$)      & $\sim -0.02$ \\
  \cite{bressan} & 1.74 & $\sim +0.07$ \\
  \cite{maeder} & 1.6   & $\sim -0.07$ \\
  \cite{baraffe} & 1.7 & $\sim +0.03$ \\
  \cite{young} & 1.6 & $\sim -0.07$ \\
\enddata
\tablenotetext{a}{Warning: these values conflate several aspects of convection; see text.}
\tablenotetext{b}{Here $\delta_\epsilon = \alpha_{\rm MLT}-5/3$ which is a measure of the consistency of the stellar models with 3D ILES simulations of ionized, stratified turbulent convection.}

\end{deluxetable*}

For most stars, such averages give $\beta \sim 1$ and
 $\alpha = \ell_d/H_P \rightarrow \Gamma_1 \sim \gamma \sim  5/3$. 
The difference  
 $\delta_\epsilon \equiv \alpha_{\rm MLT} - 5/3 $ 
is a measure of consistency between the stellar models and pure 3D simulations of deep turbulent convection, ignoring recombination (atmospheres) and pairs (very late stages). 
The variation in $|\delta_\epsilon |$ in Table~\ref{table1} is primarily an indicator of envelope/atmospheric effects, so $\alpha_{\rm MLT} \sim 5/3$ is a surprisingly good approximation.  
 
\cite{alvio87} noted that
``...one single value of the free parameter gives a very good fit to stellar radii from the main sequence to the tip of the giant branch...", so that ``MLT is not that bad''.

For extremely massive stars, which are rare, $\Gamma_1 \rightarrow 4/3$.  
In regions of partial ionization  $\Gamma_1$ may drop as low as $1.1$ \citep{young} in the envelopes, as it also does in regions of O+O burning (Table 5 in \citealt{amy09vel}) due to electron-positron pair formation. However, global hydrostatic equilibrium requires that the average over the whole star satisfies $\langle \Gamma_1 \rangle \geq 4/3$ \citep{wda96}, giving a lower bound for a quasi-static object.

The arguments above suggest (i) why this is so (MLT mimics the universality of the turbulent cascade in a stratified medium), (ii) what the value of that single parameter should be (it asymptotically approaches $\alpha_{\rm MLT} \sim \Gamma_1 \sim 5/3$ for strong stratification and complete ionization), and (iii) some possible causes of deviations (e.g., different atmospheres produce variations of 20 percent or so;  Table~\ref{table1b}).

Stellar evolutionary sequences use the simplest approach to atmospheres, with a mixing length which is constant in time and space to define the background structure, separating the structure calculation from that of a simplified photon spectrum at the surface.

\subsection{Stellar atmospheres}\label{s-art2}

In \S2 through \S4, a self-consistent treatment of convective physics in deep interiors of stars was developed, having a turbulent energy cascade whose flux was determined by anomalous dissipation \citep{onsager}.  No parameterized viscosity was needed. Our ILES represent the highest Re possible with the zoning used.  They tend toward Onsager's ``ideal turbulence'',  the inviscid limit which is the appropriate generalization of  ``universal turbulence'' 
\citep{batchelor60} for stellar scales. 
In \S5  acceleration by pressure dilatation was added to describe turbulent convection in strong stratification (plumes), and an asymptotic limit was found, suggesting 
$\alpha=\ld/H_P=\Gamma_1 \sim 5/3$, for the case of complete ionization.
 
Surface convection requires additional physics for radiation flow in an atmospheric layer, which limits the Re attainable because of the added computational load.
Attempts to infer ``effective''  MLT mixing-lengths from 3D LES of convective stellar atmospheres \citep{sn89,nsa,magic2015,magic2016,lfs99,muthsam} have focused on photospheric physics rather than fluid flow and turbulence. In practice this has required sacrifice:  crude treatment of shocks, low Re, and an inability to accurately describe the anomalous dissipation (\S2, 3). 

Fortunately the observed photon spectra can be reproduced to high accuracy using a coarse velocity field, one that barely resolves an intergranular lane (e.g., \citealt{dravins90}). While photon spectra may  require exceptional effort to replicate, observations only require coarse structure in the velocity field, such as can be provided by a low Re  representation of the flow. 
Despite their complexity, the photon spectra can only weakly constrain the deep flow; asteroseismology is necessary.

 Table~\ref{table1b}  shows  several  representative  published values of $\alpha_{\rm MLT} = \ell_{\rm MLT} / H_P$ recommended from low Re simulations of stellar atmospheres.   
 \cite{magic2015} have 
 carefully reviewed the work cited; see also \cite{magic2013,magic2013b,magic2015,magic2016,magic-solar}. 
 Using significantly higher resolution LES with the STAGGER code, they find basic agreement with \cite{tramp-stein}, 
 for calibrations with the adiabatic entropy value, and deviations up to $\pm 20 \%$ from the solar value.

  If our arguments and the 3D atmospheres are both correct, we should have agreement in the complete ionization limit ($\langle \Gamma_1 \rangle \sim 1.67$),  and should show more discrepancy at low temperatures and pressures, which they do.

\begin{deluxetable*}{llll}[h]
\tablewidth{900pt}
\tablecaption{Atmospheric estimates (at low Re) of $\alpha_{\rm MLT}=\ell_{MLT}/H_P$ \label{table1b}}
\tabletypesize{\small}

\tablehead{ \colhead{reference to code calibration} & \colhead{$\alpha_{\rm MLT}^a$} &\colhead{($\delta_\epsilon$)$^b$} 
}
\startdata
  This paper, strong stratification & $\Gamma_1 \sim 5/3 \sim 1.67$ & $\sim 0$ \\
%\\   
Some stellar atmospheres$^{c}$: \\
  \cite{lfs99}($s_{\rm ad}$, 2D)$^d$ & 1.3 to 1.75 & -0.37 to +0.08 \\
%  \cite{tramp} & 1.648 to 1.767&  -0.019 to +0.100 \\
  \cite{tramp-stein}$^{e}$ & 1.67 to 2.20 & 0.0 to +0.53 \\
  \cite{tramp-stein}$^{e}$ (Sun)& 1.76  & +0.09 \\ 
  \cite{magic2015} ($s_{\rm ad}$) &1.7 to 2.3 & 0.0 to +0.7 \\
  \cite{magic2015} ($\Delta s$)     &1.8 to 2.4 & 0.1 to +0.8 \\
  \cite{magic2015} (Sun)&1.98 & 0.31 \\
%  \\
%  Standard solar model calibrations: \\
 % \cite{serenelli}& 2.18$^c$ & $\sim +0.51$ \\
 % \cite{serenelli}& 2.11$^d$ & $\sim +0.44$ \\
  Some Combinations$^{f}$ \\
  GARSTEC-Eddington & 1.70 & +0.03\\ 
  MESA-Eddington & 1.67 & +0.00 \\
    GARSTEC-3DAtmosphere & 1.81 &  +0.14\\
  MESA-3DAtmosphere & 1.83 & +0.16 \\
\enddata
\tablenotetext{a}{Warning: these values conflate several aspects of convection and stellar structure.}
\tablenotetext{b}{Here $\delta_\epsilon = \alpha_{\rm MLT}-5/3$ is a rough measure of the consistency of the models with 3D  ILES} simulations of ionized, stratified turbulent convection; see text.
\tablenotetext{c}{To aid the reader, we use published material in an attempt to place very different and complex computations on a consistent scale. See \cite{tramp-stein,magic2015}.  Experts are invited to suggest improvements.}
\tablenotetext{d}{\cite{lfs99} use the CO5BOLD code.}
\tablenotetext{e}{\cite{tramp-stein} use ``mass-mixing length" as $\ell/H_P$; see also \cite{magic2016}.}
\tablenotetext{f}{\cite{mosum} compare a classical method (Eddington outer boundary) to fitting with 3D  atmospheres.}

\end{deluxetable*}
 
  The survey  of  \cite{magic2015}, their Fig. 2, spans a typical range of variation. 
  The low gravity-high $\log T_{\rm eff}$ edge of their Kiel diagram, for MLT parameter calibrated with the entropy of the convection zone, is consistent with our estimate of $\alpha_{\rm MLT} \sim 1.67$. 
  This boundary is roughly paralled by contours of $\alpha_{\rm MLT}$,
  down to minimum temperature and maximum gravity, for both solar metalliticity, [Fe/H] = 0, and -2.0.
 The reason for such behavior is the recombination of H ions, which gives large variation of opacity and 
 $\Gamma_1$ through the surface layers.  Here $\langle\overline{\Gamma_1}\rangle$ is not a local quantity, but a space-time average, and the general features of the Kiel diagrams of \cite{tramp-stein,magic2015,magic2016} can be so understood. 

  Experience from stellar evolution has shown that $\alpha$ appears to require an increase for giants ($>2.0$), which is also indicated in Table~\ref{table1b}.
 The Sun represents a step in this trend; for $M\gtrsim1 M_\odot$ stars spend the main sequence and the ascent of the giant branch in the $1.67 \lesssim \alpha \lesssim 1.8$ region; such values are common in stellar evolution lore (\S\ref{s-art1}).

The last four entries in Table~\ref{table1b} compare results from the MESA and the GARSTEC stellar evolution codes, with (i) a simple Eddington atmosphere, and (ii) a full 3D fluid-dynamic atmosphere \citep{mosum}. The combined differences in codes and atmospheres is less than 10\%. It appears that the inferred value of $\alpha_{\rm MLT}$ is surprisingly insensitive to turbulence, and even to the widely varying details of atmospheric physics  (\citealt{tramp}).  The simplest boundary condition (Eddington) 
gives close to our estimate for full ionization, only shifting by 10\% from the average with use of  3D atmospheres.
%This applies to statistical samples; for individual stars the larger range (20\%) estimated by \cite{magic2015} is relevant. 

%--Generally: Deeply stratified convection zones are referred to as "surface convection". We usually distinguish between surface convection (~ atmospheres) and envelop convection. Maybe you are using surface convection = envelope convection?
%--Table 5 (alpha, atmospheres): The large values of alpha are probably for RGB stars. This is something we see a lot, alpha appears require an increase for giants, $> 2.0$.
%--Sec. 6.21: This is a bit long for its purpose I think. Suggest making it 30-50\% shorter.

\subsection{The Depth Problem for Velocity}
The geometric confusion inherent in 1D models faded with the introduction of 3D fluid dynamics to the radiative transfer problem, resulting in a major advance  \citep{sn89,nsa}.
Descriptions of granule-size, flow magnitude, and line-formation and many other surface details improved considerably. Even 3D LES with MHD in a limited volume have allowed sunpot simulations to become feasible \citep{rempel,kritsuk}.

Despite successes at the surface (depth $<10^8\rm cm$) for a quiet, nonrotating Sun, there are
problems with velocity at greater depth. %convection. 
\cite{gizon-birch,hanasoge-gizon,greer15} summarize evidence suggesting  that
the amplitude and depth scaling of the convective velocity are in serious question.
%The flow model may be faulty, at least in that it does not extrapolate to depth. 
%As shown in \S\ref{s-methods}, the hyperviscosity fluid flow methods have issues
%related to prediction of spectra of turbulent kinetic energy.

Our ILES may give insight, having higher Re and using different methods.
In \S2, Fig.~2, the bottleneck effect was illustrated for the spectrum of  turbulent velocity;  while small for
our ILES, it is larger for hyperviscosity LES. The strategy for hyperviscosity methods is to add a large, high order dissipation at the grid level to avoid steep gradients. %The idea is to move complex  physics away from these edges.  
However we find that {\em  turbulent dissipation is due to mild shocks}, so the turbulent cascade would be truncated by hyperviscosity at small scales, as shown in Fig.~2. Shocks captured with pseudo-viscosity will also be smeared, requiring more zones for a given Re.

In our ILES we find strongly nonlocal flow, with threads and filaments forming with increasing {\em Reynolds} number Re (see Fig.~\ref{fig34} and movies referred to in \S\ref{S-ILES}).
We find intermittency to be a fundamental feature, 
with filamentary structure increasing with Re (Fig.~\ref{fig34}). Our ILES 
agrees fairly well with the dissipation of converged DNS  at similar Re  (Fig.~\ref{fig33}).

\cite{spruit97} suggested that a new paradigm for turbulence was needed to deal with the depth problem, and presented one inspired by experimental work of Libchaber and associates with liquid helium, which were interpreted as strongly nonlocal flow, with threads and filaments formed at the boundaries.
These experiments  \citep{libchaber,castaing,wu92} give insight into flows of large {\em Rayleigh} number (up to Ra $\sim10^{14}$),  but with nonslip boundaries.
Howerever,  stellar boundaries are ``free-slip'', and differ drastically between  external (atmospheres) and internal cases.
\cite{brandenburg} connects these inconsistencies with {\em  entropy rain} which accompanies the large negative kinetic energy flux at high stratification.
It seems likely that simulations of high Re could preserve such features.

\cite{greer16} have introduced a dynamic time delay into their analysis and infer ``a downflow rate consistent with anelastic simulations and inconsistent with a 3D cellular structure''.  
However, from their Fig. 3 \& 4, the flow may be consistent with the intermittent plume geometry which we find at high Re. 
The trajectories of these descending plumes should be arcs, as they tend to conserve specific angular momentum \citep{am10rot}  and are intermittent. Viscosity (low Re) tends to smear such plumes; high Re seems to allow more coherence (descending tendrils to a shear layer, and interaction between intermittency and rotation?).

%\cite{bekki} suggest that newly revealed effects of small-scale magnetic fields inhibit the turbulent mixing of entropy between upflow and downflows, and strengthen the plume-like nature of the flow.
%\cite{hotta} compare results with and without the solar surface in the local domain, and without the surface in the full sphere. The calculations do not include rotation and magnetic fields. They conclude that the surface region has an unexpectedly weak influence on the deep convection zone; see however \citep{spruit97,tramp-stein}. 

Perhaps problems at depth should be no great surprise; any extrapolation from Re $<100$ (STAGGER)\footnote{Re* $ < 360$ for STAGGER \citep{magic2016}.} to Re $\sim10^{14}$ (Sun) requires support.  
We may begin to extrapolate to higher Re using our presently available (but still  meager\footnote{Our ILES do extend  to include the region of ``mixing transition'' of \cite{dimotakis}, i.e., Re $\sim10^4$.
Our dissipation is not parameterized, but is automatically produced by weak shocks from turbulence itself (the ``rough'' flow of \citealt{onsager}); see also \cite{perry}.})
resolution study \citep{andrea,andrea2,321D,alvio1}.

\section{Summary}\label{Conclusion}

In this paper, and Paper I (\citealt{alvio1}), we have analyzed stellar convection theory
with our library %of 3D turbulent 
of implicit large eddy simulations (ILES) of 3D compressible turbulence (using realistic physics for stellar interiors)  which have no restrictions  for the classical \cite{bv58}  problem other than zoning.  

Our ILES method is supported by new developments  both 
 in mathematics  \citep{dls13,buckmaster,constantin,isett} and in physics \citep{eyink-comp,eyink-rel,eyink,sree,perry}, related to the Onsager conjecture \citep{onsager} and ``dissipative anomalies''.
We solve the Euler equations, appropriate to stars (Fig.~\ref{fig34}), which are  the high Reynolds number (inviscid) limit of the Navier-Stokes equations.  At the zone level, the nonlinear Riemann problem is solved by a high order shock-capturing algorithm (PPM, \citealt{cw84}), giving H\"older continuity.
% Driven by shocks from ``blowup'' of the Euler equations \citep{frisch}, 
 In our ILES, dissipation occurs by {\em nonlinear fluid dynamics}, without requiring a pseudo-viscosity or a Navier-Stokes term for molecular viscosity (\S\ref{cascade}); we term this feature ``Onsager damping".  
We find that the effective Reynolds number Re* depends only on zoning (\S\ref{s-setup}); finer zoning allows higher Reynolds numbers (lower effective viscosity).
We suggest that our ILES represent a numerical approximation to the inviscid limit, or Re*$\,\sim 10^4$.

The ILES results compare well with high resolution DNS   (Fig.~\ref{fig33}) %LES, 
and seem consistent with experimental trends.
Essential  features of real turbulence  develop automatically: intermittency, coherent structures, and turbulent boundary layers \citep{warhaft,sree}.
These features are  called ``anomalous''  because they go beyond the ``universal'' theory of a turbulent cascade (\citealt{kolmg41}), but are present in experiment, in DNS, and in our ILES. The dissipation rates agree 
with those of DNS and experiment (see also \citealt{sytine}).

Turbulence is ``self-limiting". The non-linear physics of turbulence  determines the rate of conversion of TKE into heat. This dynamically-controled limit sets the maximum TKE luminosity for a given structure and input luminosity. In turn, this sets a limit for the heat transport by turbulence, which will affect the structure of stars. Excessive luminosity causes vigorous turbulence and shocks, which cause expansion and possibly mass loss by ejection or by winds.

In the weak stratification case, we consider O+O and C+C shells (\S\ref{Ssubgrid}), in which the evolution is accelerated sufficiently for the use of time stepping at the causal (Courant) limit (\S\ref{s-methods}). No assumptions need be made about dissipation, which is calculated self-consistently with turbulence, and confirms our previous analysis. This extends the ``universal equilibrium theory'' of \cite{batchelor60} to a compressible case, in which dissipation is provided by mild shocks from ``blowup'' of the Euler equations, not NS viscosity.

For strong stratification (\S\ref{Sstratified}) we are guided by the red giant simulations of \cite{viallet2013}. Acceleration by pressure dilatation becomes important. %We extend  to 
In the asymptotic limit we find that the turbulent damping length equals the density scale height, on average. 

We conclude that  the apparent ``universality'' of the mixing-length \citep{alvio87} is a consequence of dissipation in deeply stratified convection. Our predicted value of $\alpha_{\rm MLT}=\ell/H_P \approx 5/3$ fits the data well,  if $\pm$20$\%$ effects of atmospheres are allowed for, and agrees at that level with estimates involving detailed atmospheres using hyperviscosity approximations \citep{tramp-stein,magic2015}. 

Using hyperviscosity methods to account for shocks in large Re flows, although imperfect ({\S\ref{s-methods}}),
could be adequate to provide a base upon which to build 
stellar atmospheres \citep{tramp-stein,magic2015}, but Re should be established before drawing conclusions regarding turbulence, or deeper flows. 

We suggest that for astrophysical conditions (very high Re, so $ 1/{\rm Re} \rightarrow 0$), the dominant dissipation occurs by nonlinear fluid flow (shock processes), and that the ``molecular'' viscosity $\nu_{\rm NS}$ commonly used in turbulence theory may not be relevant, its direct effects (e.g., shock thickness) being insigificant at large scales. Both DNS and ILES seem to converge to the same limit in this, the inviscid case  (Onsager's ``ideal'' turbulence; \S\ref{s-methods}).

%\begin{appendices}

\appendix

\section{Euler and Navier-Stokes equations}\label{a-E-NS}
The Euler equations are:
conservation of mass (baryon number):
%\begin{equation}
$\partial_t \rho + {\bf \nabla \cdot } \rho {\bf u } = 0,$
%\end{equation}
conservation of momentum:
%\begin{equation}
%\partial_t {\bf u } + ( {\bf u \cdot \nabla  }){\bf u} = -{ 1 \over \rho}{\bf \nabla }P - {\bf g}.
%\end{equation}
%\begin{equation}
$\partial_t  \rho {\bf u } + ( \rho {\bf u \cdot \nabla  }){\bf u} = -{\bf \nabla }P - \rho{\bf g},$
%\end{equation}
and conservation of energy in a co-moving frame:
%\begin{equation}
$dE/dt + PdV/dt = dq/dt.$
%\end{equation}
To get the Navier-Stokes equations (which often imply incompressibility) we (i) keep the conservation of mass equation as is, (ii) introduce a new term ($\bf \nu \nabla^2 \bf u$) in the momentum equation, so
\begin{equation}
\rho \Big [ \partial_t {\bf u } + ( {\bf u \cdot \nabla  }){\bf u} \Big ] = -{\bf \nabla }P - \rho{\bf g} + \nu \nabla^2 \bf u,
\end{equation}
and (iii) adjust the energy equation to include viscous heating (add $\bf u \cdot \nu \nabla^2 u $). See \cite{llfm} and
\cite{pope}. Setting the Navier-Stokes viscosity $\nu$ to zero reverts to the Euler equations, so they are the limit at infinite Re.

The Reynolds number itself is $$ {\rm Re}=  |( {\bf u \cdot \nabla  }){\bf u} /  \nu \nabla^2 \bf u | \sim  \ell  u /\nu_{\rm NS}, $$ which is the ratio of the inertial term to the dissipation term in the NS equation. There is an important structure factor which is suppressed here in the approximation on the RHS; it is the ratio of the shock (dissipation) width to the viscous damping length (\S\ref{s-euler}, \S\ref{s-setup}), which is assumed to be unity for the conventional NS equations. 

\section{The  codes}

\par The core of our multi-dimensional hydrodynamics code PROMPI is
the PROMETHEUS solver written by B. Fryxell, E. M\"uller and D. Arnett (circa 1988) 
which is based on the direct
Eulerian implementation of PPM \citep{cw84} with generalization to
non-ideal gas equation of state. PROMETHEUS also led to the FLASH and VH1 codes.
A short history of the PPM algorithm may be found in
\cite{woodward,porterILES1}.  PPM has front steepening, monotonicity 
(no Gibbs ringing) and low viscosity; the net effect on dissipation appears to be indirect, approaching the inviscid limit through an accurate representation of nonlinear fluid dynamics (\S\ref{sLars}).
The implicit MUSIC code is described in detail in \cite{viallet2011}.
Our codes solve the
Euler equations, to which we can add nuclear reactions and radiative
diffusion through an operator-split formulation.  

\par PROMPI has been adapted to many parallel
computing platforms (see below), using domain decomposition, the sharing of a
three zone layer of boundary values, and using the MPI message passing
library to manage interprocess communication.

PROMPI was designed \citep{ma07b} to explore nucleosythesis in deep convection, entrainment, and turbulence.
As we saw in \S\ref{Ssubgrid} it provides a self-calibration, and insight into dissipation as a process, not a parameter. 
Freed of the computational  burden of explicit atmospheres, higher resolution and Reynolds numbers are available.

 We may separate the physics of atmospheres  into two aspects: (1) line formation and (2) turbulence. We simply accept the first as stated, following \cite{magic2013}, and examine the second. 

\cite{freytag2013} describes an impressive list of improvements to fluid flow algorithms in CO5BOLD. At least one option (PP) resembles the algorithm in PROMPI \citep{ma07a}, so that CO5BOLD can be free of the problems of hyperviscosity seen in \S\ref{s-methods}. However the computational load of the atmospheric physics implies that useful zoning must be coarse relative to PROMPI, giving a lower effective Re. 

CO5BOLD has some notable successes. White dwarf atmospheres have required a variation of MLT, allowing small mixing lengths and shallow stratification.  This is a problem for standard MLT but not for 3D simulations (the damping length can be much smaller than a pressure scale height, \S\ref{Ssubgrid}). \cite{co5bold-wd} find
overshoot and departure from hydrostatic equilibrium are needed as well (see also \citealt{andrea,andrea2}).  Shallow convection is more sensitive to boundary effects \citep{321D} and steep gradients.

Red giants provide a different challenge: deep stratification. \cite{ludwig-rg} find a lower region dominated by convection and an upper region  by wave activity and  shocks. They find no MLT choice of mixing length and turbulent pressure which approximate their LES. This is reminscent of \cite{andrea,andrea2}, who show that for strong driving, the neglected acceleration terms  and shocks become important (\S\ref{s-added}). 

%In \S\ref{s-methods} we compared fluid dynamics methods. 
All the ``envelope''  entries in Table~\ref{table1b}  used LES, and most used the STAGGER hyperviscosity code. \cite{lfs99} used the CO5BOLD code.
As \S\ref{Ssubgrid} showed, our ILES automatically satisfy a dynamic constraint which hyperviscosity methods must parameterize.

Regarding the STAGGER code,
\cite{sn89} note: 
``Since this is compressible flow, shocks can form. These instabilities are removed and the code is stabilized by applying artificial diffusion to all the fluid equations." 
The Re at ``stabilization'' does not seem to be provided.
\cite{tramp-stein} construct a ``mass-mixing length", based on stratified flows and LES with the STAGGER hyperviscosity code. 
 Their simulations ``have as free parameters only the resolution of the computational grid used and the magnitude of the dissipation used to stabilize the calculations". 
These LES use an adjustable viscosity to
 replace the actual shock dissipation relevant for stellar scales.  
 As we have seen (\S\ref{Ssubgrid}), the turbulent dissipation rate determines flux of turbulent kinetic energy, so that they should not be connected by a freely adjustable parameter.

%\end{appendices}

%\acknowledgements
\begin{acknowledgements}
We thank Dr.-Ing. Andrea Beck, IAGD, Universit\"at Stuttgart, for insightful comments on the manuscript,
 Axel Brandenburg and Nils Haugen for providing the data from their study, used in our Fig. 2, and
 Prof. John Lattanzio for vigorous support and guidance for this project, and
 the Theoretical Astrophysics Program (TAP) at the University of Arizona, and Steward Observatory for support.
%This work was supported by resources provided by the Pawsey Supercomputing Centre with funding from the Australian Government and the Government of Western Australia.
This work used the Extreme Science and Engineering Discovery Environment (XSEDE), which is supported by National Science Foundation grant number OCI-1053575.  Some computations in our work made use of ORNL/Kraken and TACC/Stampede.

AC acknowledges partial support from NASA Grant NNX16AB25G. AC acknowledges the use of resources from the National Energy Research Scientific Computing Center (NERSC), which is supported by the Office of Science of the U.S. Department of Energy under Contract No. DE­AC02­05CH11231.

The authors acknowledge support from EU­FP7­ERC­2012­St Grant 306901. RH acknowledges support from the World Premier International Research Centre Initiative (WPI Initiative), MEXT, Japan. This article is based upon work from the €œChETEC€ COST Action (CA16117), supported by COST (European Cooperation in Science and Technology). CG, RH, and CM thank ISSI, Bern, for their support on organising meetings related to the content of this paper. CG acknowledges support from the Swiss National Science Foundation and from the Equal Opportunity Office of the University of Geneva.

This work used the DiRAC@Durham facility managed by the Institute for Computational Cosmology on behalf of the STFC DiRAC HPC Facility (www.dirac.ac.uk). The equipment was funded by BEIS capital funding via STFC capital grants ST/P002293/1 and ST/R002371/1, Durham University and STFC operations grant ST/R000832/1. This work also used the DiRAC Data Centric system at Durham University, operated by the Institute for Computational Cosmology on behalf of the STFC DiRAC HPC Facility. This equipment was funded by BIS National E-Infrastructure capital grant ST/K00042X/1, STFC capital grants ST/H008519/1 and ST/K00087X/1, STFC DiRAC Operations grant ST/K003267/1 and Durham University. DiRAC is part of the National E-Infrastructure.
We acknowledge PRACE for awarding us access to resource MareNostrum 4 based in Spain at Barcelona Supercomputing Center. The support of David
Vicente and Janko Strassburg from Barcelona Supercomputing Center, Spain, to the technical work is gratefully acknowledged.

S.W.C. acknowledges federal funding from the Australian Research Council through a Future Fellowship (FT160100046) and Discovery Project (DP190102431). This work was supported by computational resources provided by the Australian Government through NCI via the National Computational Merit Allocation Scheme (project ew6), and resources provided by the Pawsey Supercomputing Centre which is funded by Australian Government and the Government of Western Australia.
\end{acknowledgements}


\begin{thebibliography}

%
%\bibitem[Abarzhi(2010)]{abarzhi10} Abarzhi, S. I., 2010, Phil. Trans. R. Soc. A, 368, 1809
%
%\bibitem[Abarzhi \& Rosner(2010)]{ab-rosner10} Abarzhi, S. I., \& Rosner, R., 2010, Physica Scripta, T142, 014012

%\bibitem[Acton(1970)]{acton} Acton, F. S., 1970,  Numerical Methods That Work, Harper International Edition, Harper \& Row, New York
%
%\bibitem[Aerts, et al.(2010)]{aerts} Aerts, C., Christensen-Dalsgaard, J.,  Kurtz, D. W., 2010,  Asteroseismology, (Berlin: Springer)

\bibitem[Aller(1963)]{aller} Aller, L. H., 1963. Astrophysics: The Atmospheres of the Sun and Stars., 2nd. ed., Roland Press Company. New York

%%\bibitem[Andrassy, et al.(2020)]{robo} Andrassy, R., Davis, A.,  Edelmann, Goffrey, T., Guillet, T., Harpole, A., Herwig, F., Higl, J., Hirschi, R., Horst, L., LLoyd-Ronning, N., Mocak, M., Roepke, F., Vlaykov, D., Woodward, P., Zingale, M., poster, Astronomische-Gesellschaft conference, 2020

%\bibitem[Apsden, et al.(2008)]{apsden} Apsden, A. Nikiforakis, N., Dalziel,, S., \& Bell, J. B., 2008, Comm APP. Math. jand Compp. Sci., 3, 103

\bibitem[Anderson(2005)]{anderson} Anderson, J. D., 2005, Physics Today, 48, 42

\bibitem[Arnett(1968)]{wda68} Arnett, W. D., 1968, Nature, 219, 1344

%\bibitem[Arnett(1971)]{wda71} Arnett, W. D., 1971, \apj, 163, 11 %SN L(t) and preSN models


%\bibitem[Arnett \& Falk(1976)]{af76} Arnett, W. D. \& Falk, S. W., 1976, \apj, 210, 733
 
%\bibitem[Arnett(1977)]{a77Tx8} Arnett, W. D., 1977, 8th Texas Symposium on Relativistic Astrophysics, 302, Ann. N. Y. Acad. Sci., 90 % binary and/or extensive Mdot, He bare cores, Ni56

%\bibitem[Arnett(1979)]{a79} Arnett, W. D., 1979, \apjl, 230, L37

%\bibitem[Arnett(1980)]{a80} Arnett, W. D., 1980, \apj, 237, 541

%\bibitem[Arnett(1982)]{a82a} Arnett, W. D.,  1982, \apj, 253, 785

%\bibitem[Arnett(1988)]{wda88-87A} Arnett, W. D.,  1988, \apj, 331, 377

\bibitem[Arnett(1996)]{wda96} Arnett, D., 1996,  Supernovae and
Nucleosynthesis, Princeton University Press, Princeton NJ

% new windows on massive stars, 
%\bibitem[Arnett(2015)]{arnett2015} Arnett, W. D., 2015, IAUS 307, 459
%
\bibitem[Arnett, Meakin \& Young(2009)]{amy09vel} Arnett, W. D., Meakin, C., \& Young, P. A., 2009, \apj, 690, 1715 %velocity field, damping=driving but out of phase, anisotropy and KE, terms in TKE vs radius, decay of turbulence, saturation
\bibitem[Arnett \& Meakin(2010)]{am10rot} Arnett,  W. D. \& Meakin, C., 2010, IAUS 265, 106
% rotation: spun up O shell

%\bibitem[Arnett, Meakin \& Young(2010)]{amy10sub} Arnett, W. D., Meakin, C., \& Young, P. A., 2010 , \apj, 710, 1619 %subphotospheric, two parameters in MLT,
%
%\bibitem[Arnett \& Meakin(2011)]{am11real} Arnett, W. D. \& Meakin, C., 2011, \apj, 733, 78
%%realistic progenitors, 2D, shell interactions
%
\bibitem[Arnett \& Meakin(2011)]{am11turb} Arnett, W D., \& Meakin, C., 2011, \apj, 741, 33
%turbulent cells in stars: fluctuations in KE  Lorenz model, Schwarzschild cell averages
%

\bibitem[Arnett \& Meakin(2016)]{am16key} Arnett, W D., \& Meakin, C., 2016, Reports of Prog. Phys., 79, 2901
%%Key Issues Review; ropp

%\bibitem[Arnett, Meakin \&  Viallet(2014)]{amv2014} Arnett, W. D., Meakin, C. \& Viallet, M., 2014, AIP Advances, 4, 1010 %instabilities in preSN

\bibitem[Arnett, et al.(2015)]{321D} Arnett, W. D., Meakin, C. A., Viallet, M., Campbell, S. W., Lattanzio, J. C., Moc\'ak, M., 2015, \apj, 809, 30 %321D and hires; fig. 2 shows u fluctuations

\bibitem[Arnett \& Moravveji(2017)]{ehsan} Arnett, W. D., \& Moravveji, E., 2017, \apj, 836L, 19
%

\bibitem[Arnett, et al(2019)]{alvio1} Arnett, W. D., Meakin, C., Hirschi, R., Georgy, C., Campbell, S., Scott, L., Kaiser, E., Viallet, M., Moc\'ak, M., 2019,  (Paper I),  \apj, 882, 18

%\bibitem[Arnett, et al.(2019)]{wda2} W. D. Arnett, C. Meakin, R. Hirschi, A. Cristini, C. Georgy, S. Campbell, L. Scott, E. Kaiser, paper II, submitted

\bibitem[Asplund, et al.(2009)]{asplund} Asplund, M.,  Grevesse, N., Sayval, A. J., Scott, P., 2009,
\araa, 47, 481

\bibitem[Asplund, Ludwig, Nordlund \& Stein(2000)]{alns00} Asplund, M, Ludwig, H.-G., Nordlund, \AA,  Stein, R. F.. 2000, \aap, 359, 669 %numerical resolution

\bibitem[Baraffe \& El Eid(1991)]{baraffe} Baraffe, I., \& El Eid, M., 1991, \aap, 245, 548

%\bibitem[Basu \& Antia(1997)]{basu97} Basu, S. \& Antia, H.,  1997, \mnras, 287, 189

\bibitem[Batchelor(1960)]{batchelor60} Batchelor, G. K., 1960, The Theory of Homogeneous Turbulence, Cambridge University Press

%\bibitem[Bayless, et al.(2013)]{bayless} Bayless, A. J., Even, W., Frey, L. H., Fryer, C. L., Roming, P. W. A., Young, P. A. 2015, ApJ, 805, 98
%
%%%
\bibitem[Baz\`{a}n \& Arnett(1998)]{ba98} Baz\`{a}n, G., \& Arnett, D. 
1998, \apj, 494, 316
%\
%%\bibitem[Beeck, et al.(2012)]{beeck} Beeck, B., Collet, R., Steffen, M., Asplund, M., Cameron, R. H., Freytag, B., %%Hayek, W., Ludwig, H.-G., \& Sch\"ussler, M.. 2012, \aap, 539A, 121B 

%\bibitem[Bekki, Hotta \& Yokoyama(2017)]{bekki} Bekki, Y., Hotta, H., \& Yokoyama, T., 2017, \apj 851, 74

\bibitem[Benzi \& Frisch(2010)]{benzi} Benzi, R., and Frisch, U., 2010, Turbulence, Scholarpedia, 5(3):3439

%\bibitem[Bianco, et al.(2014)]{bianco} Bianco, F. B., Modjaz, M., Hicken, M., Friedman, A., Kirshner, R. P.,
%Bloom, J. S.,  Challis, P., Marion, G. H., Wood-Vasey, W. M., \& Rest, A., 2014, \apjs, 213, 19

\bibitem[Biermann(1925)]{biermann25} Biermann, L. 1925, Zs. Angew. Math. 5, 136

%\bibitem[Biermann(1932)]{biermann32} Biermann, L. 1932, Z. A. 5, 117
% astronomiche nachtrichten eq. 19 for v^3
%\bibitem[Biermann(1938)]{biermann} Biermann, L., 1938, Astron. Nachr. 264, 361

%
%\bibitem[Blondin, et al.(2013)]{sblondin} Blondin, S., Dessart, L., Hillier, D. J., Khokhlov, A. M., 2013, \mnras, 429, 2127
%
%
\bibitem[B\"ohm-Vitense(1958)]{bv58} B\"ohm-Vitense, E., 1958, \zap,
46, 108
%
\bibitem[B\"ohm-Vitense(1989)]{bv89} B\"ohm-Vitense, E., 1989, Introduction to Stellar Astrophysics, Vol. 3,  Stellar Atmospheres, Cambridge University Press % p. 185,eq.  14.46 TKE

%\bibitem[Boltzmann(1964)]{boltzmann} Boltzmann, L., Lectures on Gas Theory, trans. S.  G. Brush, University of California Press, Berkeley and Los Angeles, 1964

%\bibitem[Braginskii(1958)]{braginskii} Braginskii, S. I. {\it Soviet Physics JETP}, 6, 358  

\bibitem[Brandenburg(2016)]{brandenburg} Brandenburg, A., 2016, \apj, 832, 6

\bibitem[Bressan, et al.(2012)]{bressan} Bressan, A.; Marigo, Paola; Girardi, L\`eo; Salasnich, Bernardo; Dal Cero, Claudia; Rubele, Stefano; Nanni, Ambra; 2012, \mnras, 427, 127B

 \bibitem[Brown, Vasil \& Zweibel(2012)]{sproof} Brown, D. P., Vasil, G. M., \& Zweibel, E. G. \apj, 756, 109
 
\bibitem[Buckmaster \& Vicol(2019)]{buckmaster} Buckmaster, T., \& Vicol, V., 2019, arXiv:1901.09023v2, Convex integration and phenomenologies in turbulence

%\bibitem[Buldgen, et al.(2017)]{buldgen} Buldgen, G., Salmon, S. J. A. J., Noels, A., et al., 2017, \aap, 607, 58 
% improving SSM
%
%\bibitem[Campbell, et al.(2016a)]{simon16a}Campbell, S. W., Constantino, T. N., D'Orazi, V., Mealin, C., Stello, D., Christensen-Dalsgaard, J., Kuehn, C., De Silva, G. M., Arnett, W. D., Lattanzio, J. C., MacLean, B. T., 2016, 2015arXiv151204774C, to appear in Astronomische Nachrichten special issue " Reconstructing the Milky Way's History''
\bibitem[Campbell, Lugaro \& Karakas(2010)]{campbell10} Campbell, S. W., Lugaro, M., \& Karakas, A., 2010, \aap 522, 6

%\bibitem[Canuto \& Mazzitelli(1991)]{cm91} Canuto, V. M. \& Mazzitelli, I., 1991, \apj, 370, 295

%\bibitem[Canuto(1992)]{can92} Canuto, V. M., \apj, 392, 218

%\bibitem[Canuto, Goldman \& Mazzitelli(1996)]{can96} Canuto, V. M., Goldman, I.,
%\& Mazzitelli, I., \apj, 473, 550

%\bibitem[Canuto(2000)]{can00} Canuto, V. M. 2000, \apj, 571, L79

%\bibitem[Canuto(2002)]{can02} Canuto, V. M. 2002, \aap, 384, 1119

%\bibitem[Canuto(1999)]{can99} Canuto, V. M. 1999, \apj, 524, 311
%
%\bibitem[Canuto(2011a)]{can11a} Canuto, V. M. 2012, \aap, 528, A76
%
%\bibitem[Canuto(2011b)]{can11b} Canuto, V. M. 2012, \aap, 528, A77
%
%\bibitem[Canuto(2011c)]{can11c} Canuto, V. M. 2012, \aap, 528, A78
%
%\bibitem[Canuto(2011d)]{can11d} Canuto, V. M. 2012, \aap, 528, A79
%
%\bibitem[Canuto(2011e)]{can11e} Canuto, V. M. 2012, \aap, 528, A80

\bibitem[Castsaing, Gunaratne, Heslot,  et al.(1989)]{castaing} Castaing, B., Gunaratne, G., Heslot, F., \etal(1989)]{castaing} Castaing, B., Gunaratne, G., Heslot, F., \etal, 1989, J. Fluid Mech. 201, 1

%\bibitem[Castro, \etal(2014)]{castro} Castro, N., Fossati, L., Langer, N., Sim\'on-Diaz, S., Schneider, F. R. N., \& Izzard, R. G., 2018, \aap, 570, L13

\bibitem[Chandrasekhar(1939)]{chandra} Chandrasekhar, S., 1939, An Introduction to the study of Stellar Structure, University of Chicago Press, Chicago

%\bibitem[Chevalier(1976)]{cheval} Chevalier, R. A., 1976, \apj, 208, 826
%
%\bibitem[Chevalier \& Kirshner(1978)]{ck-casA} Chevalier, R. A. \& Kirshner, R. P., 1978, \apj, 219, 931
%
%\bibitem[Chevalier \& Fransson(2008)]{CF08} Chevalier, R. A. \& Fransson, C. 2008, \apj, 683, L135
%
%\bibitem[Clayton(1968)]{ddc} Clayton, D. D., 1968, Principles of Stellar Evolution and Nucleosynthesis, McGraw-Hill, New York
%
%\bibitem[Colgate \& McKee(1969)]{colgate} Colgate, S. A., 1969, \apj, 157, 623
%
%\bibitem[Colgate, Petschek \& Kriese(1980)]{cpk80} Colgate, S.  A.,  Petschek, A. G., \& Kriese, J. T., 1980, \apjl, 237, L133
%

\bibitem[Colella \& Woodward(1984)]{cw84} Colella, P., \& Woodward, P., 1984, J. Comp. Phys., 54, 174

\bibitem[Constantin \& Titi(1994)]{constantin} Constantin, P., \& Titi, E. S., 1994, Commun. Math. Phys. 165, 207

%\bibitem[Couch, et al.(2015)]{couch2015} Couch, S. M., Chatzopoulos, E., Arnett, W. D., \& Timmes, F. X., 2015,
%\apj, 808, 21 %Si shell and core collapse
%
%\bibitem[Cox(1980)]{cox1980} Cox, J. P., 1980, Theory of Stellar Pulsation, Princeton University Press
%
%\bibitem[Cristini, et al.(2015)]{andrea2015} Cristini, A., Hirschi, R., Georgy, C., Meakin, C., Arnett, D., \& Viallet, M., 2015, IAUS 307, 459 %C shell

%
\bibitem[Cristini, et al.(2017)]{andrea} Cristini, A., Meakin, C., Hirschi, R., Arnett, W. D., Georgy, C., Viallet, M., Walkington, I.,  2017, \mnras, 471, 279
%
%
\bibitem[Cristini, et al.(2019)]{andrea2} Cristini, A., Hirschi, R., Meakin, C., Arnett, W. D., Georgy, C., Walkington, I., 2019,  \mnras, 484, 464.
%
%\bibitem[Davidson(2001)]{davidsonmhd} Davidson, P. A., 2001, An Introduction to Magnetohydrodynamics, Cambridge University Press

%\bibitem[Davidson(2004)]{davidson} Davidson, P. A., 2004, Turbulence, Oxford University Press

\bibitem[De Lellis \& Sze'kelyhidi(2013)]{dls13} De Lellis, C., \& Sze'kelyhidi, L., Jr., 2013, Dissipative Continuous Euler Flows, Invent., Math., 193(2):377-407

%\bibitem[Dessart, et al.(2014)]{dessart14} Dessart, L., Hillier, D. J., Blondin, S., Khokhlov, A., 2014, \mnras, 441, 3249
%
%\bibitem[Dessart, et al.(2015)]{dessart} Dessart, L., Hillier, D. J., Woosley, S. E., Livne, E., Waldman, R., Yoon, S-C., Langer, N., 2015, \mnras, 453, 2189 %snIa
%
%\bibitem[Dessart, et al.(2016)]{dessart16} Dessart, L., Hillier, D. J., Woosley, S. E., Livne, E., Waldman, R., Yoon, S-C., Langer, N., 2016, \mnras, 458, 1618 %snIbc

\bibitem[Dimotakis(2005)]{dimotakis} Dimotakis, P., 2005, Ann. Rev. Fluid Mech., 37, 329

%\bibitem[Drake(2006)]{drake} Drake, R. P., 2006,  High-Energy-Density Physics, Springer, Berlin
 
 \bibitem[Dravins \& Nordlund(1990)]{dravins90} Dravins, D. \& Nordlund, \AA., 1990, \aap 228, 203
 
%\bibitem[Drazin(2002)]{drazin} Drazin, P. G., 2002,  Introduction to Hydrodynamic Stability, Cambridge University Press, Cambridge

\bibitem[Duchon \& Robert(2000)]{duchon} Duchon, J., \& Robert, R., 2000, Nonlinearity, 13, 249

%\bibitem[Eggleton(1972)]{ppe72} Eggleton, P. P., 1972, \mnras, 156, 361
%
%\bibitem[Eggleton(1973)]{eggleton} Eggleton, P. P., 1973, \mnras, 163, 279
%
%\bibitem[Eggleton, Dearborn \& Lattanzio(2008)]{edl08} Eggleton, P., Dearborn, D., \& Lattanzio, J.,
% 2008, \apj, 677, 581

\bibitem[Ekstr\"om, et al.(2012)]{ekstrom} Ekstr\"om, S., Georgy, C., Eggenberger, P., et al., 2012, \aap, 537, A146
%
 
%\bibitem[Endal \& Sofia(1976)]{endal} Endal, A. S., \& Sofia, S., 1976, \apj, 210, 184
%
% review
\bibitem[Eyink(2018)]{eyink-2018} Eyink, G., 2018, arXiv:1803.02223
% compressible fluid turbulence
\bibitem[Eyink \& Drivas(2018a)]{eyink-comp} Eyink, G. \& Drivas, T. D.,  2018, Phys. Rev. X 8.011022
% relativistic fluid
\bibitem[Eyink \& Drivas(2018b)]{eyink-rel} Eyink, G. \& Drivas, T. D.,  2018, Phys. Rev. X 8.011023
%
\bibitem[Eyink \& Sreenivasan(2006)]{eyink} Eyink, G. L. \& Sreenivasan, K. R., 2006, Rev. Mod. Phys., 78, 87
%
%\bibitem[Falk(1978)]{falk-break} Falk, S. W., 1978, \apj, 226, L133
%
%\bibitem[Falk \& Arnett(1973)]{fa73} Falk, S. W., \& Arnett, W. D., 1973, \apj, 180, L65
%
%\bibitem[Falk \& Arnett(1977)]{syd} Falk, Sydney W. \& Arnett,  W. David, 1977, \apjs, 33, 515
%%
\bibitem[Falkovich \& Sreenivasan(2006)]{falkovich}  Falkovich, G., \& Sreenivasan, K., 2006, Phys. Today, 59, 43.
%%% 
%\bibitem[Featherstone \& Hindman(2016)]{featherstone} Featherstone, N. A., \& Hindman, B. W., 2016, \apj, 830, 15
 
%\bibitem[Fesen \& Milisavljevic(2016)]{fesen2016} Fesen, R. A., \& Milisavljevic, D., 2016, \apj, 818, 17


%\bibitem[Firth, et al.(2015)]{firth} Firth, F. E., Sullivan, M., Gal-Yam, A., Howell, D. A., Maguire, K., Nugent, P.,
%Piro, A. L., Baltay, C., Feindt, U., Hadjiyksta, E., McKinnon, R., Ofek, E., Rabinoowitz, D., Walker, E. S., 2015,
%\mnras, 446, 3895re
%
%\bibitem[Freytag, Ludwig, \& Steffan(1996)]{fls96} Freytag, B.,
%Ludwig, H.-G., \& Steffan, M., 1996, \aap, 313, 497

\bibitem[Freytag(2013)]{freytag2013} Freytag, B., 2013, Mem. S.A.It. Suppl. 24, 26

%
\bibitem[Frisch(1995)]{frisch} Frisch, U., 1995,  Turbulence,
Cambridge University Press, Cambridge
%
\bibitem[Fryxell, M\"uller \& Arnett(1989)]{fma89} Fryxell, B. A., M\"uller, E., \& Arnett, W. D., 1989, 
%
\bibitem[Fujimoto, Ikeda \& Iben(2000)]{fujimoto00} Fufimoto, M., Ikeda, Y., \& Iben, Icko, Jr., 2000, \apj 529, L25

%\bibitem[Gabriel, et al.(2014)]{gabriel} Gabriel, M., Noels, A., Montalb\'an, J., Miglio, A., 2014, \aap, 569, A63

%\bibitem[Gabriel \& Belkacem(2018)]{gabriel18} Gabriel, M. \& Belkacem, K., 2018, \aap, 612, 21

%\bibitem[Gagnier \& Garaud(2018)]{gagnier-garaud} Gagnier, D. \& Garaud, P., 2018, \apj, 862, 36

%\bibitem[Garaud, Gagnier \& Verhoeven(2017)]{garaud1} Garaud, p., Gagnier, D., \& Verhoeven, J.,, 2017, \apj, 837, 133

\bibitem[Gizon \& Birch(2012)]{gizon-birch} Gizon, L., \& Birch, A. C., 2012, PNAS109:11896-97

%\bibitem[Gonzalez-Gait\'an, et al.(2012)]{gg12} Gonzalez-Gait\'an, S., Conley, A., Bianco, F. B., et al., 2012, \apj, 745, 44
%

%%
%\bibitem[Gough(1968)]{gough} Gough, D. O., 1968, \aj, 72, 799
%
%\bibitem[Gough(1977)]{gough77} Gough, D. O., 1977, in  Problems of stellar convection, Proc. 38th Coloquium, Nice, France, Aug. 1976, (A78-28526 11-90) Berlin and New York, Springer-Verlag, 1977, p. 15-56

\bibitem[Grinstein, Magolin \& Rider(2007)]{iles07} Grinstein, F.,  Magolin, L. G.,  \& Rider, W., 2007, Implicit Large Eddy Simulations, Cambridge University Press

\bibitem[Greer, et al.,(2015)]{greer15} Greer, B., Hindmsn, B., Featherstone, N., \& Toomre, J., 2015, \apj, 803, L17

\bibitem[Greer, Hindman \& Toomre(2016)]{greer16} Greer, B., Hindman, B., \& Toomre, J., 2016, \apj, 824, 128

%\bibitem[Glatzel \& Kiriakidis(1993)]{glatzel} Glatzel, W., \& Kiriakidis, M., 1993, \mnras, 263, 375

%\bibitem[Hansen \& Kawaler(1994)]{hansenkawaler} Hansen, C. J., \& Kawaler, S., 1994, Stellar Interiors: Physical Principles, Structure, and Evolution, 1st. ed.,  Springer-Verlag, New York
%

%\bibitem[Hansen, Kawaler, \& Trimble(2004)]{hansenkawaler2} Hansen, C. J., Kawaler, S., \& Trimble 2004,  Stellar Interiors: Physical Principles, Structure, and Evolution, 2nd. ed.,  Springer-Verlag, New York


%\bibitem[Hanasoge et. al(2010)]{hanasoge-apj}Hanasoge, S. M., Duvall, T. L.., \& DeRosa, M. L., 2010, \apj, 712, L98

%\bibitem[Hanasoge et. al(2012)]{hanasoge-nas}Hanasoge, S. M., Duvall, T. L.., \& Sreenivasan, K. R., 2012, Proceed. lN. A. S., 109, 11928

%\bibitem[Hanasoge \& Sreenivasan(2014)]{hanasoge} Hanasoge, S. M., \& Sreenivasan, K. R., 2014, Solar Physics, 289, 3403


\bibitem[Hanasoge, Gizon \& Sreenivasan(2016)]{hanasoge-gizon} Hanasoge, S. M., Gizon, L.,\& Sreenivasan, K. R., 2016, Ann. Rev. Fluid Mech., 48, 191

 \bibitem[Haugen \& Brandenburg(2006)]{haugen2004} Haugen, N., \& Brandenburg, Axel, 2006, Phys. Rev. E, 70, 026405
 
 \bibitem[Haugen \& Brandenburg(2006)]{haugen2006} Haugen, N., \& Brandenburg, Axel, 2006, Phys. Fluids, 18, 7, 075106

 % p-C12 in Sakurai's object
%\bibitem[Herwig, \etal(2011)]{herwig11} Herwig, F., Pignatari, M., Woodward, P. R., Poerer, P. H., Rockefeller, G., Fryer, C., Bennett, M., Hirschi, R., 2011, \apj 727, 89

% global H-ignition flash 
%\bibitem[Herwig, et al.(2014)]{herwig} Herwig, F., Woodward, P. R., Lin, P.-H., Fryer, C., 2014, \apj, 792, 3
%
\bibitem[Heslot, Castaing, \& Libchaber(1987)]{libchaber} Heslot, F., Castaing, B., \& Libchaber, A., 1987, Phys. Rev. A 36, 5870

%  blowup of incompressible euler solutions
\bibitem[Hoffman \& Johnson(2008)]{hoffman} Hoffman, J. \& Johnson, Claes, 2008, BIT Numerical Mathematics, 48, 285
%
\bibitem[Holmes, Lumley \& Berkooz(1996)]{holmes} Holmes, P., Lumley, J., \& Berkooz, G., 1996, Turbulence, Coherent Structures, Dynamical Systems and Symmetry, Cambridge University Press

%\bibitem[Hotta, Lijima \& Kusano(2019)]{hotta} Hotta, H., Lijima, H., \& Kusano, K., 2019, Sci. Adv.; %, eaau2307

%\bibitem[Humphreys(1978)]{roberta} Humphreys, R. M., 1978, \apjs, 38, 309

%\bibitem[Humphreys \& Davidson(1979)]{lbv-limit}  Humphreys, R. M., \& Davidson, K., 1979, \apj, 232, 409 %HD limit

%\bibitem[Humphreys \& Davidson(1994)]{lbv}  Humphreys, R. M., \& Davidson, K., 1994, \pasp, 106, 1025 %LBV's

%\bibitem[Humphreys \& Stanek(2005)]{most}  Humphreys, R. M., \& Stanek, K., Z., 2005,  \pasp, 332 %Most massive stars
\bibitem[Isett(2019)]{isett} Isett, P., 2019, H\"older Continuous Euler Flows in Three Dimensions with Compact Support in Time, Princeton Unversity Press

\bibitem[Iyer, et al.(2018)]{iyer4096} Iyer, K. P., Schumacher, J., Sreenivasan, K. R., Yeung, P. K., \prl, 121, 26, id.264501

%\bibitem[Hoeflich, Khoklov \& Mueller(1991)]{hoeflich} Hoeflich, P., Khoklov, A., \& M\"uller, E., 1991, \aap, 248, L7
%
\bibitem[P. Johnson(2021a)]{perry} Johnson, P. L., 2021, Phys. Rev. Letters 124, 104501

\bibitem[P. Johnson(2021b)]{perry2} Johnson, P. L., 2021, Physics Today, 74, 47

\bibitem[Jones, et al.,(2017)]{jones17} Jones, S., R. Andrassy, S. Sandalski, A. Davis, P. Woodward and F. Herwig, 2017, \mnras, 465, 2991

%in press,  	arXiv:1605.03766v2 %O-shell 4pi

%\bibitem[Joyce \& Chaboyer(2018)]{joyce-chaboyer} Joyce, M., \& Chaboyer, B., 2018, \apj, 856, 10

\bibitem[Kaneda, etal(2003)]{kaneda2003} Kaneda, Y., Ishihara, T., Yokokawa, K., \& Uno, A., 2003, Physics of Fluids, 15, L21

\bibitem[Kippenhahn \& Weigert(1990)]{kippen} Kippenhahn, R. \& Weigert, A. 
1990,  Stellar Structure and Evolution, Springer-Verlag

%\bibitem[Kirshner \& Chevalier(1977)]{kc-casA} Kirshner, R. P. \& Chevalier, R. A., 1977, \apj, 218, 142

%
%\bibitem[Kitiashvili, et al.(2016)]{kitiashvili} Kitiashvili, I N., Kosovichev, A. G., Mansour, N. N., \& Wray, A. A. 2016, arXiv:1512.07298v1
%
\bibitem[Kolmogorov(1941)]{kolmg41} Kolmogorov, A. N., 1941, Dokl. 
Akad. Nauk SSSR, 30, 299
%
%\bibitem[Kolmogorov(1962)]{kolmg}  Kolmogorov, A. N., 1962, J. Fluid 
%Mech., 13, 82
%
\bibitem[Kritsuk, et el.(2011)]{kritsuk} Kritsuk, A., Nordlund, A., Collins, D., Padoan, P., Norman. M., Abel, T., Banerjee. R., Federrath, C., Flock, M., Lee, D., Li, P. S., M\"uller, W.-C., Teyssier, R., Ustyugov, S., Vogel, C., Xu, H., 2011, \apj, 735, 1


\bibitem[Landau \& Lifshitz(1959)]{llfm} Landau, L. D. \& Lifshitz, E. M., 1959, Fluid Mechanics,
Pergamon Press, London. 
%

%\bibitem[Landau \& Lifshitz(1969)]{llsp} Landau, L. D. \& Lifshitz, E. M., 1969, Statistical Physics, 2nd Ed.,
%  Addison--Wesley, Reading MA.

%\bibitem[Lantz et al.(2004)]{Lantz04} Lantz, B., Aldering, G., Antilogus, P., et al.\ 2004, \procspie, 5249, 146 
%
%\bibitem[Lecoanet \& Jeevanjee(2019)]{lecoanet2019} Lecoanet, D. \& Jeevanjee, N., Quat. Journ. Royal Meteorol. Soc., in press..

\bibitem[Leveque(2002)]{leveque} Leveque, R. J., 2002, Finite Volume Methods for Hyperbolic Problems, Cambridge University Press, Cambridge,  UK

%\bibitem[Lighthill(1978)]{lighthill} Lighthill, J., 1978, Waves in Fluids, Cambridge University Press, Cambridge, UK

%\bibitem[Magee et al.(1995)]{magee95} Magee, N.~H., Abdallah, J., Jr.,
%  Clark, R.~E.~H., Cohen, J.~S., Collins, L.~A., Csanak, G., Fontes,
%  C.~J., Gauger, A., Keady, J.~J., Kilcrease, Merts, A.~L. 1995, in
%  ASP Conf. Ser. 78, Astrophysical Applications of Powerful New
%  Databases, ed. S. J. Adelman \& W. L. Wiese (ASan Francisco, CA:
%  ASP). 51
%
%Landau, L. D. \& Lifshitz, E. M., 1969, Statistical Physics, 2nd Ed.,
%Addison--Wesley, Reading MA. 
%

%\bibitem[Lorenz(1963)]{lorenz} Lorenz, E. N., 1963, Journal of Atmospheric
%Sciences, 20, 130

\bibitem[Ludwig, Freytag, \& Steffen(1999)]{lfs99} Ludwig, H.-G., Freytag, B., \& Steffen, M., 1999, \aap, 346, 111
%calibration of ML
%
\bibitem[Ludwig \& Kucinskas(2012)]{ludwig-rg} Ludwig, H.-G. \& Kucinskas, A., 2012, \aap, 547, A118
%
\bibitem[Maeder(1999)]{maeder} Maeder, A., 1999,  Physics, Formation and Evolution of Rotating Stars,
Springer, Berlin
%
%\bibitem[Maeder, Meynet, \& Hirschi(2005)]{mmh} Maeder, A., Meynet, G., \& Hirschi, R., 2005, in  The Fate of the Most Massive Stars, ed. R. Humphreys \& K. Stanek, \pasp Conf. 332, 3

% magic1
\bibitem[Magic, et al.(2013)]{magic2013} Magic, Z.; Collet, R.; Asplund, M.; Trampedach, R.; Hayek, W.; Chiavassa, A.; Stein, R. F.; Nordlund, A.; 2013, \aap, 557, 26 
% magic1b
\bibitem[Magic, Collet, Hayek, \& Asplund(2013)]{magic2013b} Magic, Z.; Collet, R.; Hayek, W.; Asplund, M., 2013b, \aap, 560, 8
% magic2
\bibitem[Magic, Weiss, \& Asplund(2015)]{magic2015} Magic, Z., Weiss, A. \& Asplund, M., 2015, \aap, 573, 89
% magic3
 \bibitem[Magic, Weiss, \& Asplund(2016)]{magic2016} Magic, Z., Weiss, A. \& Asplund, M., 2016, \aap, 586, 88
% solar
\bibitem[Magic \& Weiss(2016)]{magic-solar} Magic, Z.; \& Weiss, A.; 2016, \aap, 592, A26

%\bibitem[Manneville(2010)]{manneville} Manneville, P., 2010,  Instabilities, Chaos, and Turbulence, Imperial College Press, London

%
%\bibitem[Meakin \& Arnett(2006)]{ma06} Meakin, C. A., \& Arnett, W. D., 2006, \apj, 637, 53
%%C and O shells in 2D
%
%\bibitem[Meakin \& Arnett(2007c)]{ma07c} Meakin, C. A., \& Arnett, W. D., 2007, IAUS, 239, 296
%%IAU Si,O,Ne,C in 2D
%
\bibitem[Meakin \& Arnett(2007a)]{ma07a} Meakin, C. A., \& Arnett, W. D., 2007, \apj, 665, 690
%%anelastic and compressible O+O
%
\bibitem[Meakin \& Arnett(2007b)]{ma07b} Meakin, C. A., \& Arnett, W. D., 2007, \apj, 667, 448
% first 3d O+O shell paper
\bibitem[Meakin \& Arnett(2010)]{ma10apss} Meakin, C. A., \& Arnett, W. D., 2010, \apss, 328, 22
%apss driving regions top and bottom

\bibitem[Miesch, et al.(2009)]{miesch09} Miesch, M. S., Browning, M.K., Brun, A.S., Toomre, J., Brown, B.P., 2009.  Solar-Stellar Dynamos as Revealed by Helio- and Asteroseismology, eds. M. Dikpati, T. Arentoft, I. Gonz\'alez Hern\'andez, C. Lindsey, F. Hill, \pasp\, Conference Series, 416

%\bibitem[Miesch, et al.(2012)]{miesch} Miesch, M., Featherstone, N., Rempel, M., \& Trampedach, R., 2012, \apj, 757, 128

%
%:
%
%\bibitem[Moc\'ak et al.(2011)]{mocak11} Moc\'ak, M., Meakin, C.,  M\"uller, E., \& Siess, L., 2011, \apj, 743, 55
%
%\bibitem[Modjaz, et al.(2008)]{modjaz} Modjaz, M., Li, W., Butler, N., et al., 2009, \apj, 702, 226
%
%
% \bibitem[M\"uller, Fryxell, B. \& Arnett, D.(1991)]{mfa91} M\"uller, E., Fryxell, B. \& Arnett, D., 1991, \aap, 251, 505
% 
%\bibitem[Munari, et al.(2013)]{munari} Munari, U., et al., 2013, New Astronomy, 20, 30
%
\bibitem[Moc\'ak et al.(2010)]{mocak10} Moc\'ak, M., Campbell, S W..,  M\"uller, E., \& Kifonidis, K., 2010, \aap, 510, 114

\bibitem[Moc\'ak, et al.(2014)]{miro2014} Moc\'ak, M., Meakin, C., Viallet, M., \& Arnett, D., 2014, arXiv, 1401.5176 %RANS details

\bibitem[Moc\'ak, et al.(2018)]{miro} Moc\'ak, M., Meakin, C., Arnett, D., \& Campbell, S., 2018, \mnras, 481, 2918 %O-Ne shells

%\bibitem[Montalb\'an, et al.(2013)]{montalban} Montalb\'an, J., Miglio, A., Noels, A., et al. 2013, \apj, 766, 118

%\bibitem[Moravveji, et al.(2016)]{ehsan16a}Moravveji, E. 
%%
\bibitem[Mosumgaard, et al.(2018)] {mosum}Mosumgaard, J. R., Ball, W. H., Aquirre, V. S., Weiss,  A., Christensen-Dalsgaard, J., 2018, \mnras, 478, 5650

\bibitem[Muthsam, Kupka \& L\"ow-Baselli, \etal(2010)]{muthsam} Muthsam, H. J., Kupka, F., L\"ow-Daselli, B., \etal, 2010, New A. 15, 460

%\bibitem[Nadyozhin(1994)]{nadyozhin} Nadyozhin, D. K., 1994, \apjs, 92, 527
%
%\bibitem[Nugent, et al.(2011)]{nugent2011} Nugent, P. E.,  Sullivan, M., Cenko, S. B., et al., 2011, Nature, 480, 344

\bibitem[Nordlund, Stein, \& Asplund(2009)]{nsa} Nordlund, A., Stein, R., \& Asplund, M., 2009,
\url{http://ww.livingreviews.org/lrsp-2009-2}
%
%\bibitem[Olling, et al.(2015)]{keplersn} Olling, R., Mushotzky, R., Shaya, E., Rest, A., Garnavich, P., Tucker, B.,
%Kasen, D., Margheim, S., \& Filippenko, A., 2015, Nature, 521, 332

%\bibitem[O'Mara, et al.(2018)]{omara} O'Mara, B., Miesch, M. S., Featherstone, N. A., Agustson, K. C., 2018, arXiv:1603.06107v1,
%Advances in Space Research, submitted

\bibitem[Onsager(1949)]{onsager} Onsager, L., 1949, {\it Statistical Hydrodynamics}. Suppl. al Il Nuovo Cimento, 6, 279

% convective free-fall
%\bibitem[Orvendahl, et al.(2018)]{orvendahl} Orvendahl, R., Calkins, M., Featherstone, N., \& Hindman, B., 2018, \apj, submitted

%
%\bibitem[Owocki(2005)]{owocki} Owocki, S., 2005,  The Fate of the Most Massive Stars, ed. R. Humphreys \& K. Stanek, \pasp Conf. 332, 169


\bibitem[Parker(1979)]{parker} Parker, E. N., 1979, Cosmic Magnetic Fields, Clarendon Press, Oxford

%\bibitem[Pereira, et al.(2013)]{sn2011fe} Pereira, R., Thomas, R. C., Aldering, G., et al., 2013, \aap, 554, A27
%
%\bibitem[Phillips(1993)]{phillips93} Phillips, M., 1993, \baas, 182, 2907
%
%\bibitem[Pinto \& Eastman(2000a)]{pe00a} Pinto, P. A., \& Eastman, R., 2000a, \apj, 530, 744
%
%\bibitem[Pinto \& Eastman(2000b)]{pe00b} Pinto, P. A., \& Eastman, R., 2000b, \apj, 530, 757
%
%\bibitem[Pinto \& Eastman(2000c)]{pe00c} Pinto, P. A., \& Eastman, R., 2000c, New Astronomy, 6, 307
%%

\bibitem[Pope(2000)]{pope} Pope, S. B., 2000,  Turbulent Flows, 
Cambridge University Press, Cambridge, GB

%\bibitem[Popper(1959)]{popper} Popper, Karl, 1992, The Logic of Scientific Discovery, Routledge, London (Routledge Classics)

\bibitem[Porter, Anderson \& Woodward(1997)]{porterRG} Porter, D. H., Anderson, S. E., and P. R. Woodward. 1997, Access, NCSA,
\url{http://www.lcse.umn.edu/research/RedGiant/}

\bibitem[Porter \& Woodward(2000)]{porterwoodward} Porter, D. H. and P. R. Woodward. 2000, \apjs, 127, 159

\bibitem[Porter \& Woodward(2007)]{porterILES1} Porter, D. H., \& Woodward, P. R., 2007, in  Implicit
Large Eddy Simulantions, ed. F. F. Grinstein, L. G. Margolin, \& W. J.
Rider, Cambridge University Press, p. 245

\bibitem[Porter \& Woodward(2007)]{porterILES} Porter, D. H., \& Woodward, P. R., 2007, in  Implicit
Large Eddy Simulantions, ed. F. F. Grinstein, L. G. Margolin, \& W. J.
Rider, Cambridge University Press, p. 439 

\bibitem[Pratt, et al.(2017)]{pratt} Pratt, J., Baraffe, Goffrey T., Constantino, T., Viallet, M., Popov, M. V., Walder, R., Folini, D., 2017, \aap, 604, id.A125

%\bibitem[Prat, et al.(2016)]{prat-sheer} Prat, V., Guilet, J., Viallet, M., \& M\"uller, E., 2016,
%\aap, in press

%\bibitem[Prandtl(1925)]{prandtl} Z. Angew. Math. Mech. 5:136-139

%\bibitem[Radice, Couch \&  Ott(2015)]{radice15} Radice, D., Couch, S., \& Ott, C., 2015, Computational Astrophysics and Cosmology, 2, 7 (arXiv:1501.03169)

\bibitem[Rempel, \etal(2009)]{rempel} Rempel, M., Sch\"ussler, M., \& Kn\"olker, M., 2009\apj 691, 640

\bibitem[Renzini(1987)]{alvio87} Renzini, A., 1987, \aap, 188, 49

\bibitem[Richtmeyer \& Morton(1967)]{richtmeyer}  Richtmeyer, R. D., \& Morton, K. W., 1967, Difference Methods for Initial-value Problems, 2nd ed., Interscience

%
%\bibitem[Riess, et al.(1999)]{riess} Riess, A., 1999, \aj, 118, 2675 
%
%\bibitem[R\"opke, et al.(2012)]{roepke} R\"opke,  F. K., Kromer. M., Seitenzahl, I.  R., et al., 2012, \apjl, 750:L19

%
%\bibitem[Schindler, et al.(2015)]{jt2015} Schindler, J.-T., Green, E. M., \& Arnett, W. D., 2015, \apj, 806, 178
%%sdB stars
%\bibitem[Seitenzahl, Taubenberger \& Sim(2009)]{ivo2009} Seitenzahl, I. R., Taubenberger, S., \& Sim, S. A.. 2009, \mnras, 400, 531

\bibitem[Scott, et al.(2020)]{laura} Scott, L. J. A., Hirschi, R., Georgy, C., Arnett, W. D., Meakin, C., Kaiser, E. A., Ekstr\"om, S., Yusof., H., 2020, \mnras, in press
%
%\bibitem[Smith \& Arnett(2014)]{nathan2014} Smith, Nathan, \& Arnett, W. D., 2014, \apj, 785, 82
%% connections to MLT and Lorenz
%
%\bibitem[Smith(2017)]{nathan} Smith, Nathan, arXiv:1612.02006, to be published in the Supernova Handbook
%
%\bibitem[Smitka, et al.(2016)]{smitka} Smitka, M., et al., in preparation
%
%\bibitem[Soderberg et al.(2008)]{soderberg} Soderberg, A. M., Berger, E., Page, K. L, et al. 2008, Nature, 453, 469

\bibitem[Spruit(1997)]{spruit97} Spruit, H., 1997, Memorie della Societa Astronomica Italiana 68, 397

%\bibitem[Spiegel(1971)]{spiegel71}  Spiegel, E. A., 1971, \araa, 9, 323

%\bibitem[Spiegel(1972)]{spiegel72}  Spiegel, E. A., 1972, \araa, 10, 261

%\bibitem[Spitzer(1962)]{spitzer} L. Spitzer,  {\it Physics of Fully Ionized Gases}, Second Edition, Interscience Publishers, NY 

\bibitem[Sreenivasan(1979)]{sree79} Sreenivasan, K. R., Antonia, R. A. \& Britz, D., 1979, J. Fluid Mech. {\bf 94}, 745-775

\bibitem[Sreenivasan(2019)]{sree} Sreenivasan, K. R., 2019, PNAS, 116 (37) 18175-18183, 
\url{https://doi.org/10.1073/pnas.1800436115} 

% topology of convection beneath solar surface
\bibitem[Stein \& Nordlund(1989)]{sn89} Stein, R. F., \& Nordlund, A., 1989, \apj, 342, 95
%
\bibitem[Stein, Nordlund \& Kuhn(1989b)]{sn89s} Stein, R. F., Nordlund, A., \& Kuhn, J. R.,1989,  p. 318, Solar and Stellar Granulation, R. J. Rutten and G. Severino (eds.), Kluwer Academic Publishers
%
%\bibitem[Stein \& Nordlund(1998)]{sn98} Stein, R. F., \& Nordlund, A., 1998, \apj, 499, 914

%\bibitem[Stein \& Nordlund(2000)]{sn00} Stein, R. F., \& Nordlund, A., 2000, Sol. Phys. 34:91-108

%%
%\bibitem[Swisher, et al.(2015)]{swisher} Swisher, N. C., Kuranz, C. C., Arnett, D., Hurricane, O., Remington, B. A., Robey, H. F., \& Abarshi, S. I., 2015, Physics of Plasmas, 22, 102707
%%
\bibitem[Sytine, et al.(2000)]{sytine} Sytine, I. V., Porter, D. H., Woodward, P. R., Hodson, S. W., Winkler, K.-H., Journal of Computational Physics 158 (2000), 225
%
%\bibitem[Tartaglia, et al.,(2016)]{tartaglia} Tartaglia, L., Fraser, M., Sand, D. J., Valenti, S., Smartt, S. J., McCully,, C., et al., 2016, arXiv: 1611.00419v1
%%
%
\bibitem[Taylor(1937)]{gitaylor} Taylor, G. I., 1937, Journ. Aeronautical Sci., 4, 311

%%
\bibitem[Tennekes \& Lumley(1972)]{tennekes} Tennekes, H., \& Lumley, 
J. L., 1972,  A First Course in Turbulence, MIT Press, Cambridge MA
% 150 x 150 x 82 = 1,845,000  82^{4/3} ~ 356
\bibitem[Trampedach \& Stein(2011)]{tramp-stein} Trampedach, R., \& Stein, R.. F., \apj, 731, 78 
%
\bibitem[Trampedach, et al.(2014)]{tramp} Trampedach, R., Stein, R.. F., Christensen-Dalsgaard, J., Nordlund, A., Asplund, M.  2014, \mnras, 
% T(tau) relations

\bibitem[Tremblay, et al.(2015)]{co5bold-wd} Tremblay, P.-E., Ludwig, H.-G., Freytag, B., Fontaine, G., Steffen, M., and Brassard, P., 2015, \apj, 799, 142
%
%\bibitem[Tritton(1988)]{tritton} Tritton, D. J.,  Physical Fluid
%Dynamics, 2nd ed., Oxford University Press, Oxford UK
%

%\bibitem[Truran, Arnett \& Cameron(1967)]{tac67} Truran, J. W., Arnett, W. D., \& Cameron, A. G. W., 1967, Can. J. Phys., 45, 231


%\bibitem[Turner(1973)]{turner} Turner, J. S. 1973, Buoyancy Effects in Fluids, Cambridge University Press, Cambridge, UK

%\bibitem[Tzeferacos, et al.(2018)]{dqlB} Tzeferacos, P. et al., 2018, Nature Communications,  DOI: 10.1038/s41467-018-02953-2
%
%\bibitem[Unno, et al.,(1989)]{unno}Unno, W., Osaki, Y., Ando, H., Saio, H., \& Shibahashi, H. 1989, Nonradial Oscillations of Stars (2nd ed.; Tokyo: Univ. of Tokyo Press)

%\bibitem[Valenti, et al.(2008)]{valenti} Valenti, S., Benetti, S., Cappellaro, E., Patat, F., Mazzali, P., Turatto, M., Hurley, K., Maeda, K., Gal-Yam, A., Foley, R. J., Filippenko, A. V., Pastorello, A., Challis, P., Frontera, F., Harutyunyan, A., Iye, M., Kawabata, K., Kirshner, R. P., Li, W., Lipkin, Y. M., Matheson, T., Nomoto, K., Ofek, E. O., Ohyama, Y., Pian, E., Poznanski, D., Salvo, M., Sauer, D. N., Schmidt, B. P., Soderberg, A., \& Zampieri, L., 2008, \mnras, 383, 1485

\bibitem[Vassilicous(2015)]{vassilicous} Vassilicous, J. C., 2015, Ann. Rev. Fluid Mechanics, 47:95-114

\bibitem[Van Dyke(1982)]{vandyke} Van Dyke, M., 1082, An Album of Fliud Motion, Parabolic Press, Stanford, CA

%MUSIC code explication
\bibitem[Viallet, et al.(2011)]{viallet2011} Viallet, M., Baraffe, I., \& Walder, R., 2011,
\aap, 531, 86
%RG plus medres O paper II
\bibitem[Viallet, et al.(2013)]{viallet2013} Viallet, M., Meakin, C., Arnett, D., Moc\'ak, M., 2013, \apj, 769, 1

%\bibitem[Viallet, et al.(2015)]{viallet2015} Viallet, Maxime, Meakin, C., Prat, V., \& Arnett, D., 2015, \aap, 580, 61 %Prandl number

%\bibitem[Vinyoles, et al.(2017)]{serenelli} Vinyoles,, N., Serenellii, A., et. al., 2017, \apj,  835, 202

%\bibitem[Vitense(1953)]{vitense53} Vitense, E., 1953, \zap, 32, 135
%

%\bibitem[Waxman et al.(2007)]{W07} Waxman, E., M?esza?ros, P., \& Campana, S. 2007, \apj, 667, 351

\bibitem[Warhaft(2002)]{warhaft} Warhaft, Z., 2002, PNAS 99, 2481 

%\bibitem[Weinberg(1972)]{weinberg} Weinberg, S., Gravitation and Cosmology, John Wiley \& Sons, N.Y.
%
%
%\bibitem[Woodward \& Colella(1984}]{wood-colella}Woodward, P. R. \& Colella, P, 1984, \J. Comput. Phys. 54, 115

%\bibitem[Woodward et al.,(2001)]{woodward01} Woodward, P. R., Porter, D., Sytine, I., Anderson, S. E.,
%Mirin, A. A., Curtis, B. C., Cohen, R. H., Dannevik, W. P., Dimits, A. M., Eliason, D. E.,
%Winkler, K-H., and Hodson, S., Turbulent Flows, World Scientific, 2001

\bibitem[Woodward(2007)]{woodward} Woodward, P. R., 2007, in  Implicit
Large Eddy Simulantions, ed. F. F. Grinstein, L. G. Margolin, \& W. J.
Rider, Cambridge University Press, p. 130 
%
%PPB He-flash AGB stars
\bibitem[Woodward et al.,(2009)]{woodward09} Woodward, P. R., Porter, D., Herwig, F., Fuchs, T., Nowatzki, A., Pignatari, M., 2007, AIPC, 990, 300, First Stars II Conference %arXiv:0711.2091  
% 

\bibitem[Woodward et al.(2015)]{woodward2015} Woodward, P. R., Herwig, F., Lin, P-H.,
2015, \apj, 798, 49

\bibitem[Woodward(2006)]{ppb} Woodward, P. R., 2006, PPB, the Piecewise-Parabolic Boltzmann Scheme for Moment-Conserving Advection in 2 and 3 Dimensions, \url{lcse.umn.edu}

%\bibitem[Woosley \& Weaver(1986)]{ww86} Woosley, S. E. \& Weaver, T. A., 1986, \araa, 24, 205

%\bibitem[Wood \& Arnett(2011)]{wood11} Wood, P., \& Arnett, D., 2011, Astron. Soc. of the Pacific Conf. 445, 183
%% pulsating stars

\bibitem[Wu \& Libchaber(1992)]{wu92} Wu, X.-Z., \& Libchaber, A., 1992, Phys. Rev A. 45, 842

\bibitem[Young, et al.(2001)]{young} Young, P. A., Mamajek, E. E., Arnett, D., Liebert, J., \apj, 556, 230

%\bibitem[Zheng, et al.(2013)]{2013dy} Zheng, W., et al., \apjl, 778, L15

%\bibitem[Zingale, et al.(2011)]{zingale-maestro11} Zingale, M., Nonaka, A., Almgren, A. S., Bell, J. B., Malone, C. M., \& Woosley, S. E., 2011, \apj, 740, 8

\end{thebibliography}
\end{document}